\documentclass[review]{elsarticle}
\usepackage[margin=2.0cm]{geometry}
\usepackage{lineno,hyperref}
\usepackage{subfig}
\usepackage{longtable}
\usepackage{caption}
\modulolinenumbers[5]
\usepackage{algorithmic}
\usepackage{algorithm}
\usepackage{amsmath}
\usepackage{longtable}
\usepackage{multicol}
\usepackage{multirow}
\usepackage{graphicx}
\usepackage{lscape}

\usepackage{fancyhdr}
\journal{Journal of \LaTeX\ Templates}

\makeatletter \def\ps@pprintTitle{  \let\@oddhead\@empty  \let\@evenhead\@empty  \def\@oddfoot{\hfill\thepage}  \def\@evenfoot{\thepage\hfill}} \makeatother

\begin{document}

\begin{frontmatter}

\title{A Dependable Hybrid Machine Learning Model for Network Intrusion Detection}

\author[1]{Md. Alamin Talukder}
\ead{mdalamintalukdercsejnu@gmail.com}
\author[2]{Khondokar Fida Hasan}
\ead{fida.hasan@qut.edu.au}
\author[1]{Md. Manowarul Islam}
\ead{manowar@cse.jnu.ac.bd}
\author[1]{Md Ashraf Uddin}
\ead{ashraf@cse.jnu.ac.bd}
\author[1]{Arnisha Akhter}
\ead{arnisha@cse.jnu.ac.bd}
\author[4]{Mohammad Abu Yousuf}
\ead{yousuf@juniv.edu}
\author[3]{Fares Alharbi}
\ead{faalhrbi@su.edu.sa}
\address[1]{Department of Computer Science and Engineering, Jagannath University, Dhaka, Bangladesh}
\address[2]{Information Security Discipline, School of Computer Science, Queensland University of Technology (QUT), 2 George Street, Brisbane 4000, Australia}
\address[4]{Institute of Information Technology, Jahangirnagar University, Savar, Dhaka, Bangladesh}
\address[3]{ Computer Science Department, Shaqra University, Shaqra 15526, Saudi Arabia}

\author[5]{Mohammad Ali Moni}
\ead{m.moni@uq.edu.au}

\address[5]{Artificial Intelligence \& Data Science, School of Health and Rehabilitation Sciences, Faculty of Health and Behavioural Sciences, The University of Queensland St Lucia, QLD 4072, Australia.}

\cortext[mycorrespondingauthor]{Mohammad Ali Moni}
\cortext[mycorrespondingauthor]{Md Manowarul Islam}

\begin{abstract}
Network intrusion detection systems (NIDSs) play an important role in computer network security. There are several detection mechanisms where anomaly-based automated detection outperforms others significantly. Amid the sophistication and growing number of attacks, dealing with large amounts of data is a recognized issue in the development of anomaly-based NIDS. However, do current models meet the needs of today's networks in terms of required accuracy and dependability? In this research, we propose a new hybrid model that combines machine learning and deep learning to increase detection rates while securing dependability. Our proposed method ensures efficient pre-processing by combining SMOTE for data balancing and XGBoost for feature selection. We compared our developed method to various machine learning and deep learning algorithms in order to find a more efficient algorithm to implement in the pipeline. Furthermore, we chose the most effective model for network intrusion based on a set of benchmarked performance analysis criteria. Our method produces excellent results when tested on two datasets, KDDCUP'99 and CIC-MalMem-2022, with an accuracy of 99.99\% and 100\% for KDDCUP'99 and CIC-MalMem-2022, respectively, and no overfitting or Type-1 and Type-2 issues.
\end{abstract}


\end{frontmatter}

\section{Introduction}
\label{intro}

Internet-based computer networks are becoming more vulnerable to security threats. The constant emergence of new types of threats makes developing dependable and adaptable security strategies a critical issue. Important data is always a target for attackers, making it vulnerable to concentrated network attacks. The process by which an attacker gains access to a system or system server and then sends malicious packets to the user system in order to steal, modify, or corrupt any sensitive or vital data is referred to as intrusion. An attack is defined as the unauthorized transmission of network packets or malicious conduct over a network. Existing system vulnerabilities, such as user error, misconfiguration, or software flaws, may allow the intrusion to occur on the server or system. An intelligent intrusion can also be carried out by combining various system vulnerabilities.

In a global network, a vast number of online services and millions of massive servers are active. As a result, as these networks become more appealing to attackers, they require intrusion detection systems (IDS) to secure their network systems. According to the 2020 Trustwave Global Security Report, phishing and social engineering attacks accounted for 50\% of incidents that affected corporate networks, e-commerce (22\%) systems, cloud (20\%) systems, and point-of-sale (5\%) systems. 86\% of intrusion detection occurrences in 2019 were related to external events or internet usage \citep{2020}.

IDS are typically classified into different categories based on the user's perspective. For example, IDS can be host-based (HIDS) or network-based (NIDS) \citep{sarker2020cybersecurity}. NIDS detects infiltration by monitoring network traffic and hosts connecting to the network. HIDS examines device calls, file system modifications, application logs, and host activity to detect an intrusion.

On the other hand, the type of analysis performed determines whether an intrusion detection system is signature-based (SIDS) or anomaly-based (NIDS). Signature-based schemes (also known as misuse-based schemes) look for predefined patterns called signatures in the analyzed data. For this reason, a prior signature database corresponding to known assaults is specified. On the other hand, anomaly-based detectors attempt to estimate the protected system's "normal" behavior and emit an anomaly warning whenever the difference between a given observation at a given instant and the typical behavior exceeds a predetermined threshold. Another option is to model the system's "abnormal" behavior and set off an alarm when the difference between observed and expected behavior falls below a certain threshold.
SIDS detects intrusions based on specific patterns in malware transmission, such as the number of bytes and the order of malicious blueprints. The most recent malware attack patterns, on the other hand, are frequently unknown and unpredictable. As a result, SIDS models trained on out-of-date datasets are unable to detect new malware attacks.
AIDS, on the other hand, detects unknown malware by comparing incoming packets to the model’s previous knowledge and classifying them as suspicious or not. Machine learning has become much more popular for developing models. However, it is becoming increasingly difficult to accommodate large amounts of additional information \citep{dash2019big, hasan2021security} in many fields, including machine learning, data analysis, and text mining \citep{wei2019text, humayn2021explainable}.

Machine Learning is an effective method for obtaining pattern information from well-defined inputs without depending on an algorithmic method to develop meaningful outputs \citep{mandal2020improved,hasan2019cognitive}. Machine Learning-based intrusion detection systems aid in detecting anomalous behavior of internet-connected devices and are gaining popularity. 
Several processes and methods using machine learning are already presented for detecting network intrusion.
And due to their superior ability to detect intrusions and perform generalizations, effective IDS are typically created nowadays by utilizing machine learning-based approaches. 
However, the implementation of such systems is inherently difficult. The system's intrinsic challenges can be divided into different problem groups based on competence, accuracy, and dependability. 
For example, in comparison with earlier detection methods that relied on malicious signatures, machine learning-based IDS, and particularly those strategies that are based on anomaly detection, often show a larger percentage of false-positive occurrences. 
In addition, compared to existing approaches, the system's learning process takes a significant amount of training data and is quite sophisticated. 
Another issue is the irrelevant and excessive information that increases the complexity of the dimension and hinders accurate classification predictions, which negatively affects the performance of the system \citep{tomar2021dimensionality, ayesha2020overview}. Such impact of big data on dimensionality causes poor performance, a high frequency of false positives, and other issues that reduce the performance and reliability of the model \citep{garg2018hyclass}.
Overall, new and evolving threats to these applications are always appearing, necessitating the need for reliable and advanced protection solutions. The classic machine learning-based IDS must be updated to meet the security needs of the present sustainable environment due to the quick proliferation of the network's security needs and changing threat kinds.

This paper presents a dependable IDS model that outperforms a number of current methods using a hybrid approach utilizing machine learning (ML) and deep learning (DL) algorithms. 
In this hybrid technique, a deep learning-based feature selection has been used to reduce dimensionality. By eliminating superfluous features and filtering out unnecessary data, it reduces the complexity of the data and its dimensions. Additionally, it helps to lessen the workload associated with computing while simultaneously enhancing detecting capabilities \citep{li2021lnnls}. 
Additionally, to balance the dataset and improve the effectiveness of minority incursions, which are motivated by a significant class imbalance in the intrusion dataset \citep{tan2019wireless, ahmed2022network}, synthetic minority oversampling technology (SMOTE) \citep{gonzalez2020synthetic} is utilized.
Overall, the suggested model makes use of dimensionality reduction, feature selection, and feature extraction to compress the extracted information. 
In order to reduce the dimension of imbalanced datasets, we use pre-processing techniques in our study to develop a productive and embodied architecture. Normalization, data balancing, and dimension reduction are necessary for an unbalanced dataset. In order to achieve our computations in less time with adequate prediction performance for both binary and multilabel classifications, SMOTE is used to balance our datasets, and XGBoost is used as a feature selection strategy to lower the dimension.

A variety of ML classifiers, including Random Forest (RF), Decision Tree (DT), K-Nearest Neighbor (KNN), Multilayer Perceptron (MLP), Convolution Neural Network (CNN), and Artificial Neural Network (ANN), were used to analyze the performance of our proposed model. 
For each classifier's training, a small subset of the dataset's most crucial features was chosen using the XGBoost algorithm. 
The performance parameters used to assess the model include accuracy, precision, recall, f1-score, AUC score, ROC Curve, MAE, MSE, and RMSE. The performance analysis showed that the proposed model could detect attacks with an accuracy above 99.9\% for all ML classifier algorithms. 

The overall contributions of the paper are summarized as follows:
\begin{itemize}

\item We proposed a novel hybrid approach and showed how reliable it is for detecting network intrusion by interpreting the dependability in metrics of accuracy, availability, and scalability of the model.

\item XGBoost for feature selection, SMOTE for data balancing, and proper preprocessing, we developed our own hybrid approach that uniquely outperforms the state-of-the-art models for network intrusion detection.

\item Finally, we used a number of performance indicators to assess how well the model can perform; the results show that our hybrid model is superior when compared with the existing model in detecting intrusions, resulting in lower type-1 (False Positive) and type-2 (False Negative) rates.
\end{itemize}

The succeeding units provide a summary of the existing work, our proposed model, and the results. Initially, the related work is presented in Section 2. The proposed methodology is described in Section 3, followed by presenting the details of our datasets, experiments, and findings in this paper in Section 4. Finally, the conclusion and proposed future work are presented in Section 5.

\section{Related Works}
\label{Related}

In recent years, a number of intrusion detection and prevention techniques have been proposed. To solve the problem of class imbalance as well as improve intrusion detection efficacy. In \cite{tan2019wireless}, the authors developed a strategy for balancing the data while using the SMOTE, subsequently training with the RF algorithm. The experiments were performed on a benchmark KDDCUP'99 intrusion dataset, and the RF algorithm's accuracy was 92.39\%, which was greater than that of other comparable techniques. The RF combined with the SMOTE was used to have an accuracy of 92.57\% upon resampling the minority samples.

To increase the effectiveness and correctness of IDS, the authors \citep{bhati2021improved} proposed an ensemble-based IDS leveraging XGBoost. They demonstrated that using XGBoost alongside ensemble-based IDS could produce superior performance since XGBoost depends on tree-boost machine learning techniques that aid in a gentler "bias-variance" barter. The study was carried out using the KDDCUP'99 data, and the suggested strategy's accuracy was determined to be 99.95\%.

A feature selection strategy for identifying DoS and DDoS attacks that uses insertion and union processes on subsets by the top 50\% Information Gain (IG) and Gain Ratio (GR) features \citep{nimbalkar2021feature}. With the help of a JRipclassifier, the suggested technique was tested and validated on the IoT-BoT and KDDCup'99 datasets. On the IoT-BoT and KDDCup'99 datasets, using 16 and 19 features, they outperformed the original feature set and typical IDSs with an accuracy of 99.9993\% and 99.992\%, respectively.

To recognize the attacks in IoT a Deep Neural Network (DNN) was developed by \citep{choudhary2020analysis}. The performance of DNN to detect the attacks had been assessed using the three most commonly used datasets such as KDDCUP'99, NSL-KDD, and UNSW-NB15. The proposed method using DNN showed that the accuracy rate was only 91.50\% with each dataset. In \citep{norwahidayah2021performances}, the particle swarm optimization (PSO) feature selection technique and ANN classifier were utilized to detect intrusion as normal and abnormal activities. They selected 20 features using PSO and built their model to produce better performance and achieved 98.00\% accuracy in KDDCUP'99 dataset. 

In \citep{li2021feature}, the authors introduced a method that combines feature correlation (CR) to choose the key features and a DNN classifier to create an IDS model for detecting the intrusion in order to enhance the effectiveness of the network security model. The KDDCUP'99 dataset was utilized in the study, and only 30 attributes were chosen using the CR approach and built the intrusion detection model. With an accuracy rating of 99.40\%, their IDS model is capable of detecting intrusion.

An innovative method of intrusion detection proposed by \citep{narayanasami2021biological} that combines ANN with optimized BAT to improve the detection rate of network attacks. The adopted BAT used to pick the 25 best traits in order to preserve the greatest excellence and remove unimportant attributes from the attack. Utilizing the KDDCUP'99 benchmark dataset as well as the SVM algorithm, the experiment achieved an attack detection rate of 94.12\%.

To detect unusual traffic data in a network, the authors \citep{hu2021identification} has introduced an element identification method based on the Convolutional attention LSTM network model. They extracted the shallow features using CNN and combined them with the attention-LSTM to gain recognition and classification of different networks' behavior. They performed experiments on the KDDCUP'99 dataset and achieved 98.48\% accuracy rate.

In \cite{alqahtani2020cyber}, the authors employed several ML algorithms to identify intrusion on the KDDCUP'99 dataset to compare the effectiveness of these classifiers. K-fold cross-validation, where k=10, was applied to separate the training and testing part of the dataset to enhance the performance. They got the highest accuracy rate of 94.00\% in DT than other algorithms.


In \cite{kumar2020statistical}, the authors proposed a misuse-based IDS to protect networks from modern attacks, namely DOS, Exploit, Probe, Generic, etc. The UNSW-NB15 dataset was used to examine the act of numerous classification models such as CART, C5, CHAID, and QUEST and found the accuracy, IDR, and FAR of each model. Feature selection was performed using IG to select 13 features from 47. The accuracy rate for the proposed C5 model was 99.37\%.

A network forensic mechanism \citep{koroniotis2017towards} has been developed based on network flow identifiers to mistrustful road events of botnets. The ML classifiers such as DT C4.5, ARM, ANN, and NB with the UNSW-NB15 dataset were used to perceive botnet attacks. The performance measurement was done on the Weka tool for analyzing the classification accuracy and FAR, where default parameters and 10-fold cross-validation were employed. The accuracy was only 86.45\%, 93.23\%, 72.73\%, 63.97\% and FAR was  13.55\%, 6.77\%, 27.27\%, 36.03\% for ARM, DT, NB and ANN respectively.

In \cite{kasongo2020performance}, the authors developed a filter-based feature-dropping method using the XGBoost algorithm and applied ML algorithms such as DT, ANN, KNN, support vector machine (SVM), and LR for accuracy prediction on the UNSW-NB15 IDS dataset. They confirmed that their developed method increased binary classification accuracy from 88.13\% to 90.85\%. The overall accuracy was only 90.85\%, 84.39\%, 77.64\%, 84.46\%, 60.89\% for binary and 67.57\%, 77.51\%, 65.29\%, 72.30\%, 53.95\% for multiclass of DT, ANN, LR, KNN, SVM respectively.

In \cite{salo2019dimensionality}, the authors introduced a hybrid model to reduce dimension, which combines the IG and principal component analysis (PCA) techniques and an ensemble classifier based on SVM, instance-based learning algorithms (IBK), and MLP. The performance was assessed on ISCX 2012, NSL-KDD, and Kyoto 2006+ datasets. They built the model for binary classification and found the  DR (99.10\%), accuracy rate (99.01\%), and lowest FAR (0.01\%) in the ISCX 2012 data set,  and the accuracy rate was 98.24\% and 99.95\%; DR was 98.20\% and 99.80\%; FAR was 0.017\% and  0.021\% in the both NSL-KDD and Kyoto 2006 respectively.

A feature selection technique based on linear correlation coefficient (FGCC) and cuttlefish algorithm (CFA) method designed by \cite{mohammadi2019cyber} to detect network intrusion using Decision Tree (DT) algorithm. They applied their approach on KDDCUP'99 dataset to evaluate the performance and got an accuracy rate of 95.03\%. 

In \cite{kshirsagar2021efficient}, the authors introduced an algorithm for the reduction of the feature based on filter-based algorithms such as the Input Gain Ratio (IGR), Correlation (CR), and ReliefF (ReF). It produced feature subsets based on the average weight for each classifier and an additional Subset Combination Strategy (SCS). The number of features was reduced from 77 to 24 for CIC-IDS2017 and 41 to 12 for the KDDCUP’99 dataset. It produced an accuracy rate of 99.96\% with the rule-based classifier Projective Adaptive Resonance Theory (PART) in 133.66 sec for CIC-IDS2017 and the KDDCUP’99 dataset, the accuracy rate was 99.32\%, and the required time was 11.22\%.

In \cite{mugabo2021intrusion}, the authors implemented an IDS scheme based on MapReduce to procure a small and beneficial number of features from large datasets to enhance the accuracy of detecting anomalies by the intrusion. The popular KDDCUP'99 was used for performance evaluation, and for classifying normal and abnormal in mobile cloud computing (MCC) activities RF algorithm was used. The adaptive effective feature selection (EFS) was used to minimize the training set and make the input data parallel. They selected 15 features to evaluate the performance of their model, which achieved 93.90\% accuracy.

In \cite{talita2021naive}, the authors included an IDS problem solution Method where Particle Swarm Optimization (PSO) as feature selection and Naive Bayes as a KDDCUP'99 dataset classification algorithm. The dataset presented more than 40 features and over four hundred thousand records. PSO has been used to pick 38 features from 40+ features to prevent further calculation or memory consumption. With fewer times and greater accuracy than other features, the accuracy rate reached 99.12\%.

In \cite{zhao2018filter}, the authors introduced a new redundant penalty-by-feature (RPFMI) mutual information algorithm to choose efficient malware detection features. The KDDCUP'99 and Kyoto 2006+ datasets for intrusion detection were used for the experimentation. They have shown a higher accuracy rate of the proposed algorithm than others. The accuracy rate was 99.77\% (DOS), 96.19\%(U2R), 91.07\%(R2L) for KDDCUP'99 and 97.74\% for Kyoto 2006+ dataset.

In \cite{mahhizharuvi2021effective}, the authors created an EMRFT (Enhanced Multi Relational Fuzzy Decision Tree) for categorizing network intrusion relying on genetic optimization. The classifier's efficacy was increased by using the K-Nearest Neighbor approach to fill in missing values in the data and the Fast Correlation-based feature selection technique to minimize the dimensional space. The study was carried out using the KDDCUP'99 dataset, and the findings showed that EMRFT performed better, with a binary classification accuracy rate of 98.27\% and a multilabel classification accuracy of 96.56 percent.

In \cite{indrasiri2022malicious}, the authors suggested an extra boosting forest (EBF) model that uses a stacked ensemble strategy to combine the extra tree (ET) classifier, gradient boosting (GB) classifier, and RF models, with the goal of accurately identifying malicious traffic. They used two datasets, namely UNSW-NB15 and IoTID20, which have data on local network traffic and IoT-based traffic, respectively. Utilizing PCA to reduce the number of features in each dataset to 30. The findings demonstrate that EBF scored noticeably better and received the maximum accuracy score of 98.5 and 98.4 in the multilabel of four classes for UNSW-NB15 and IoTID20, respectively.

The VolMemLyzer, among the foremost up-to-date memory feature extractors for learning environments, has been modified to emphasize disguised and obfuscated malware and combined with a stacked ensemble machine learning paradigm to develop a framework for quickly recognizing malware \citep{icissp22}. A malware memory dataset (MalMemAnalysis2022) was also constructed to test and assess the framework, with the goal of closely emulating legitimate obfuscated malware. The study revealed that by employing memory feature engineering, the suggested technique could identify obfuscated and disguised malware incredibly quickly, yielding accuracy and F1-Scores of 99.00\% and 99.02\%, correspondingly.

In \cite{dener2022malware}, the authors performed detection mechanisms utilizing a variety of machine and deep learning techniques in a large data set with memory data. Pyspark was used to conduct this investigation on the Apache Spark big data platform in the Google Colaboratory. On the equitable CIC-MalMem-2022 dataset, tests were conducted. Various machine learning and deep learning methods were used to achieve the identification of malware. The effectiveness of the employed algorithms has been contrasted, and the outcomes have been assessed using several performance measures. The study's greatest accurate method for detecting malware using memory analysis, the Logistic Regression (LR) technique, had an acceptable accuracy of 99.97\%. As a result, memory research data is extremely helpful in identifying malware.

In \cite{louk2022tree}, the authors applied a number of tree-based ensemble ML algorithms to the Portable Executable (PE) malware detection. To show how well the algorithms perform across a range of scenarios, the study uses three open-source datasets, including BODMAS, Kaggle, and CIC-MalMem-2022. According to the test results, all tree-based ensembles worked adequately, and there were no statistically relevant performing variations among the various algorithms. According to this research, the performance rates of the Gradient Boosting Machine (GBM), XGBoost, and RF are higher than those of other tree-based ensemble models. They got the accuracy rate of 99.39\%, 99.96\%, and 100\% for the Kaggle, BODMAS, and CIC-MalMem-2022 datasets, correspondingly.

The literature review is summarised in the following Table \ref{tab:related-work}.

\begin{table}[]
\centering
\resizebox{\textwidth}{!}{
\begin{tabular}{lllll}
\hline
SL.No. & Authors & Dataset & Algorithm & Accuracy (In \%) \\ \hline
1 & \cite{tan2019wireless} & KDDCUP'99 & SMOTE + RF & 92.57 \\
2 & \cite{bhati2021improved} & KDDCUP'99 & XGBoost & 99.95 \\

\multirow{2}{*}{3} & \multirow{2}{*}{\cite{nimbalkar2021feature}} & IoT-BoT & \multirow{2}{*}{IG + GR+ JRipclassifier} & 99.99 \\
 &  & KDDCUP'99 &  & 99.57 \\

\multirow{3}{*}{4} & \multirow{3}{*}{\cite{choudhary2020analysis}} & NSL-KDD & \multirow{3}{*}{DNN} & 91.50 \\
 &  & KDDCUP'99 &  & 91.50 \\
 &  & UNSWNB-15 &  & 91.50 \\
 
5 & \cite{norwahidayah2021performances} & KDDCUP'99 & PSO + ANN & 98.00 \\
6 & \cite{li2021feature} & KDDCUP'99 & CR + DNN & 99.40 \\ 
7 & \cite{narayanasami2021biological} & KDDCUP'99 & BAT + SVM  & 94.12 \\  
8 & \cite{hu2021identification} & KDD-CUP’99 & CNN + LSTM & 98.48 \\
9 & \cite{alqahtani2020cyber} & KDD-CUP’99 & DT & 94.00 \\

10 & \cite{kumar2020statistical} & UNSWNB-15 & C5 & 99.37 \\
\multirow{4}{*}{11} & \multirow{4}{*}{\cite{koroniotis2017towards}} & \multirow{4}{*}{UNSWNB-15} & DT & 86.45 \\
 &  &  & C4.5 & 93.23 \\
 &  &  & ARM & 72.73 \\
 &  &  & ANN & 63.97 \\
\multirow{2}{*}{12} & \multirow{2}{*}{\citep{kasongo2020performance}} & \multirow{2}{*}{UNSW-NB15} & \multirow{2}{*}{XGBoost + DT, ANN, LR, KNN, SVM} & 90.85 (DT), 84.39 (ANN), 77.64 (LR), 84.46 (KNN), 60.89 (SVM) -Binary \\
 &  &  &  & 67.57 (DT), 77.51 (ANN), 65.29 (LR), 72.30 (KNN), 53.95 (SVM) -Multilabel \\
\multirow{3}{*}{13} & \multirow{3}{*}{\citep{salo2019dimensionality}} & ISCX 2012 & \multirow{3}{*}{Hybrid Model (IG+PCA+SVM+IBK+MLP)} & 99.01 \\
 &  & NSL-KDD &  & 98.24 \\
 &  & Kyoto 2006 &  & 99.95 \\
 
14 & \cite{mohammadi2019cyber} & KDDCUP'99 &FGCC+CFA + DT & 95.03 \\  
\multirow{2}{*}{15} & \multirow{2}{*}{\cite{kshirsagar2021efficient}} & CIC-IDS2017 & \multirow{2}{*}{PART} & 99.95 \\
 &  & KDDCUP'99 &  & 99.32 \\
16 & \cite{mugabo2021intrusion} & KDDCUP'99 & RF & 93.90 \\
17 & \cite{talita2021naive} & KDDCUP'99 & PSO + NB & 99.12 \\
\multirow{2}{*}{18} & \multirow{2}{*}{\cite{zhao2018filter}} & KDDCUP'99 & \multirow{2}{*}{RPFMI} & 99.77 (DoS), 96.19 (U2R), 91.07 (R2L) \\
 &  & Kyoto 2006+ &  & 97.74 \\

\multirow{2}{*}{19} & \multirow{2}{*}{\cite{mahhizharuvi2021effective}} & \multirow{2}{*}{KDDCUP'99} & \multirow{2}{*}{EMRFT} & 98.27 -Binary \\
 &  &  &  & 96.56 -Multilabel \\
 
\multirow{2}{*}{20} & \multirow{2}{*}{\cite{indrasiri2022malicious}} & UNSW-NB15 & \multirow{2}{*}{EBF} & 98.5 \\
 &  & IoTID20 &  & 98.4 \\
 
21 & \cite{icissp22} & CIC-MalMem-2022 & VolMemLyzer + Stacked Ensemble & 99.00 \\ 
22 & \cite{dener2022malware} & CIC-MalMem-2022 & LR & 99.97 \\ 
\multirow{3}{*}{23} & \multirow{3}{*}{\cite{louk2022tree}} 
    & Kaggle & GBM & 99.39\% \\
 &  & BODMAS & XGBoost & 99.96\% \\
 &  & CIC-MalMem-2022 & RF &100\%\\ \hline

\hline
\end{tabular}
}
\caption{Performance summary of different proposed works.}
\label{tab:related-work}
\end{table}

\section{Proposed Methodology}
\label{sec:Method}

\subsection{Problem Statement}
The intrusion detection system (IDS) is a critical security component for ensuring network usability. In recent years, automated intrusion detection within IDS has advanced significantly through the use of artificial intelligence and how effectively and efficiently ML/DL models can be applied is a recognized approach in terms of innovation in this research area. However, as attackers become more sophisticated, network security has become more challenging nowadays. In addition to that, dealing with large amounts of data has always been a challenge when developing security components. 

In this research work, we particularly dealt with the following three problems in AI-enabled automation.

Firstly, we discovered from state-of-the-art research that when working on imbalanced datasets, the majority of the IDS model evaluates their performance in terms of precision, recall, and f1-score. And in most cases, accuracy outperforms precision, recall, and f1-score \citep{nimbalkar2021feature, talita2021naive}. In addition to that, most papers did not use a confusion matrix to determine type-1 and type-2 errors, which can be critical in evaluating the detection performance of the model. To address these, first of all, we use SMOTE, which handles the data imbalance problem and ensures that the precision, recall, and f1-score are unaffected. This gives high-performance results, like accuracy. We also evaluate our model by developing a confusion matrix that gives more confidence in and reliability of our model by outlining type-1 and type-2 errors.

Secondly, one of the issues that many authors face is the efficient reduction of the dimension \citep{choudhary2020analysis, man2021residual, shetty2020comparison, hu2021identification}. Dimension reduction can be an important consideration while working with relevant features in removing irrelevant or less important features to reduce time and speed up the process. In our proposed method, we deal with that problem effectively by using XGBoost to reduce dimensions and select the best features to reduce computational costs.

Finally, we addressed the issue of dependability - how dependable is the model? As a result, we conduct dependability analysis using key performance evaluation metrics to validate our model's efficiency, availability, and scalability.

\subsection{Hybrid Model with Machine and Deep Learning}
In addressing the aforementioned research issues, we propose a hybrid approach that combines an efficient pre-processing technique with the handling of missing values, data balancing using SMOTE, feature scaling using standardization, and label encoding to prepare the datasets. XGBoost is then used to select the optimal features to feed into ML and DL algorithms in order to construct the models. We analyzed the performance of multiple ML and DL algorithms, including RF, DT, KNN, MLP, CNN, and ANN, in order to recommend the optimal algorithm for detecting network intrusion.

In this section, we introduce the major building blocks of the proposed hybrid model.

\subsubsection{Data Balancing using SMOTE}
Data balancing is a process of rebalancing data from imbalanced data. SMOTE is a familiar approach for dealing with unbalanced data \citep{chawla2002smote}. Whenever the class distribution is biased forward into a specific category, unbalanced data issues occur. To overcome the unbalanced dataset issue, the SMOTE strategy produces imitation data to achieve a balance over the minority and majority category sizes \citep{rustam2020predicting}. To rebalance the minority and majority categories, a parameter is supplied to the SMOTE technique to establish a specified threshold for simulated samples \citep{he2009learning}. SMOTE picks relative entries as well as modifies them one column at a time by adding a random number based on the difference between the neighbouring records.  \citep{barandela2004imbalanced} whenever the majority or minority ratio is very high, it is required to oversample the minority class rather than undersample it to equalize the minority class ratio. In our hybrid approach, the SMOTE plays a signification role as the datasets which are imbalanced, like KDDCUP’99, need to balance to overcome the overfitting and less realistic prediction models.

\subsubsection{Feature Selection using XGBoost}
Feature selection is a method of selecting a subset of the underlying features in order to minimize the feature space to the smallest possible size based on some criteria. Feature extraction is a technique for creating a new set of features that can be utilized alone or in combination \citep{motoda2002feature}. Moreover, it can locate and choose the far more beneficial properties inside data. It's an important stage in the machine learning workflow since it assists in minimizing the fitting problems, reducing adaptation efficiency on the testing data, reducing training duration, and reducing model interpretability \citep{medium_fs}. There are three main kinds of feature selection methods: filter-based, wrapper-based, and embedded feature selection \citep{tang2022motor}. Build-in feature selection is available in the embedded feature selection method, which helps to build a model without applying any additional feature selection method. To choose features, the filter-based feature technique employs assessment criteria, including information analysis as well as distance assessment. The wrapper-based feature selection approach builds a subset of features in a particular way before evaluating feature selection using the findings of classifiers. Using the embedded feature selection approach, certain properties can be dynamically removed within classifier construction, allowing feature selection and classification to be done simultaneously \citep{tang2022motor}. Some examples of embedded methods are RF, Lasso, XGBoost, and LGBoost algorithms. The detection of IDS can be efficiently solved by using feature selection approaches\citep{kharwar2022ensemble, mojtahedi2022feature, vaidya2022analysis}.
\\
In our hybrid approach, we used XGBoost to select the optimal features as we already know, feature selection refers to selecting a particular number of features from all available features. Feature selection is essential to reduce computation costs and enhance the model's performance \citep{devan2020efficient}. Extreme Gradient Boosting (XGBoost) is a systematic gradient boosting method that utilizes extra precise approximate to determine the best model tree. It uses a number of methods to discover the important features from the dataset, primarily structured data. It involves the following: (i) calculating second-order gradients, i.e., second-sectional loss function derivatives (similar to Newton's), which provides more insight into the development of gradients and how to achieve the lowest possible loss function. To minimize the error of the entire model, while the regular gradient increase uses the loss function of our model base (e.g., the Decision tree), the 2nd derivative of XGBoost is used as an approximation. (ii) Advanced regularization (L1 \& L2) improving the generalization of the model and training is so profligate and can be spread parallel across clusters \citep{john2010elements}.

XGBoost offers a resourceful and effective enactment of the stochastic gradient boosting algorithm to improve accuracy \citep{farrugia2020detection}. It provides an inbuilt process to directly get feature importance for feature selection by making a relationship between multiple variables and feature importance. It is made up of decision trees, and the prediction is accomplished by using these trees. To generate trees, XGBoost employs the loss and regularization function \citep{dhaliwal2018effective} to generate the main objective function as follows:

\begin{equation}
L(\phi)=\sum_{i=1}^{n}l(y_{i},\hat{y_{i}})+\sum_{k=1}^{n}\Omega(f_{k})
\label{eq:objective_function}
\end{equation}
where, l is a distinctive convex loss function measuring the difference between $\hat{y_{i}}$ (target) and $y_{i}$ (actual) prediction. $\Omega$ (penalizes the complexity of the model) is the regularization function in Equ.(\ref{eq:regularization}) which controls the complexity of a model and prevents the model from overfitting.

\begin{equation}
\Omega(f)=\gamma ^{^{T}}+ \frac{1}{2}\lambda \sum_{j=1}^{T} ||w||^{2} 
\label{eq:regularization}
\end{equation}
where $\lambda$ is projected to reduce the prediction's sensitivity, T represents the number of terminal nodes,$\gamma$ represents encouraging pruning, and $w$ represents the leaf weights.
To optimize the model of Equ.(\ref{eq:objective_function}) in an additive manner, we can do as follow:
\begin{equation}
L^{(t)}=\sum_{i=1}^{n} l(y_{i},\hat{y_{i}}^{t-1} +f_{t}(x_{i}))+\Omega(f_{t})\\
\end{equation}
where, $\hat{y_{i}}^{t}$ be the $i-th$ instance prediction in $t-th$ iteration and $f_{t}$ enhance our model.

To enhance the objective, $2^{nd}$ order approximation can be used as follows:
\begin{equation}
L^{(t)} {\simeq} \sum_{i=1}^{n}[l(y_{i},\hat{y_{i}}^{t-1})+ g_{i}f_{t}(x_{i})+\frac{1}{2}h_{i}f_{t}^{2}(x_{i})]+ \sum_{k=1}^{n}\Omega(f_{k})
\label{equ:2nd_order}
\end{equation}
where, 
$g_{i}=\partial_{\hat{y}^(t-1)} l(y_{i},\hat{y_{i}}^{t-1})$
$h_{i}=\partial^{2}_{\hat{y}^(t-1)} $ are the loss function of 1st and 2nd order gradient statistics.

By eliminating the constant part of Equ.(\ref{equ:2nd_order}) we can get the abridged objective \citep{chen2016xgboost} as follows:
\begin{equation}
\tilde{L}^{(t)} =\sum_{i=1}^{n}[g_{i}f_{t}(x_{i})+\frac{1}{2}(h_{i}f_{t}^{2}(x_{i}))]+ \Omega (f_{t})
\label{equ:modified_objective}
\end{equation}

\begin{table}[]
\centering
\resizebox{\textwidth}{!}{%
\begin{tabular}{ll}
\hline
Dataset & \multicolumn{1}{c}{Features with Ranking} \\ \hline
KDDCUP’99 (Binary) & 22, 12, 25,   7, 38, 5, 37, 9, 36, 34, 1, 2, 4, 35, 3, 28, 39, 32, 31, 40, 33, 0, 26, 15,   11, 16, 23, 29, 24, 30, 13, 18, 27, 21, 6, 8, 17, 10, 19, 14, 20 \\
KDDCUP’99 (Multilabel) & 21, 10, 22,   23, 2, 1, 4, 37, 32, 13, 5, 34, 36, 14, 29, 16, 9, 35, 3, 0, 7, 30, 18, 31,   11, 39, 28, 40, 12, 8, 15, 33, 25, 26, 38, 17, 24, 27, 19, 6, 20 \\
CIC-MalMem-2022 & 45, 48, 7, 8, 52, 27,   12, 30, 11, 0, 46, 5, 32, 23, 42, 19, 4, 15, 51, 14, 37, 43, 6, 2, 1, 20, 24,   41, 29, 39, 13, 21, 17, 26, 28, 18, 16, 10, 9, 50, 49, 3, 44, 22, 47, 25, 53,   31, 33, 34, 35, 36, 38, 40, 54 \\ \hline
\end{tabular}%
}
\caption{XGBoost-based feature ranking.}
\label{tab:feature_ranking}
\end{table}

\begin{figure*}[!htbp]
\centering
\includegraphics[width=0.9\textwidth]{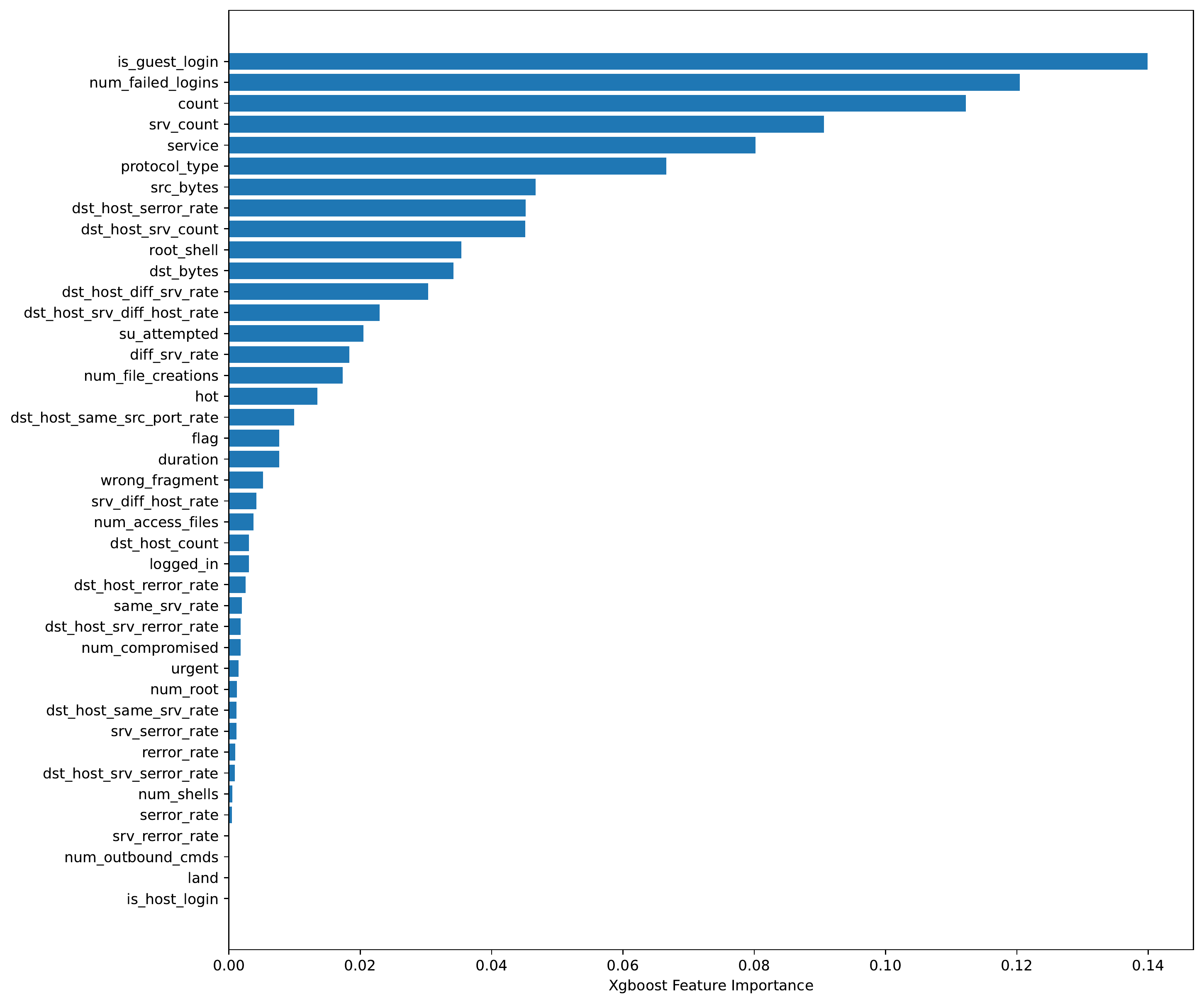}
\caption{Feature importance graph by XGBoost algorithm.}
\label{fig:feature_importance_graph}
\end{figure*}

Table \ref{tab:feature_ranking} shows the ranking of features of the KDD-CUPP'99 dataset generated from our proposed XGBoost algorithm. Figure \ref{fig:feature_importance_graph} shows the important features in the horizontal bar chart of the KDD-CUPP'99 dataset generated from our proposed XGBoost algorithm. 

To select the best features, we have conducted an extensive experiment using four different ML algorithms, including RF, DT, KNN, and MLP. Firstly, the output features are sorted in descending order, known as ranking features, for our feature selection process. After that, we calculate the accuracy of each of the algorithms using various subsets of features in the ranking features, which are the features that are generated from the XGBoost algorithm. Secondly, we calculate the accuracy of each of the algorithms using the k number of features where we set $k=N$ initially, where N is the number of features. If the accuracy results for all the algorithms satisfy an accuracy threshold $Th_{acc}$, this set of features will be nominated as a candidate features set. Then, we reduce the value of $k$ by $2$ and calculate the accuracy until $k>0$. For instance, if we have a dataset of features size is 40 and if we take k=1, the time complexity of completing the feature selection process will be O(N) means it takes much more time with is equal to the number of features. If we take k=10, then its time complexity will be O(N/10) which is less complexity, but we can’t easily find out the minimum number of features that can produce better accuracy as it produces a decrement of k by 10, which is 40, 30, 20, 10 features. So, we take k=2 so that we can get minimum time complexity which is O(N/2), and the minimum number of better feature sets of decrements of k by 2, which is 40, 38, 36, 34...N features without taking more or less than k=2. Moreover, this selection process helps us to clearly find out the variation of the accuracy rate while decreasing the number of features.

\begin{figure*}[!htbp]
\centering
\includegraphics[width=0.7\textwidth]{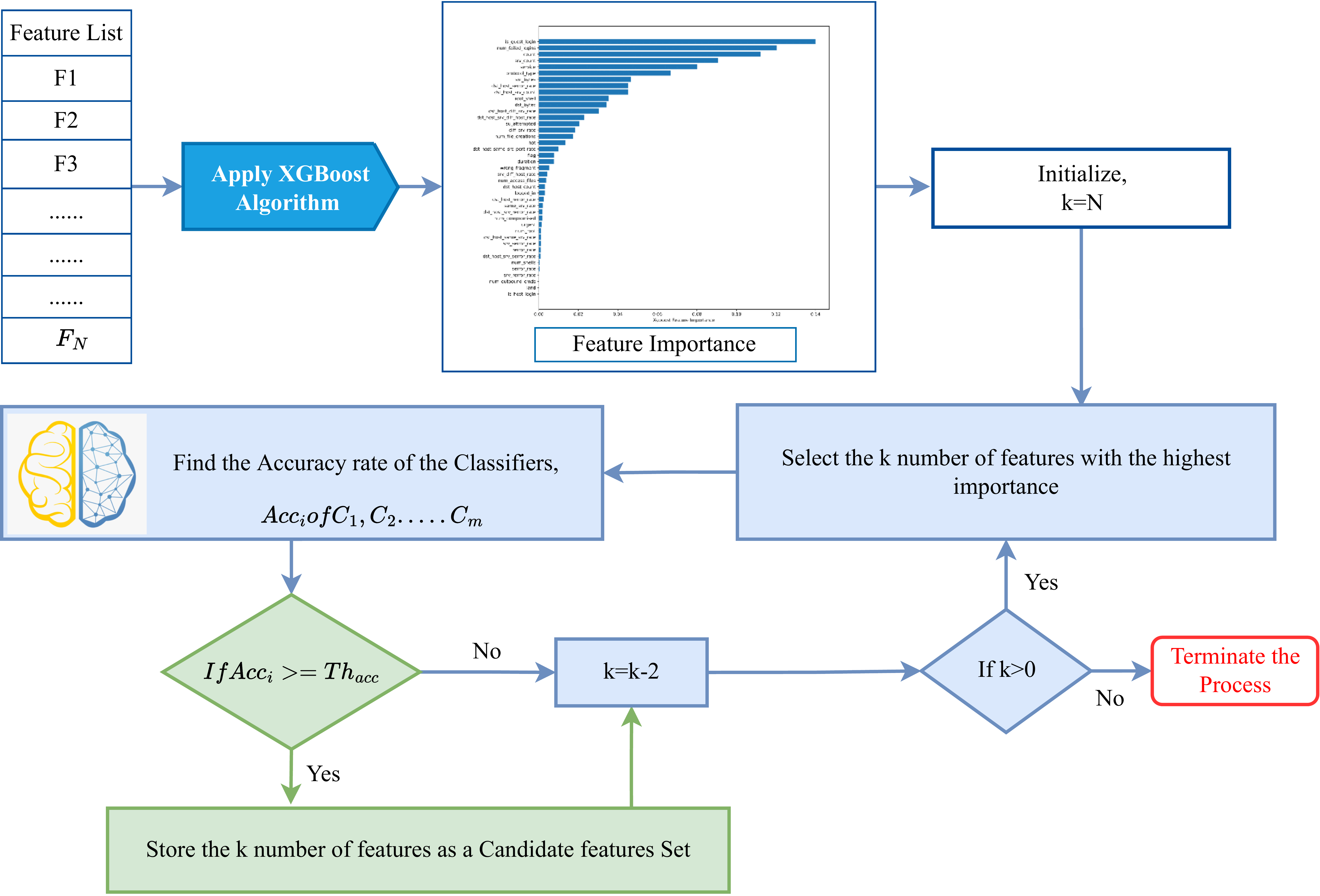}
\caption{Feature selection process.}
\label{fig:feature_selection}
\end{figure*}

A set of features will be a candidate set if all the algorithms satisfy to achieve the accuracy greater or equal to the accuracy threshold $Th_{acc}$. Finally, from all the candidate sets of features, we choose the smallest set for which the accuracy of all algorithms can provide accuracy greater than $Th_{acc}$, which is 99.95\%. The feature selection process is illustrated in the following graph \ref{fig:feature_selection}. In the graph, we take the features and apply the XGBoost algorithm and initialize k=N, where N is the size of the features. Then we apply ML classifiers and check the accuracy rate of each classifier. If all the classifiers achieve the threshold hold accuracy rate, then those features are selected and stored as a candidate set of features; if not, then it will deduce the value of k by 2 to get the minimum number of features that will produce a better performance with the half of time complexity of total feature size (O(N/2)) and repeat the process.

\begin{figure*}[!htbp]
\centering
\includegraphics[width=0.7\textwidth]{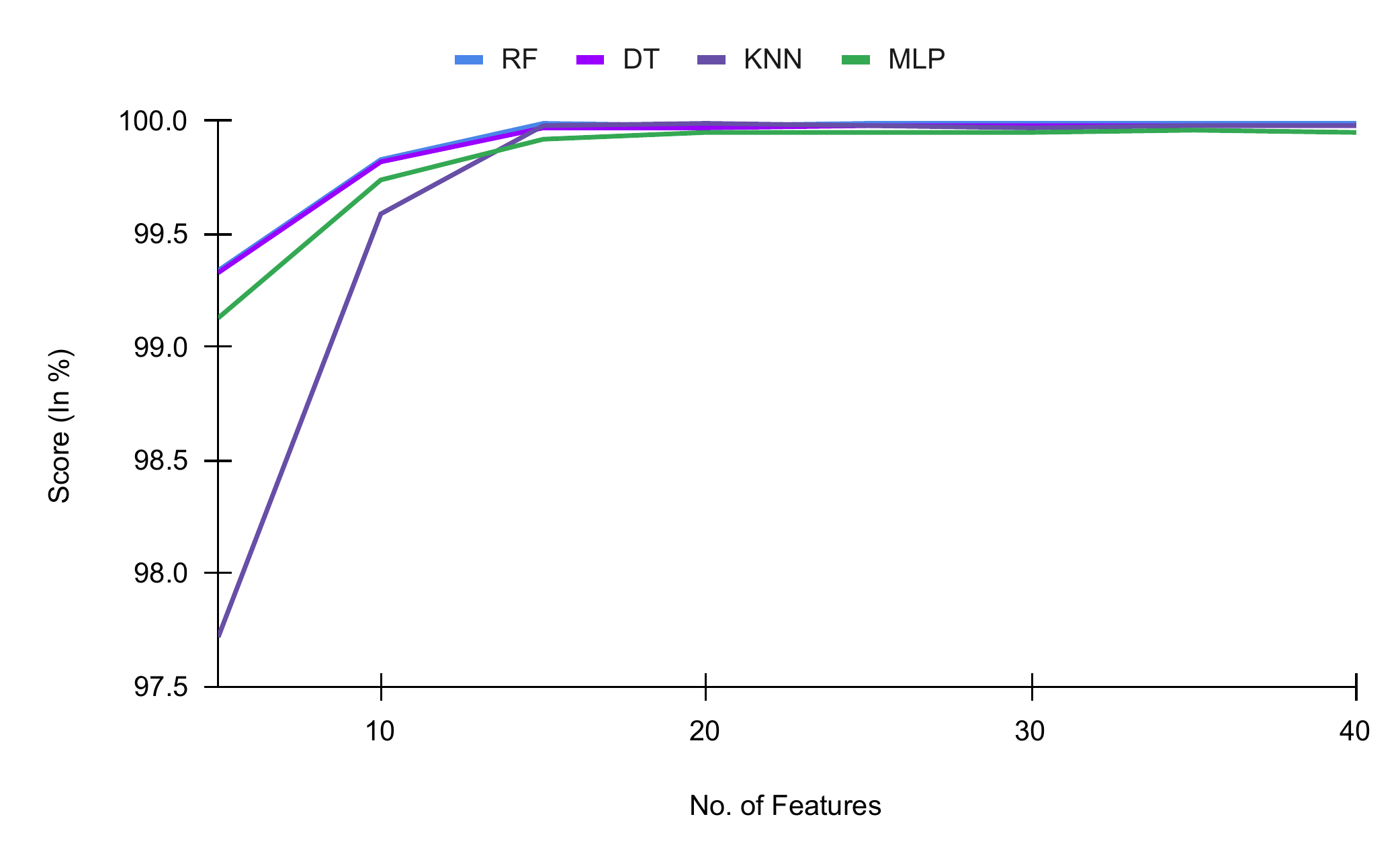}
\caption{Feature importance results by XGBoost algorithm.}
\label{fig:feature_importance_results}
\end{figure*}

The accuracy performance rate of different subsets of features is shown in Fig. \ref{fig:feature_importance_results}. After analyzing all the accuracy results for all the ML algorithms for each subset, we find that after 20 feature selections, all ML algorithms satisfy the accuracy threshold value, which is 99.95\%. So, we nominated these first top 20 features as a candidate feature set by which we produced in our proposed work.

\subsubsection{Machine and Deep learning algorithms}
In this work, we have used several ML and DL algorithms which are listed below, to perform binary and multilabel classification on our datasets.

\begin{itemize}
    \item{Random Forest (RF):}
    An RF is a meta-approximation that employs averaging to enhance the accuracy level. Multiple decision tree classifiers are fitted to various sub-trials of the data set to prevent over-fitting \cite{alkhatib2020predictive}. It combines several uncut DTs derived from different bootstrap samples of the training data, and each attribute subset is sampled separately from the actual feature space \citep{breiman2001random}. Each tree and the class estimate a class that the majority of trees forecast becomes our model prediction \citep{ahmad2021intrusion}. It creates several independent decision trees from the training data set's original features and then votes to combine them into a single classification model \citep{pan2014predicting}. 
    
    \item{Decision Tree (DT):}
    Decision trees are a versatile tool that can be used in a variety of fields, including machine learning, image processing, and pattern recognition. DT is mainly used for grouping purposes. Furthermore, in Data Mining, DT is a commonly used classification model \citep{gavankar2017eager}. The root node, branches, and leaf nodes are the three main components of DT. The whole dataset is signified by the root node, which is alienated into two or more homogeneous sets, branches are the combination of features or attributes, and the output nodes are known as leaf nodes, where apartheid is halted \citep{dey2016machine}. Because of their ease of analysis and precision across a variety of data types, decision trees have a broad range of applications \citep{mrva2019decision}.
    
    \item{K-Nearest Neighbour (KNN):}
    KNN classifies unlabeled data instances by assigning them to another class of similar marked examples according to the closest measured distance between the data instance and the groups. Even lacking prior information about data distribution, KNN is simple to implement since it employs the Euclidean distance equation to calculate connections. It also generates precise end-user suggestions based on the simple application of similarity or distance for classifications \citep{ahmed2021machine, jahan2021automated}.

    \item {Multilayer Perceptron (MLP):}
    MLP is a common ANN architecture that consists of a series of layers made up of neurons and their connections. It can measure the weighted sum of its inputs before applying an activation function to generate a signal that will be sent to the next neuron \citep{castro2017multilayer}. It has one or more hidden layers between the input and output layers. The neurons are arranged in layers, connections are often guided from lower to upper layers, and the neurons in the same layer are not linked \citep{ramchoun2016multilayer}. In the input layer, the number of neutrons equals the number of measurements for the pattern query, while the number of neurons in the output layer equals the number of classes \citep{talukder2022machine}.
    
    \item {Convolution Neural Network (CNN):}
    A CNN is a type of Artificial Neural Network (ANN) that is usually used to evaluate visual representations in deep learning \citep{valueva2020application}. Relying on the sharable structure of the convolution kernels or filters which move across input properties and give translational substance replies referred to as data maps, they're also defined as Shift Invariant or Space Invariant Artificial Neural Networks (SIANN) \citep{zhang1988shift, zhang1990parallel}. The connecting arrangement among neurons matches the arrangement of the vertebrate neural activity, which was motivated by biological processes \citep{fukushima1982neocognitron, yamashita2018convolutional, matsugu2003subject}. It is also a regularized variant of the multilayer perceptron \citep{lin2013network}, with a distinctive strategy for regularization: it uses the hierarchical structure in information and assembles motifs of evolving utilizing down into simpler patterns engraved in their filtering \citep{arvinth2021weed, ahamed2021deep}.
    
    \item {Artificial Neural Network (ANN):}
    ANNs, sometimes known as neural networks (NNs), are computer systems modelled after the biological neurons seen in animal brains \citep{bland2020advances}. They are a set of interconnected systems or nodes in an ANN that roughly model neurons in a neocortex \citep{gorgun2022characterization}. so each link, unlike synapses in the human brain, has the ability to send a response to neighboring neurons. An artificial neuron obtains the signals, analyses them, and afterwards sends signals to the neurons to which this is attached. Each neuron's signal is produced by certain non-linear functions of the combination of its inputs, and the "signal" at a link is a true number \citep{bland2020advances}. The interconnections are referred to as edges. The weight of synapses and edges varies as training progresses. The weight affects the signal power at a connector. Synapses may well have a criterion at which a signal is just transmitted if indeed the cumulative activity exceeds it. Neurons are generally organized into layers. On their inputs, various layers may conduct distinct modifications. Signals go from the very first layer (the input layer)to the last layer (the output layer) \citep{feldmann2019all, hossain2022automatic}.
\end{itemize}

\subsection{Proposed Architecture}

\begin{figure*}[!h]
\centering
  \includegraphics[width=1.0\textwidth]{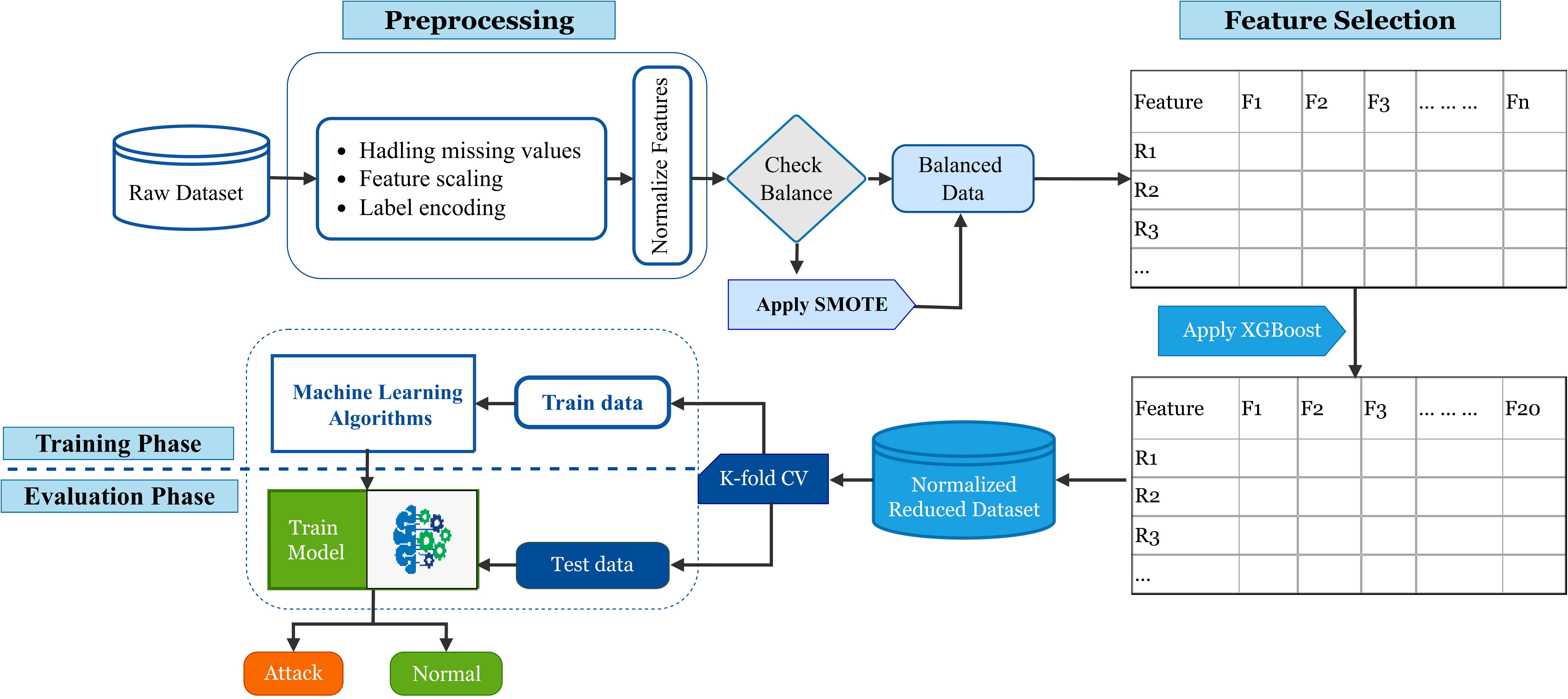}
\caption{Proposed framework for feature selection and intrusion detection.}
\label{fig:proposed model}
\end{figure*}

Figure \ref{fig:proposed model} depicts the schematic block diagram of the proposed approach. The approach consists of five sequential stages discussed as follows:

\begin{itemize}
    \item {Stage-1: } In this stage, data preprocessing has been accomplished by handling missing values, scaling features, and encoding attribute values.
    
    \item {Stage-2} In this stage, we check the dataset and apply SMOTE if the dataset is imbalanced to balance our datasets and handle the data imbalance problem.
    
    \item {Stage-3: } To exclude the less correlated features with the class label, the XGBoost algorithm is applied to obtain the most relevant features from the datasets.
    \item {Stage-4: } This stage splits the processed dataset to form training and testing dataset using K-fold cross-validation.
    \item {Stage-5: } In this phase, the algorithms are trained and tested to evaluate the performance in terms of several performance parameters, including accuracy, precision, recall, and f1-score. Afterwards, based on the performance, the highest-performed model is selected as a recommended model to detect network intrusion, and then the model is compared with other existing models.
\end{itemize}

\section{Model Implementation and Evaluation}

We have implemented a novel hybrid approach to detect network intrusion by utilizing SMOTE for data balancing and XGBoost for feature selection and applying several ML and DL algorithms to analyze and select the best model. The approach is validated and assessed using extensive experiments on various datasets. In the following section, we described the dataset descriptions followed by the data preparation and training process.

\subsection{Dataset Description}
This paper tests several ML and DL algorithms on the KDDCUP'99 and CIC-MalMem-2022 (MalMemAnalysis-2022) datasets.

\subsubsection{KDDCUP'99 Dataset}
\begin{figure*}[!htbp]
	\centering
	\subfloat[Before SMOTE]{\includegraphics[scale=.4]{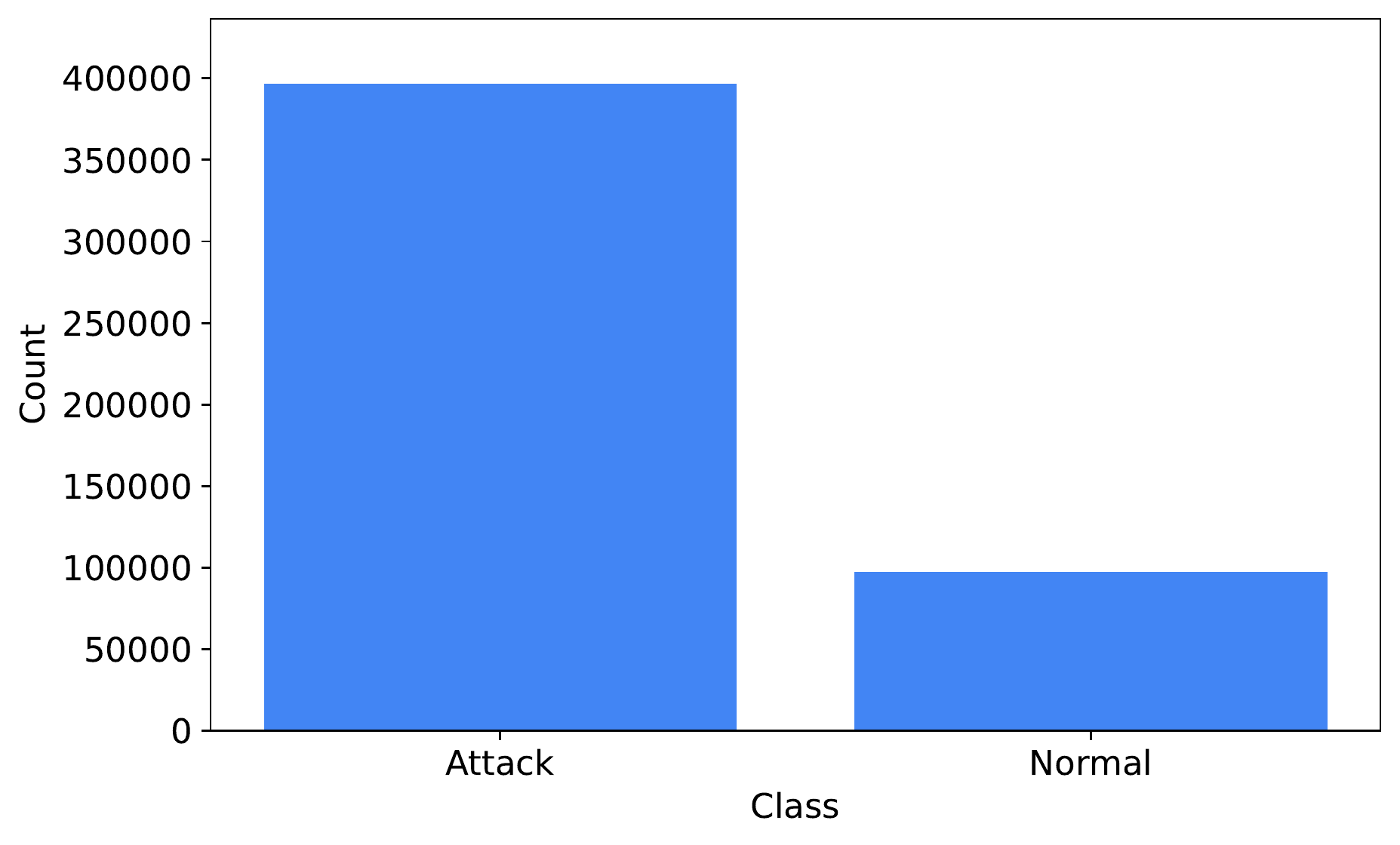}}\hspace{0.1cm}
	\subfloat[After SMOTE]{\includegraphics[scale=.4]{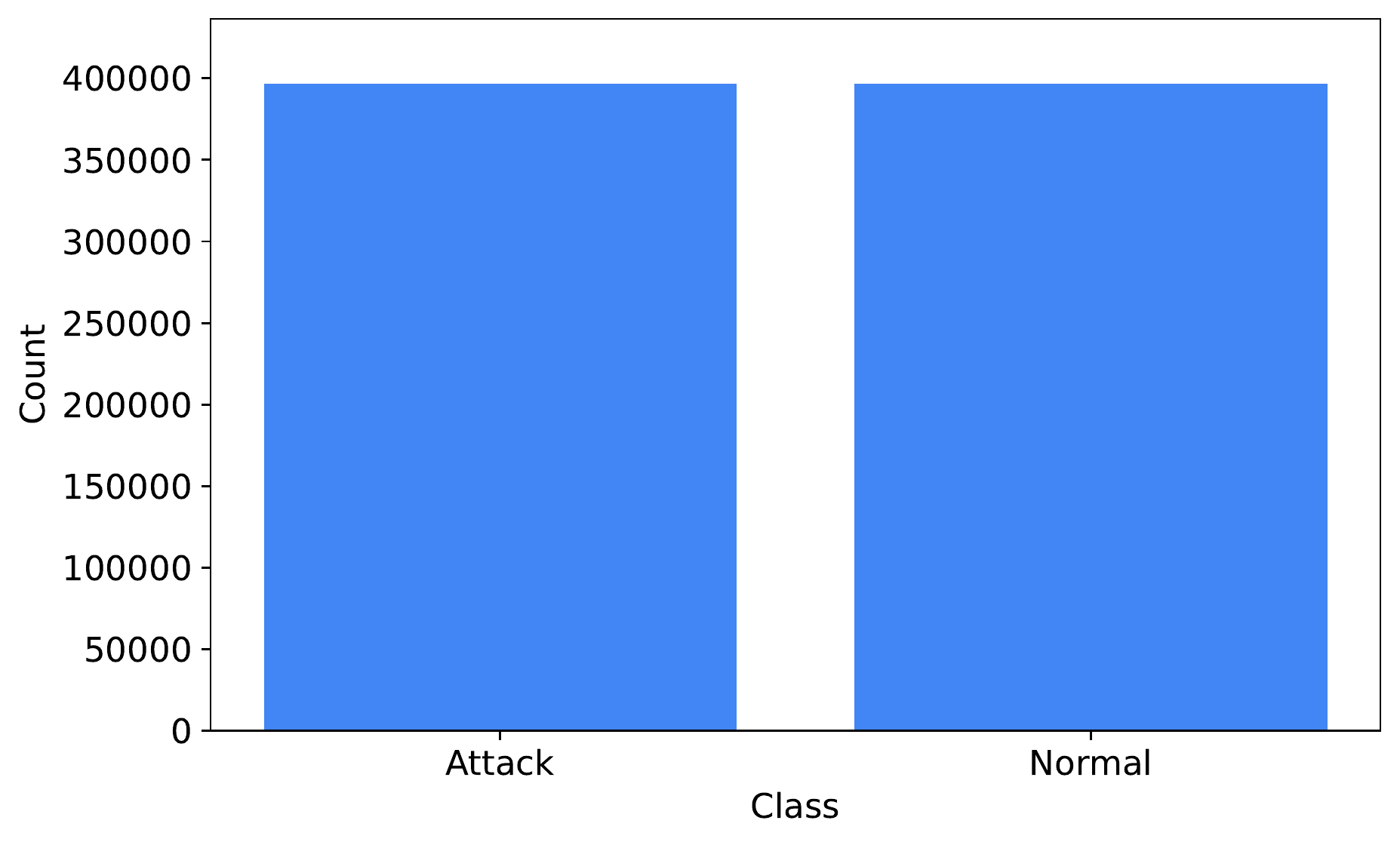}}
	\caption{Before and After the distribution of attacks in KDDCUP'99 for binary classification.}
	\label{fig:data1_binary}
\end{figure*}

\begin{figure*}[!htbp]
	\centering
	\subfloat[Before SMOTE]{\includegraphics[scale=.4]{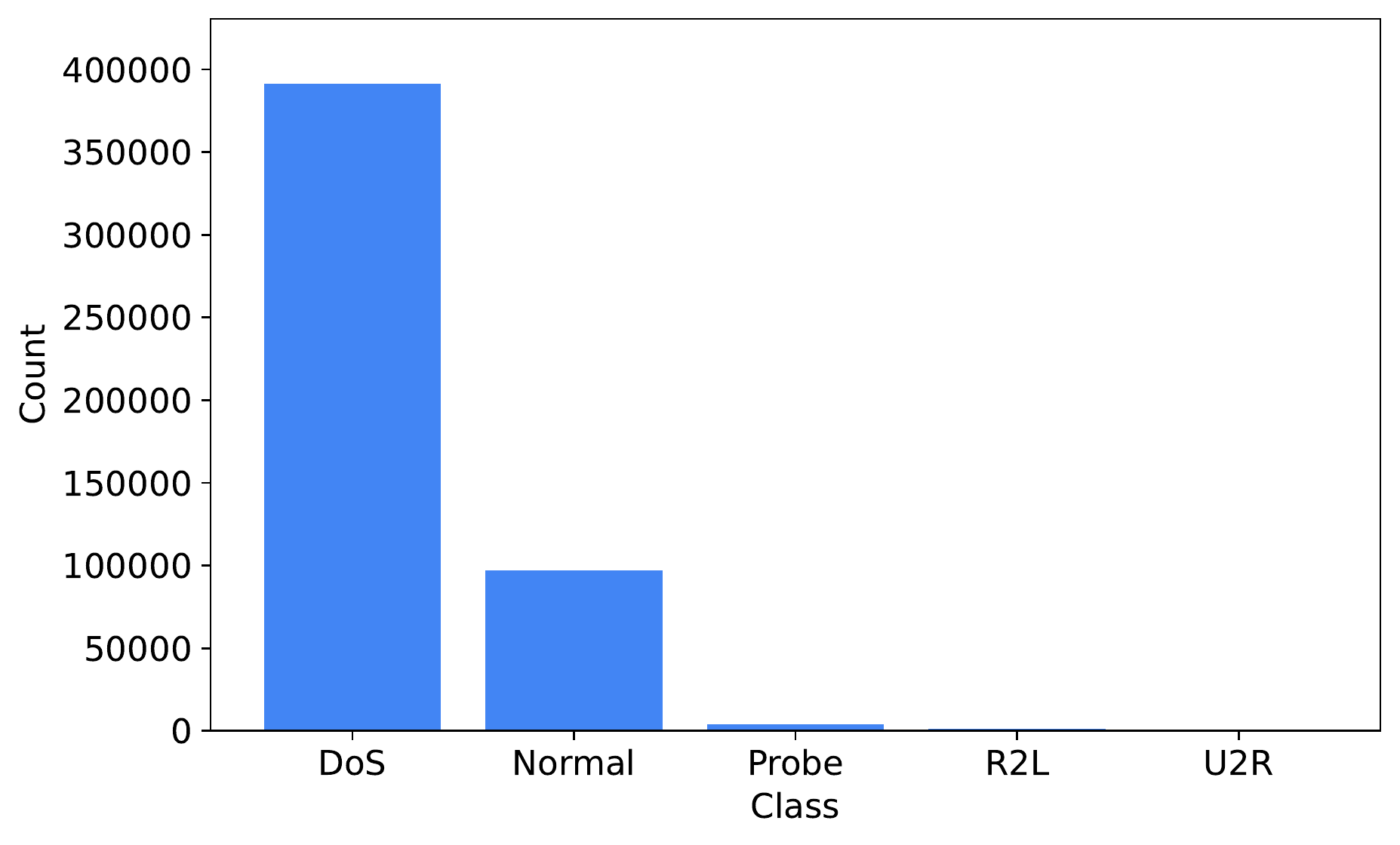}}\hspace{0.1cm}
	\subfloat[After SMOTE]{\includegraphics[scale=.4]{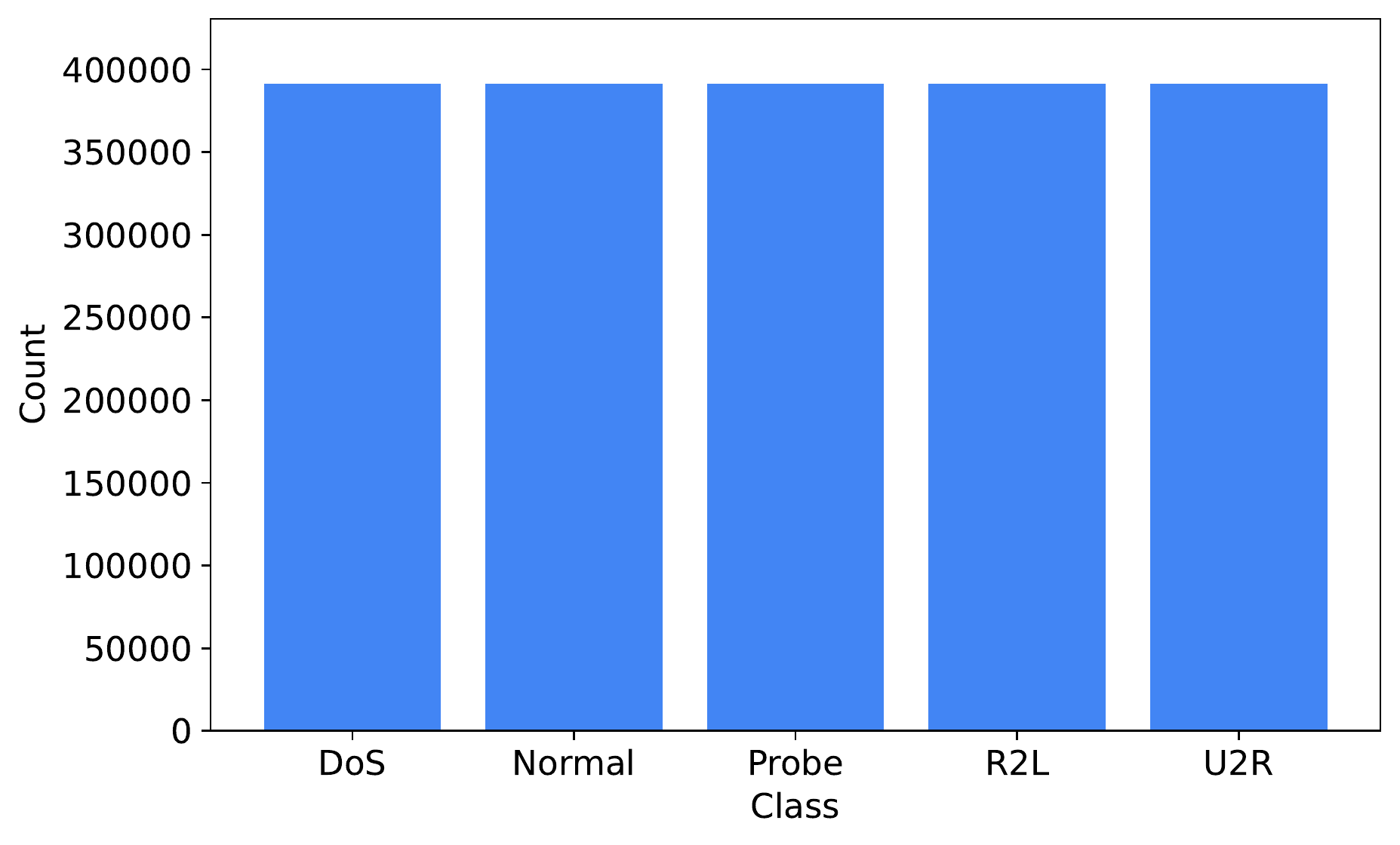}}
	\caption{Before and After the distribution of attacks in KDDCUP'99 for multilabel classification.}
	\label{fig:data1_multilabel}
\end{figure*}

Knowledge Discovery and Data Mining (KDD)Cup
1999 dataset, \citep{kddcup99} which is a benchmark dataset released by the US Department of Defense Advanced Research Projects Agency (DARPA). The dataset was created to forecast a model differentiating between bad connections, known as intrusions or attacks, and regular connections. The dataset comprises a standard set of auditable data covering a wide range of virtual intrusions in a military network environment. A communication link is defined as a series of TCP packets sent from the source IP address to the target IP address over a fixed period with data being transmitted from the source IP address to the target IP address according to a pre-configured protocol \citep{zhong2021sequential}. This is the most broadly used dataset for network intrusion detection assessment \citep{siddique2019kdd}, and it has been available for a long time. In our experiment, we used 10\% of the KDDCUP'99 dataset (‘kddcup.data\_10\_percent.gz’) to reduce the experimental time and cost. It has 41 features(input), 23 subcategories (outcome), 5 categories (attack\_type) and 2 labels of attacks (label) with 44 attributes. 
Table \ref{tab:subcategory} represents the different sub-categories of the various attacks, while Figure \ref{fig:data1_binary} and \ref{fig:data1_multilabel}depicts the distribution of attack categories before and after SMOTE using a bar chart both for binary and multilabel classification. All the features  are described in the following Table \ref{tab:44feature}.

\begin{table}[]
\resizebox{\textwidth}{!}{
\begin{tabular}{llll}
\hline
Category & Description & Sub-category & Count \\ \hline
dos & Turn off a network rendering it unreachable   to its target purposes & back, land, Neptune, pod, smurf, and teardrop & 391458 \\ 
u2r & Illegal access from a distant computer & buffer\_overflow, loadmodule, Perl and rootkit & 52 \\ 
r2l & Illegal access to superuser (root) credentials on the   local machine & ftp\_write, guess\_passwd, imap,   multihop, phf, spy, aware client, and ware master & 1126 \\
probe & Sends packets to a remote machine to acquire local access. & ipsweep, Nmap, port sweep, and satan & 4107 \\ 
normal & Normal packets & normal & 97278 \\ \hline
\end{tabular}
}
\caption{Sub-categories in each class.}
\label{tab:subcategory}
\end{table}

\begin{table}[]
\centering
\begin{tabular}{llllll}
\hline
SI.No. & Feature &  Type & SI.No & Feature  &  Type \\ \hline
0 & Duration & int64 & 22 & Count & int64 \\ 
1 & protocol\_type & object & 23 & srv\_count & int64 \\ 
2 & Service & object & 24 & serror\_rate & float64 \\ 
3 & Flag & object & 25 & srv\_serror\_rate & float64 \\ 
4 & src\_bytes & int64 & 26 & rerror\_rate & float64 \\ 
5 & dst\_bytes & int64 & 27 & srv\_rerror\_rate & float64 \\ 
6 & Land & int64 & 28 & same\_srv\_rate & float64 \\ 
7 & wrong\_fragment & int64 & 29 & diff\_srv\_rate & float64 \\ 
8 & Urgent & int64 & 30 & srv\_diff\_host\_rate & float64 \\ 
9 & Hot & int64 & 31 & dst\_host\_count & int64 \\ 
10 & num\_failed\_logins & int64 & 32 & dst\_host\_srv\_count & int64 \\ 
11 & logged\_in & int64 & 33 & dst\_host\_same\_srv\_rate & float64 \\
12 & num\_compromised & int64 & 34 & dst\_host\_diff\_srv\_rate & float64 \\ 
13 & root\_shell & int64 & 35 & dst\_host\_same\_src\_port\_rate & float64 \\ 
14 & su\_attempted & int64 & 36 & dst\_host\_srv\_diff\_host\_rate & float64 \\ 
15 & num\_root & int64 & 37 & dst\_host\_serror\_rate & float64 \\ 
16 & num\_file\_creations & int64 & 38 & dst\_host\_srv\_serror\_rate & float64 \\ 
17 & num\_shells & int64 & 39 & dst\_host\_rerror\_rate & float64 \\ 
18 & num\_access\_files & int64 & 40 & dst\_host\_srv\_rerror\_rate & float64 \\ 
19 & num\_outbound\_cmds & int64 & 41 & Outcome & object \\ 
20 & is\_host\_login & int64 & 42 & attack\_type & object \\ 
21 & is\_guest\_login & int64 & 43 & label & object \\ \hline
\end{tabular}
\caption{Features in the KDD-CUP'99 dataset.}
\label{tab:44feature}
\end{table}

\subsubsection{CIC-MalMem-2022}

\begin{figure*}[!htbp]
	\centering{\includegraphics[scale=.4]{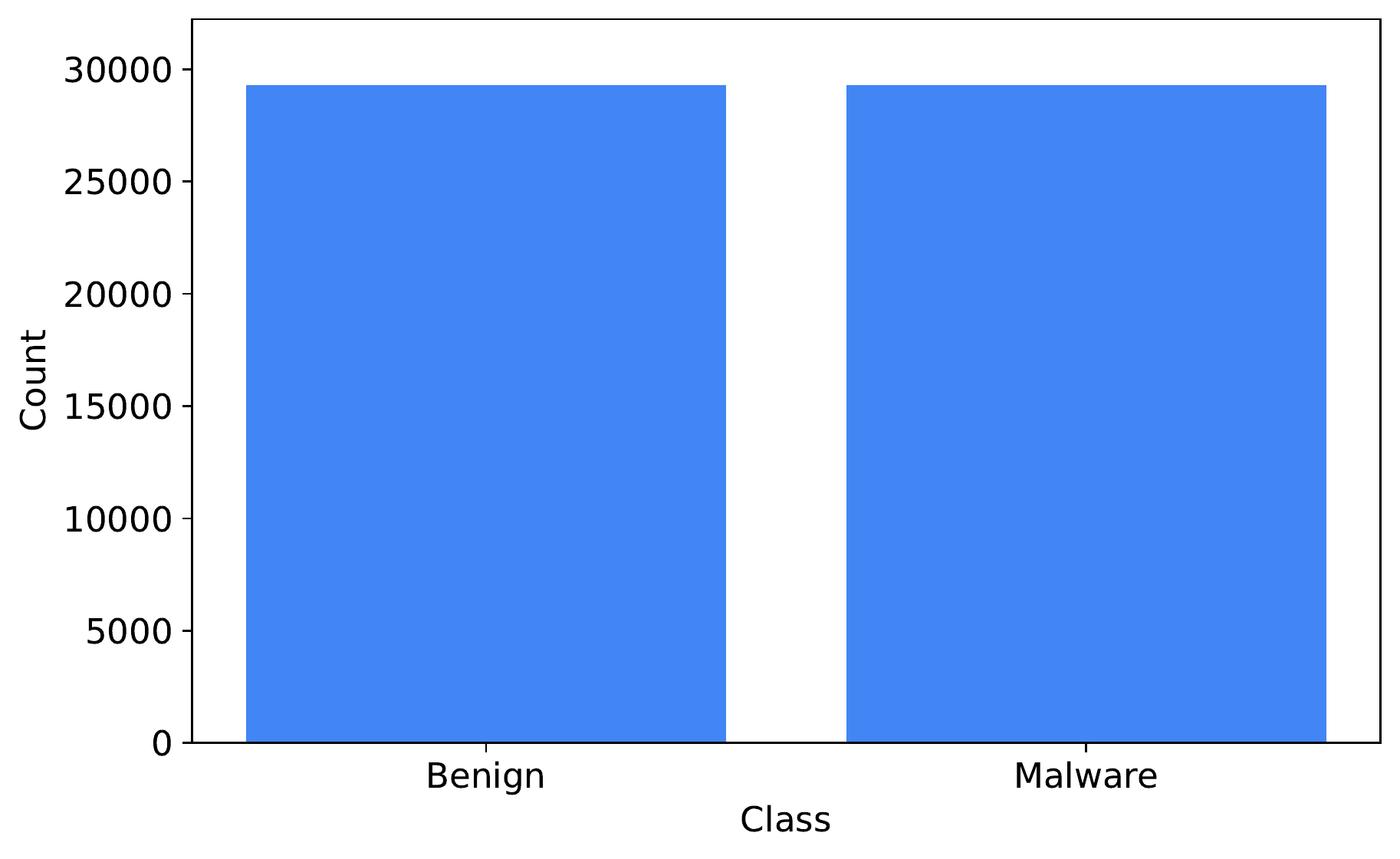}}\hspace{0.1cm}
	\caption{Distribution of attacks in CIC-MalMem-2022}
	\label{fig:data2_binary}
\end{figure*}

Malware that conceals to escape detection and elimination is known as obfuscated malware. The Obfuscated dataset \citep{icissp22} is a memory-based evaluation of obfuscated malware identification algorithms. It is a Malware Memory Analysis-2022 dataset. The data was intended to mimic a real scenario as closely as feasible by utilizing malware widely used in the real world. It's a fair dataset composed of Spyware, Ransomware, including Trojan Horse malware which may be employed to evaluate obfuscated malware monitoring systems. The dataset is balanced, with half of it conjured up of malignant memory dumps while half of it is composed of benign memory dumps. There are 58,596 entries in total, comprising 29,298 benign as well as 29,298 harmful entries. Figure \ref{fig:data2_binary} illustrates the distribution of attack categories using a bar chart for binary classification. All the features are described in the following Table \ref{tab:data2_description}.

\begin{table}[]
\centering
\begin{tabular}{llllll}
\hline
SI.No & Feature & Type & SI.No & Feature & Type \\ \hline
0 & Category & object & 29 & malfind .protection & int64 \\ 
1 & pslist.nproc & int64 & 30 & malfind.uniqueInjections & float64 \\ 
2 & pslist.nppid & int64 & 31 & psxview.not\_in\_pslist & int64 \\ 
3 & pslist.avg\_threads & float64 & 32 & psxview.not\_in\_eprocess\_pool & int64 \\ 
4 & pslist.nprocs64bit & int64 & 33 & psxview.not\_in\_ethread\_pool & int64 \\ 
5 & pslist.avg\_handlers & float64 & 34 & psxview.not\_in\_pspcid\_list & int64 \\ 
6 & dlllist.ndlls & int64 & 35 & psxview.not\_in\_csrss\_handles & int64 \\ 
7 & dlllist.avg\_dlls\_per\_proc & float64 & 36 & psxview.not\_in\_session & int64 \\ 
8 & handles.nhandles & int64 & 37 & psxview.not\_in\_deskthrd & int64 \\ 
9 & handles.avg\_handles\_per\_proc & float64 & 38 & psxview.not\_in\_pslist\_false\_avg & float64 \\ 
10 & handles.nport & int64 & 39 & psxview.not\_in\_eprocess\_pool\_false\_avg & float64 \\ 
11 & handles.nfile & int64 & 40 & psxview.not\_in\_ethread\_pool\_false\_avg & float64 \\ 
12 & handles.nevent & int64 & 41 & psxview.not\_in\_pspcid\_list\_false\_avg & float64 \\ 
13 & handles.ndesktop & int64 & 42 & psxview.not\_in\_csrss\_handles\_false\_avg & float64 \\ 
14 & handles.nkey & int64 & 43 & psxview.not\_in\_session\_false\_avg & float64 \\ 
15 & handles.nthread & int64 & 44 & psxview.not\_in\_deskthrd\_false\_avg & float64 \\ 
16 & handles.ndirectory & int64 & 45 & modules.nmodules & int64 \\ 
17 & handles.nsemaphore & int64 & 46 & svcscan.nservices & int64 \\ 
18 & handles.ntimer & int64 & 47 & svcscan.kernel\_drivers & int64 \\ 
19 & handles.nsection & int64 & 48 & svcscan.fs\_drivers & int64 \\ 
20 & handles.nmutant & int64 & 49 & svcscan.process\_services & int64 \\ 
21 & ldrmodules.not\_in\_load & int64 & 50 & svcscan.shared\_process\_services & int64 \\ 
22 & ldrmodules.not\_in\_init & int64 & 51 & svcscan.interactive\_process\_services & int64 \\ 
23 & ldrmodules.not\_in\_mem & int64 & 52 & svcscan.nactive & int64 \\
24 & ldrmodules.not\_in\_load\_avg & float64 & 53 & callbacks.ncallbacks & int64 \\ 
25 & ldrmodules.not\_in\_init\_avg & float64 & 54 & callbacks.nanonymous & int64 \\ 
26 & ldrmodules.not\_in\_mem\_avg & float64 & 55 & callbacks.ngeneric & int64 \\
27 & malfind.ninjections & int64 & 56 & Class & object \\ 
28 & malfind.commitCharge & int64 &  &  &  \\ \hline
\end{tabular}
\caption{Features for CIC-MalMem-2022.}
\label{tab:data2_description}
\end{table}

\subsection{Data Preparation}

The preprocessed dataset is fed into the ML and DL models. Preprocessing is accomplished by handling missing values, scaling features, and selecting a particular number of features.

\subsubsection{Handling Missing values}
The missing value occurs in the dataset due to data corruption or not recording data properly. Handling missing values is crucial in the data preprocessing phase, as many ML and DL algorithms trained with such data output incorrect results. We adopt simple methods to handle the missing value of the attributes. For example, missing values are handled by removing rows containing nan (null value) in our dataset, -inf, inf, and duplicate entries.

\subsubsection{Feature Scaling using Standardization}
Feature scaling refers to a procedure for normalizing the value of features within a certain range. We have employed a standardization method to scale the value of features in the dataset. Standardizing the features' value increases the accuracy of the model. In many use cases, the model's accuracy depends on the nature of the dataset utilized for training the model. The accuracy of ML and DL models is significantly reduced if the features' values have different measurement units or imbalanced range differences. To handle this issue, the standardization method is used to make attributes’ values within a certain balanced range. Each value of an attribute is normalized by subtracting the mean and dividing by the standard deviation of that attribute \citep{featurest}. Features' values are normalized using the equation\ref{eq:standarization}.  

\begin{equation}
X_{st} =\frac{x-mean(x)}{std(x)}
\label{eq:standarization}
\end{equation}

Here, $X_{st}$ is the standardized value,  $x$ is the actual value of an attribute, ${mean(x)}$ is the mean of actual value and ${std(x)}$ is the standard deviation of actual value.

\subsubsection{Label Encoding}
Label encoding is the procedure of converting category data to numerical values. To develop a model in an ML method, we must transform or encode categorical features into quantitative data in order to enter the data into the training module and develop a model. By replacing the categorical values with the value 0 to the total number of classes (n)-1 \citep{talukder2022machine}. We can use 0, 1, 2, 3, and 4 in place of all these various classes if the categorical data contain 5 different classes. The label encoding technique for binary and multilabel for the KDDCUP'99 dataset is shown in Tables \ref{tab:label_encode_binary} and \ref{tab:label_encode_multilabel}.

\begin{table}[!htbp]
\centering
\begin{tabular}{lc}
\hline
Attack types & Label encoding \\
\hline
Attack & 0 \\ 
Normal & 1 \\ \hline
\end{tabular}
\caption{Label encoding process.}
\label{tab:label_encode_binary}
\end{table}

\begin{table}[!htbp]
\centering
\begin{tabular}{lc}
\hline
Attack types & Label encoding \\
\hline
DoS & 0 \\ 
Normal & 1 \\ 
Probe & 2 \\ 
R2L & 3 \\ 
U2R & 4 \\ \hline
\end{tabular}
\caption{Label encoding process.}
\label{tab:label_encode_multilabel}
\end{table}

\subsection{Training Process}

The training process has been done using HP 250 G5 Notebook PC with Microsoft Windows 10 Pro 64bit (10.0.19042 build 19042) OS, Intel(R) Core (TM) i3-6006U CPU @ 2.00GHz with 8GB RAM. We have used the Jupyter Notebook 6.4.6 tool, and for the programming language, Python 3.8.5 has been used to implement the models. Pandas 1.3.4 and NumPy 1.19.5 frameworks have been used for data cleaning, extraction, feature selection and Matplotlib 3.5.0, Seaborn 0.11.2 framework for data visualization, and finally, Scikit-learn 0.24.1 package for data analysis. To evaluate the performance of our proposed method, we have used some metrics, including accuracy, precision, recall, f1-score, RMSE, and ROC Curve.

\section{Result Analysis}
The extension results are analyzed on various performance metrics to find the best model to detect network intrusion. We analyzed the performance based on all the features, selected features, and proposed features. The finding showed that the performance of our proposed feature outperforms the other two features.

\begin{figure*}[!htbp]
\centering
  \includegraphics[width=0.75\textwidth]{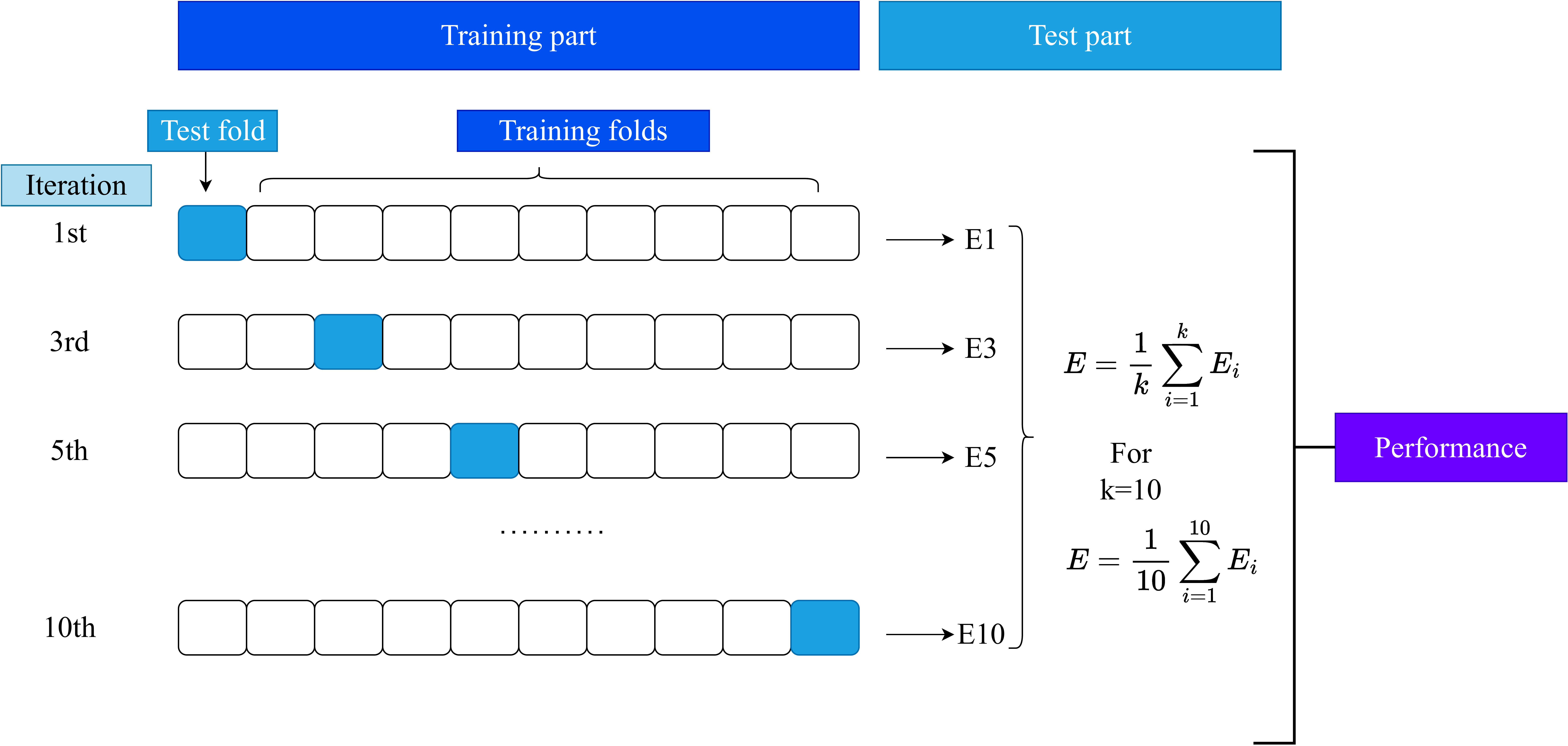}
\caption{K-fold cross-validation.}
\label{fig:kfold}
\end{figure*}

We have conducted experiments for multiclass and binary class-based intrusion detection. We have used ${k}$-fold cross-validation with the value of k=10 shown in Figure \ref{fig:kfold}. While 80\% of the dataset’s data have been selected for training, and the remaining 20\% of data are being tested, dividing the whole data into 10 different folds and taking a different fold every time during the experiments.

\subsection{Performance Analysis of KDDCUP'99 Dataset}

\begin{figure*}[!htbp]
	\centering
	\subfloat[Accuracy]{\includegraphics[scale=.50]{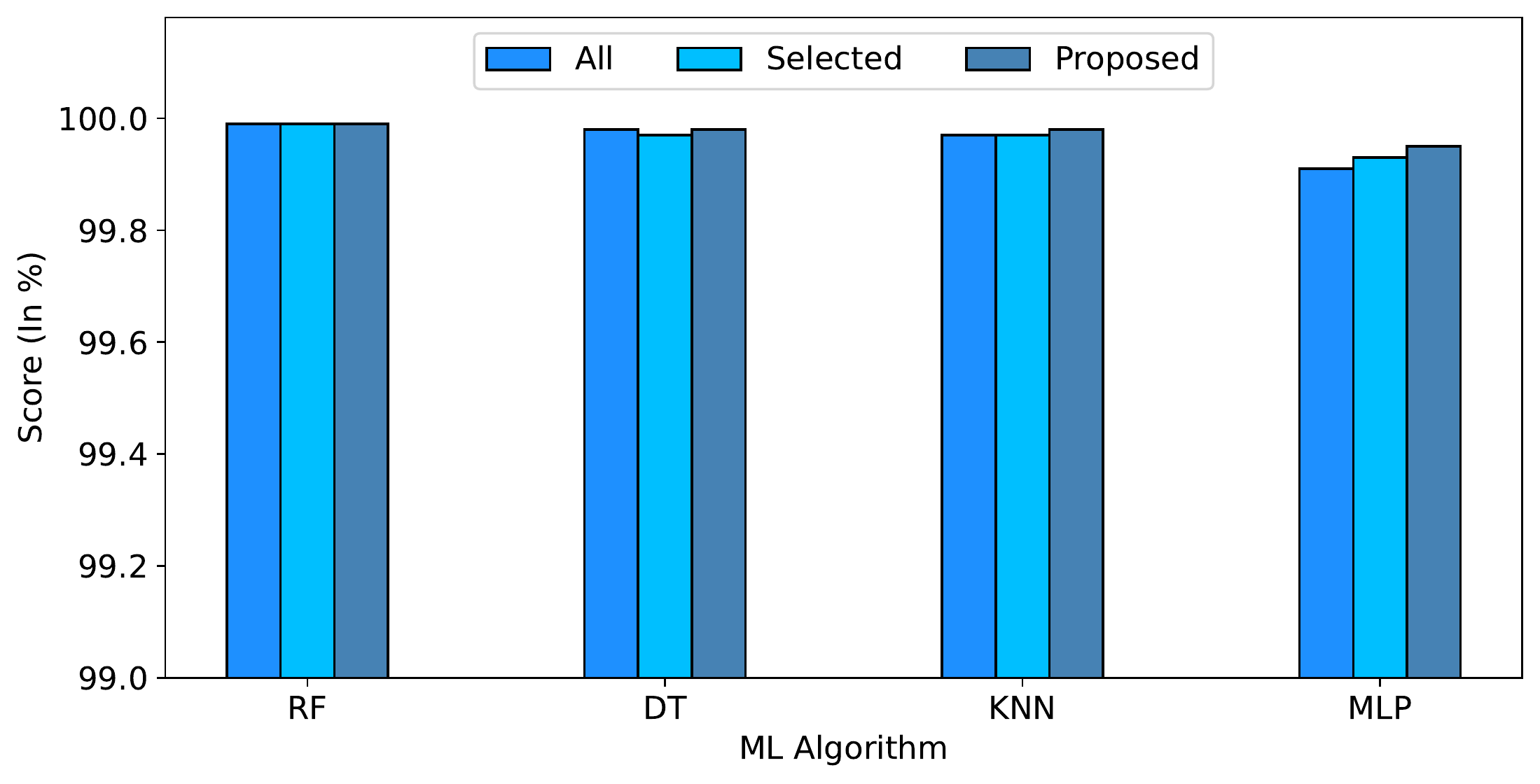}}\hspace{0.1cm}
	\subfloat[RMSE]{\includegraphics[scale=.50]{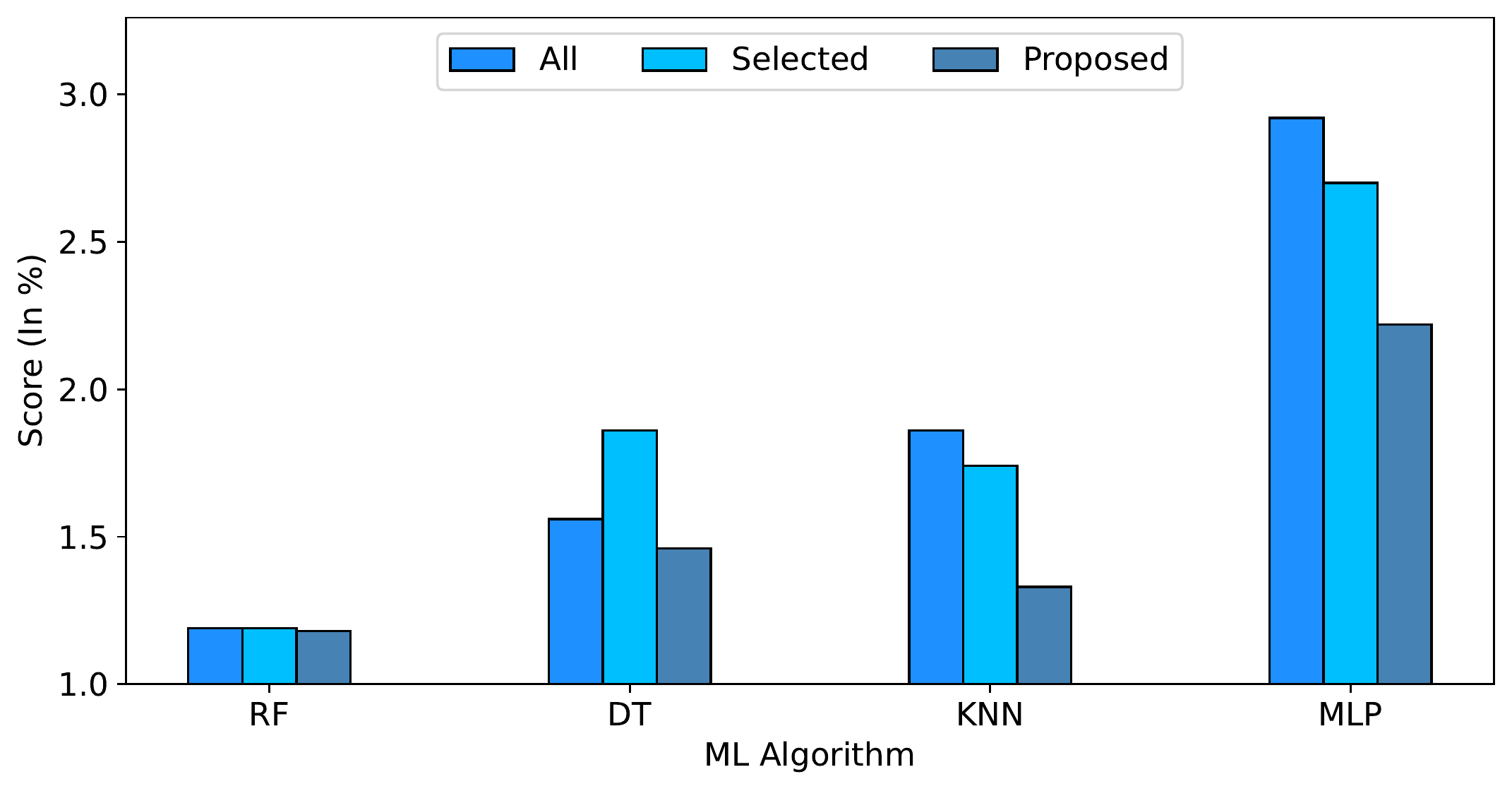}}
	\caption{Performance comparison graphs for binary classification.}
	\label{fig:kdd_comparision_binary_graph}
\end{figure*}

The performance comparison results for binary classification for the KDDCUP'99 dataset are illustrated in Table \ref{tab:kdd_comparision_binary} and Fig. \ref{fig:kdd_comparision_binary_graph} in tabular and bar chart format, respectively. The experiment results for considering all features, 20 features without applying smote, and the proposed model’s performance is represented. Here, all indicate that the features that are not selected and not applied smote are just pre-proposed, scaled, and then applied to the machine learning algorithms to build models and evaluate the performance. On the other hand, the proposed model indicates all the proposed methodology processes by which evaluation has been measured.

\begin{table}[]
\centering
\resizebox{\textwidth}{!}{%
\begin{tabular}{llllllllllllllll}
\hline
\multirow{2}{*}{ML} & \multicolumn{3}{c}{Accuracy} & \multicolumn{3}{c}{Precision} & \multicolumn{3}{c}{Recall} & \multicolumn{3}{c}{F1-score} & \multicolumn{3}{c}{RMSE}  \\
                    & All  & Selected  & Proposed  & All   & Selected  & Proposed  & All  & Selected & Proposed & All  & Selected  & Proposed  & All & Selected & Proposed \\ \hline
RF  & 99.99 & 99.99 & 99.99 & 99.97 & 99.97 & 99.99 & 99.99 & 99.99 & 99.99 & 99.98 & 99.98 & 99.99 & 1.19 & 1.19 & 1.18 \\
DT  & 99.98 & 99.97 & 99.98 & 99.96 & 99.94 & 99.98 & 99.97 & 99.95 & 99.98 & 99.96 & 99.95 & 99.98 & 1.56 & 1.86 & 1.46 \\
KNN & 99.97 & 99.97 & 99.98 & 99.94 & 99.95 & 99.98 & 99.95 & 99.95 & 99.98 & 99.95 & 99.95 & 99.98 & 1.86 & 1.74 & 1.33 \\
MLP & 99.91 & 99.93 & 99.95 & 99.87 & 99.87 & 99.95 & 99.86 & 99.9  & 99.95 & 99.87 & 99.89 & 99.95 & 2.92 & 2.7  & 2.22 \\ \hline
\end{tabular}%
}
\caption{Performance results for binary classification.}
\label{tab:kdd_comparision_binary}
\end{table}

The accuracy rate of our proposed scheme in binary classification is substantially higher, as shown in the Table and Bar Chart. The accuracy rising rate among the three bar graphs is high, and the RMSE rate is significantly lower than all selected features.

From the binary comparison graph fig \ref{fig:kdd_comparision_binary_graph}, the accuracy of the proposed model for RF, DT, KNN, and MLP is 99.99\%, 99.98\%, 99.98\%, and 99.95\%, respectively. From the graph fig \ref{fig:kdd_comparision_binary_graph}(a), it is clear that the accuracy increases by considering all the features with the proposed model are 0\%, 0\%, 0.01\%, 0.04\%; by considering selected features with the proposed model are 0\%, 0.01\%, 0.01\%, 0.02\% for RF, DT, KNN, MLP respectively that proves the effectiveness of the proposed models with a smaller number of features. Fig \ref{fig:kdd_comparision_binary_graph}(b) shows a variation in the RMSE values for different features and the proposed model. By considering all the features with the proposed models the reduced RMSE error are 0.01\%, 0.1\%, 0.53\%, 0.70\%; considering selected features with the proposed models the reduced RMSE error are 0.01\%, 0.40\%, 0.41\%, 0.48\% for RF, DT, KNN, MLP respectively.

\begin{table}[]
\centering
\begin{tabular}{llllllll}
\hline
ML & Accuracy & Precision & Recall & F1-score & MAE & MSE & RMSE \\ \hline
CNN & 99.95 & 99.95 & 99.95 & 99.95 & 0.05 & 0.05 & 2.27 \\ 
ANN & 99.93 & 99.93 & 99.93 & 99.93 & 0.07 & 0.07 & 2.68 \\ \hline
\end{tabular}
\caption{Performance analysis for binary classification.}
\label{tab:kdd_binary_nn}

\end{table}

We also explore two neural network models, namely ANN and CNN, in our approach, where in Table \ref{tab:kdd_binary_nn} shows that the performance of CNN is greater than ANN, which is 0.02\%.

\begin{figure*}[!htbp]
	\centering
	\subfloat[Performance]{\includegraphics[scale=.50]{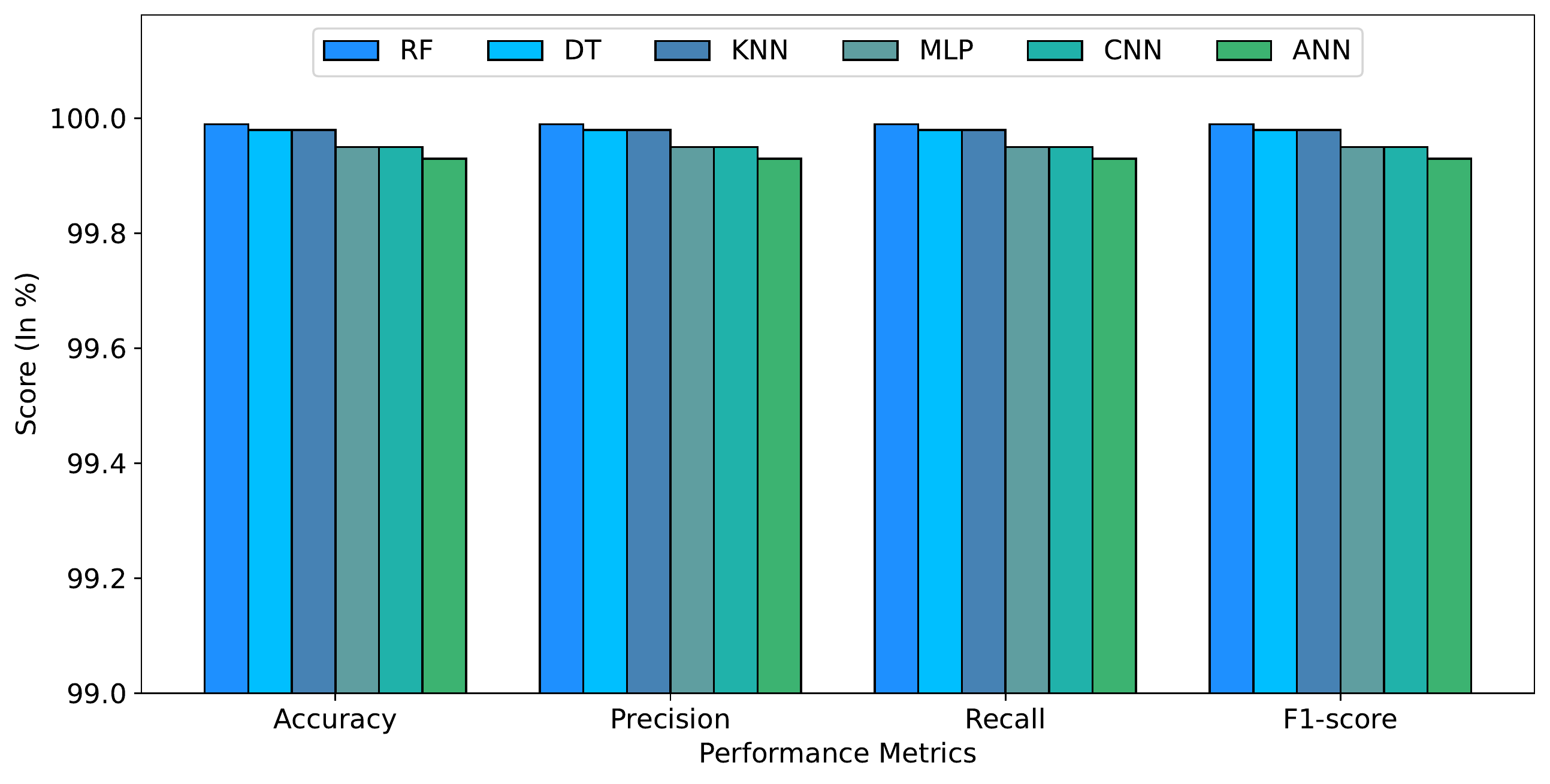}}\hspace{0.1cm}
	\subfloat[Error]{\includegraphics[scale=.50]{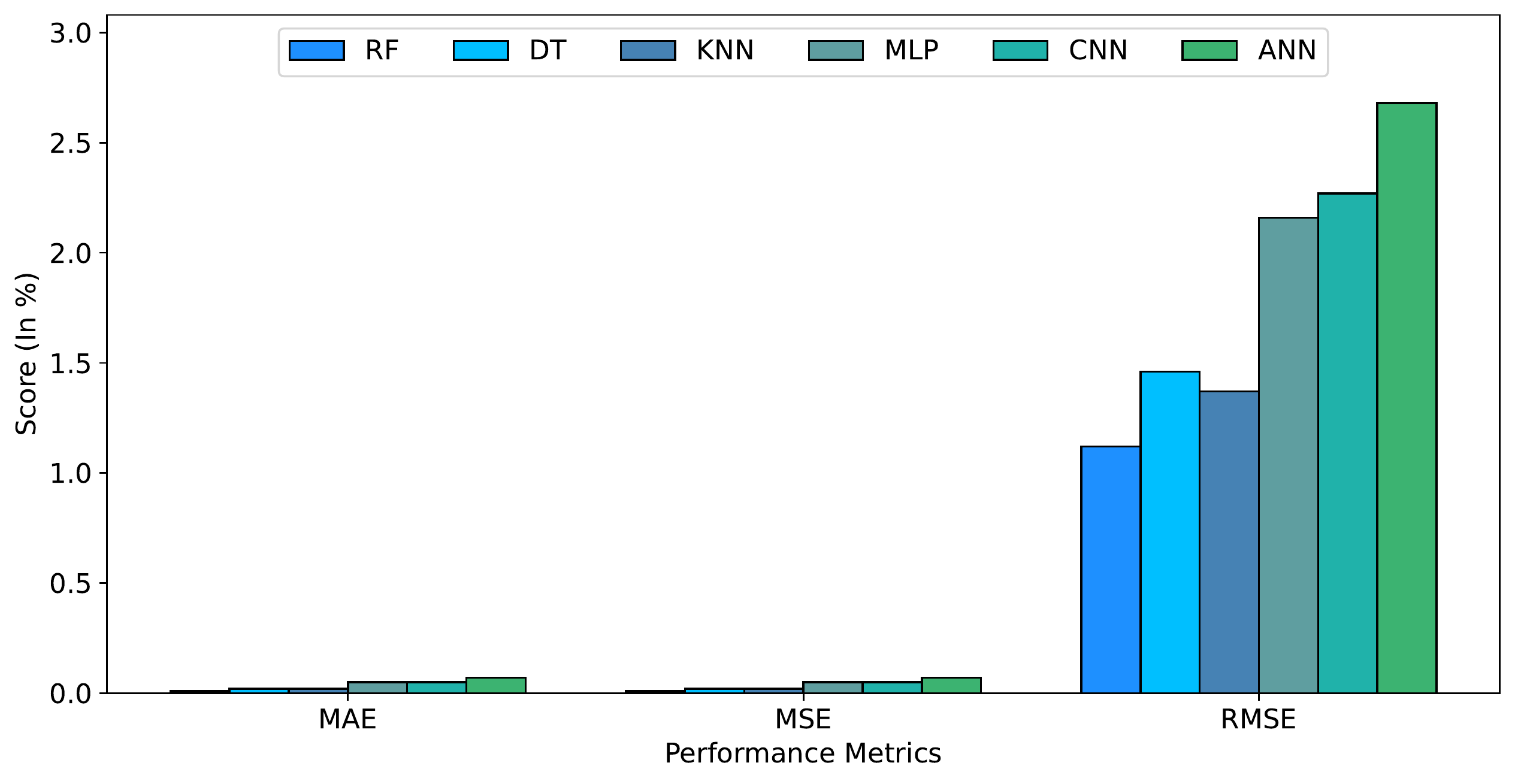}}
	\caption{Performance analysis graphs for binary classification.}
	\label{fig:kdd_performance_binary_graph}
\end{figure*}

The Performance analysis graphs for binary classification are shown in fig \ref{fig:kdd_performance_binary_graph} where the accuracy rate among all the algorithms RF gives the better performance, which is 99.99\%, and ann gives lower accuracy, which is 99.93\%. The RMSE error rate for RF is 1.18\%, and ANN is 2.68\%.

\begin{figure*}[!htbp]
	\centering
	\subfloat[RF]{\includegraphics[scale=.385]{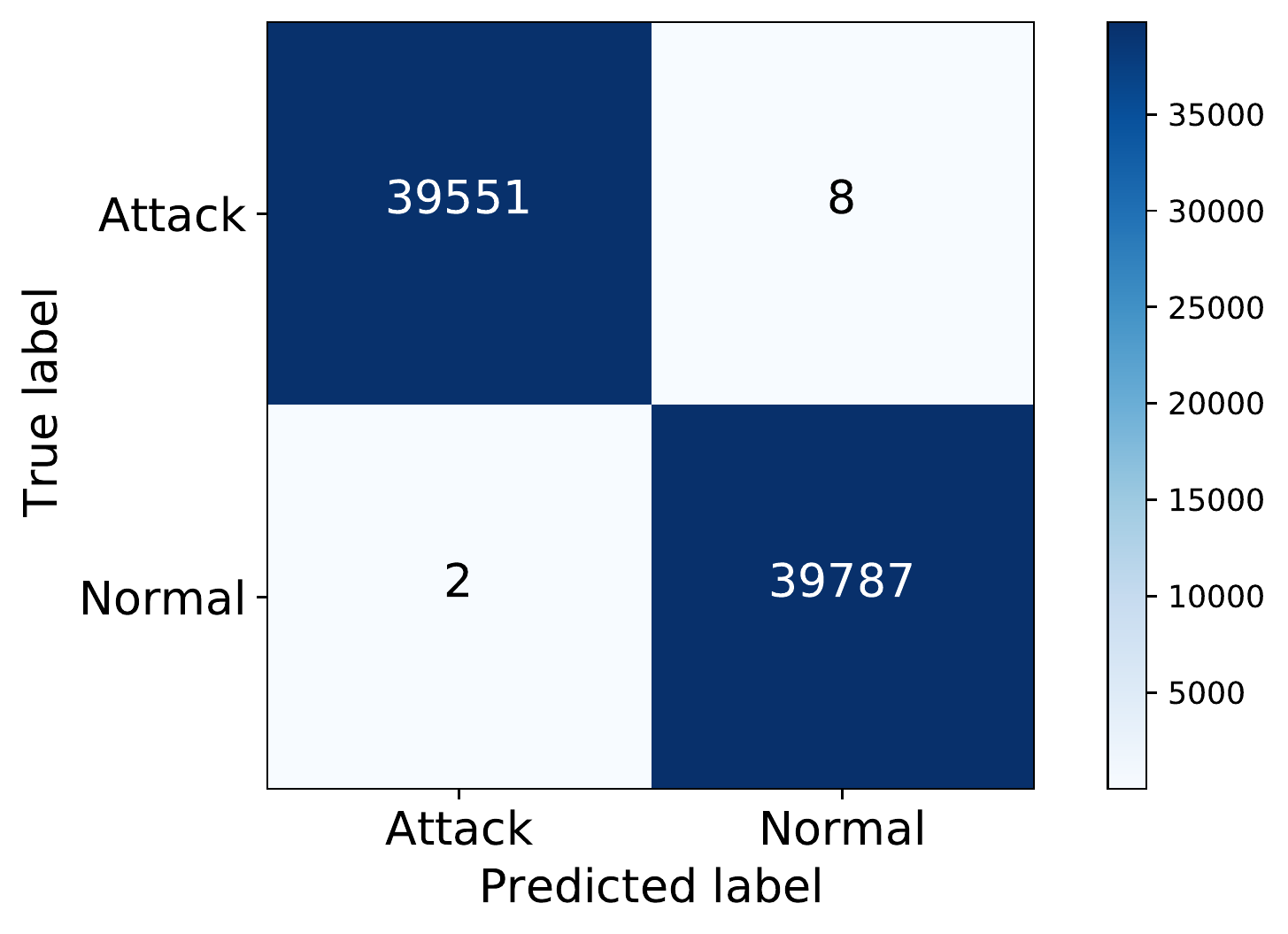}}\hspace{.1cm}
	\subfloat[DT]{\includegraphics[scale=.385]{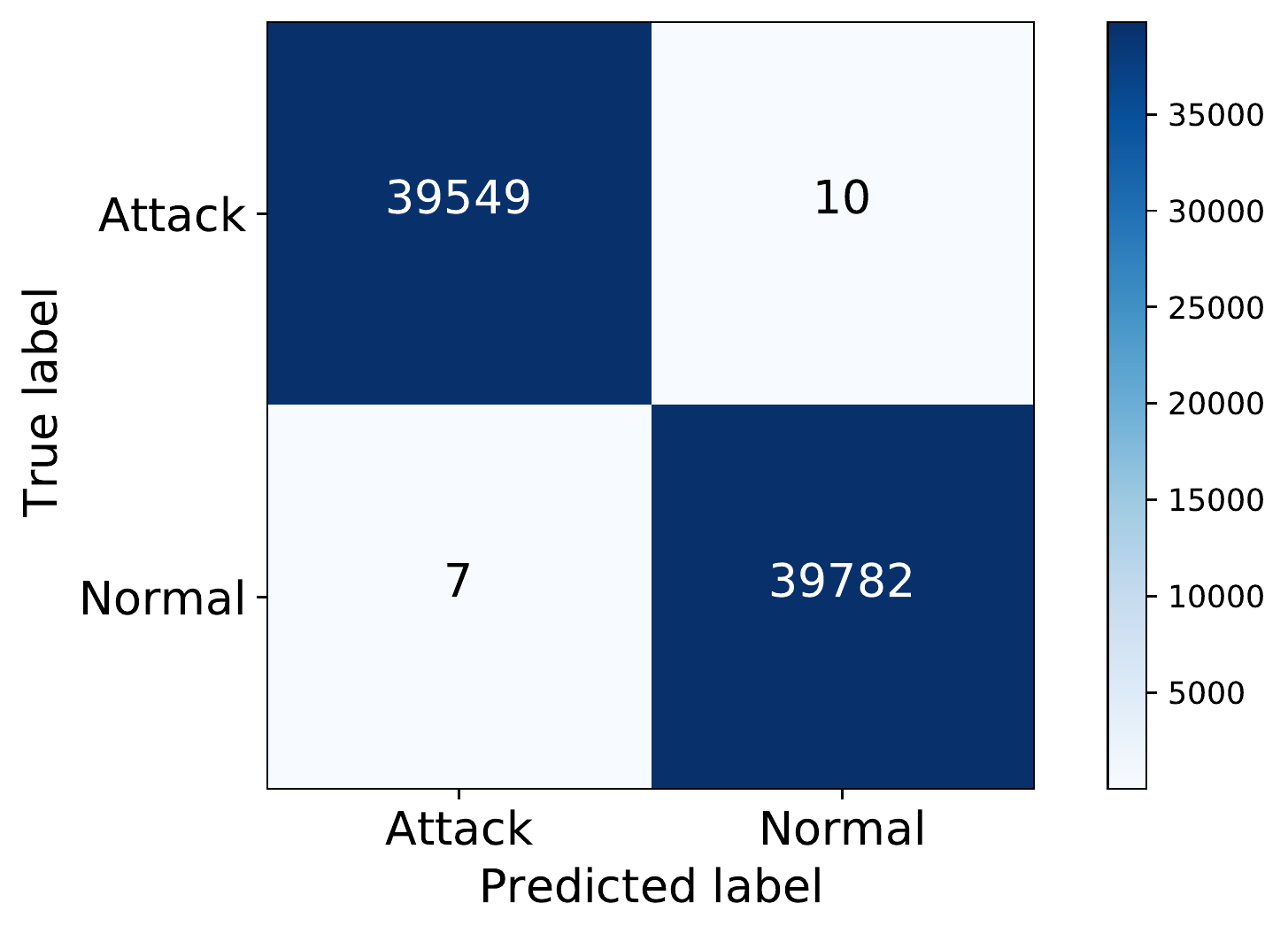}}\hspace{.1cm}
	\subfloat[KNN]{\includegraphics[scale=.385]{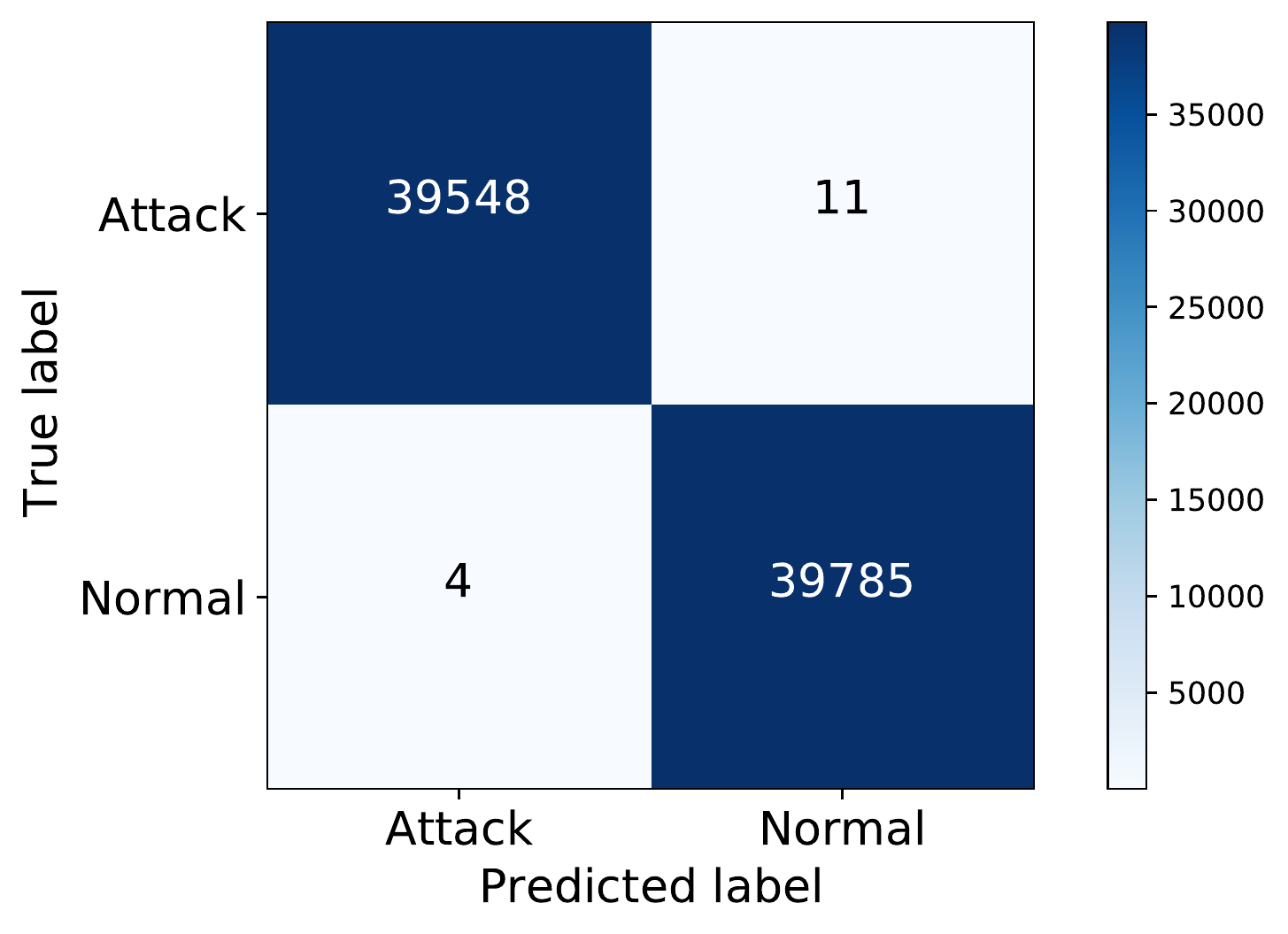}}\hspace{.1cm}
	\subfloat[MLP]{\includegraphics[scale=.385]{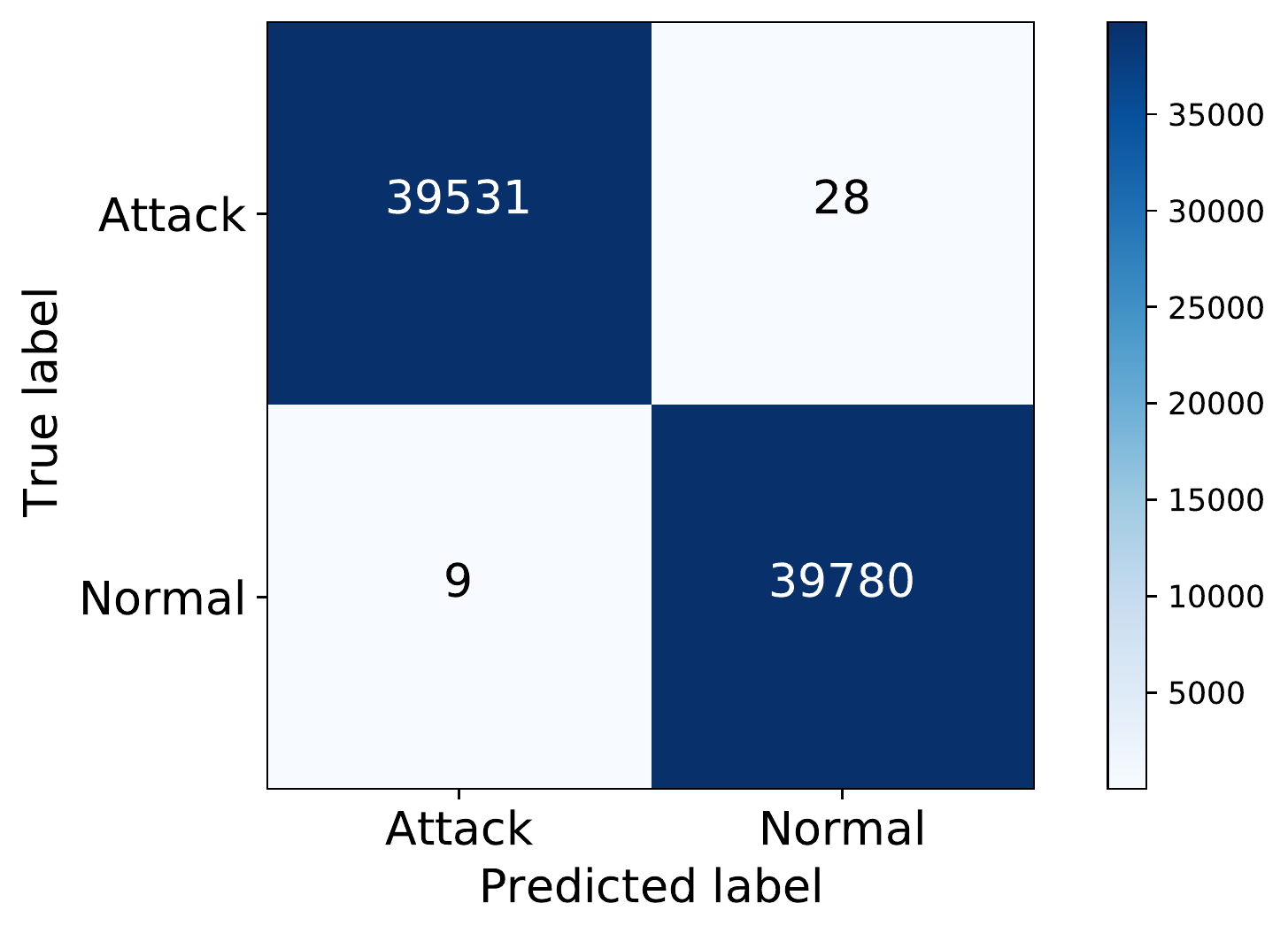}}\hspace{.1cm}
	\subfloat[CNN]{\includegraphics[scale=.385]{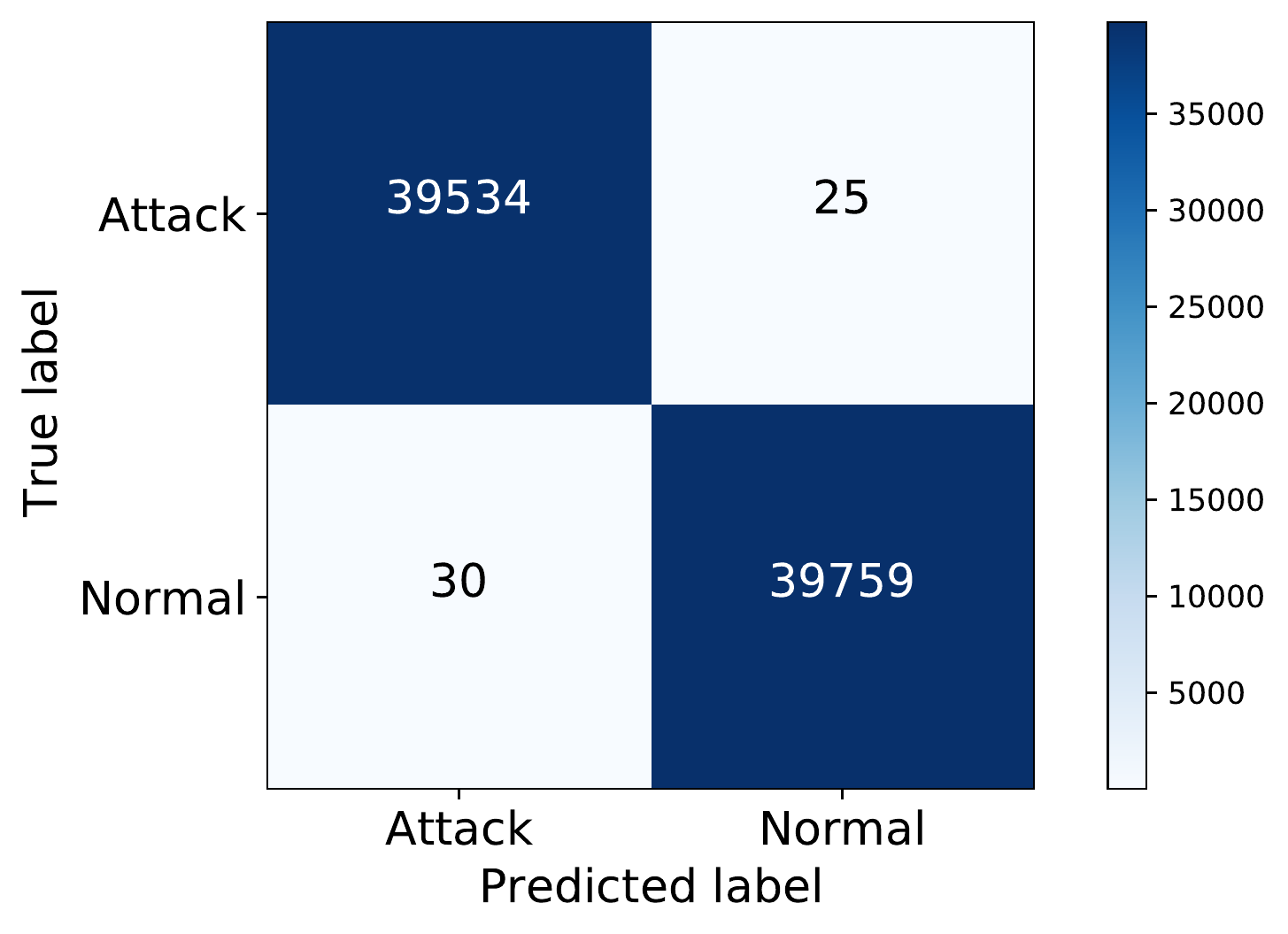}}\hspace{.1cm}
	\subfloat[ANN]{\includegraphics[scale=.385]{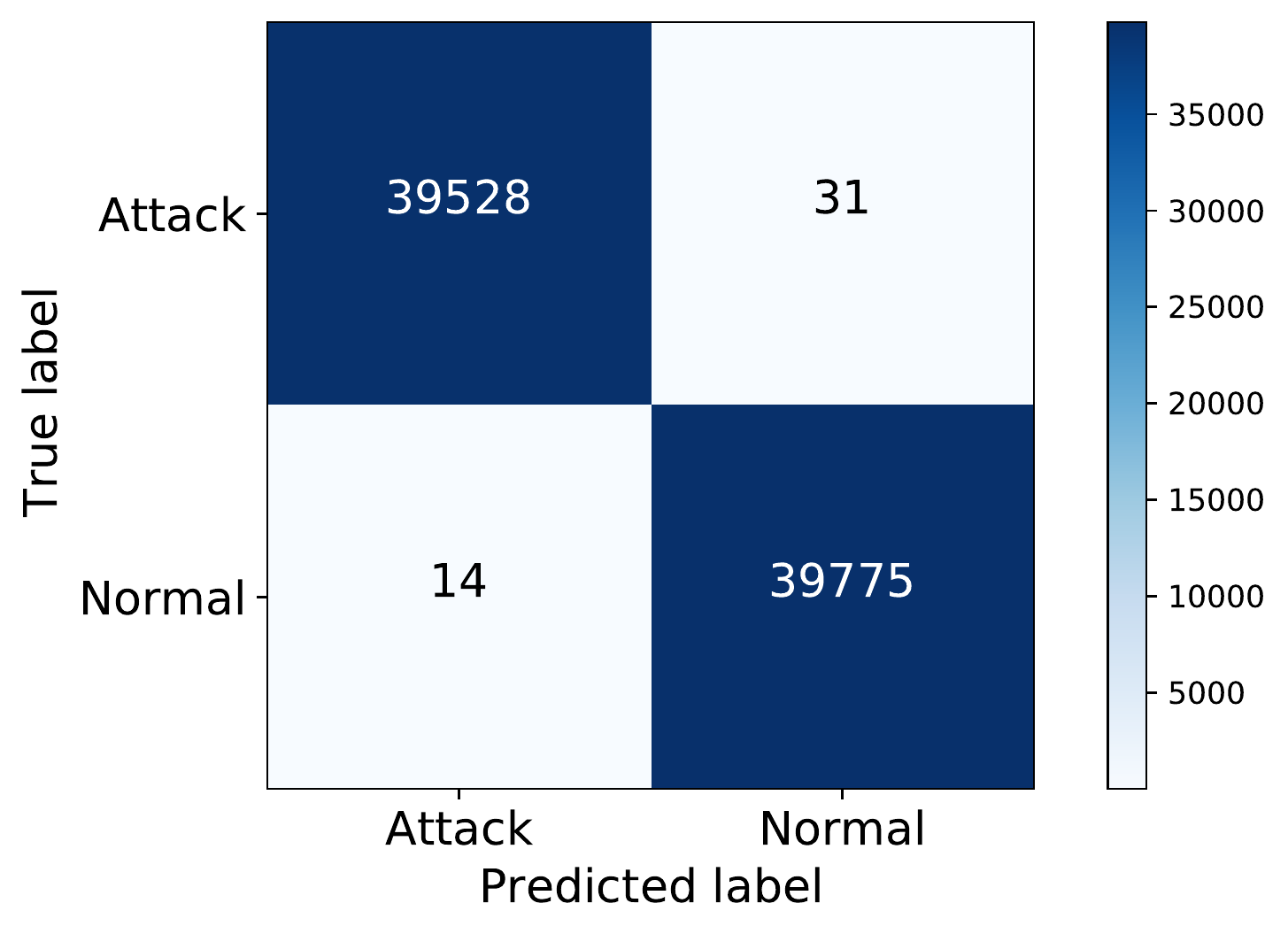}}\hspace{.1cm}
	
	\caption{Confusion matrix for binary classification of KDDCUP'99.}
	\label{fig:kdd_binary_confusion}
\end{figure*}

Fig \ref{fig:kdd_binary_confusion} shows the confusion matrix for all the proposed algorithms. The TP, TN, FP, FN rates are 49.84\%, 50.14\%, 0\%, 0.01\%; 49.84\%, 50.14\%, 0.01\%, 0.01\%; 49.84\%, 50.14\%, 0.01\%, 0.01\%; 49.82\%, 50.13\%, 0.01\%, 0.04\%; 49.82\%, 50.12\%, 0.02\%, 0.03\%; 49.8\%, 50.13\%, 0.01\%, 0.06\% for RF, DT, KNN, MLP, CNN and ANN respectively. From the confusion matrix results, it is visible that RF outperforms the other algorithms using 20 features with higher rates of TP and TN with very lower rates of FP and FN. A random forest is essentially a set of individual decision trees that work together to form an ensemble. Thus, multiple learners' algorithms can be trained with the knowledge to achieve the highest accuracy.

\begin{figure*}[!htbp]
	\centering
	\subfloat[ROC Curve]{\includegraphics[scale=.47]{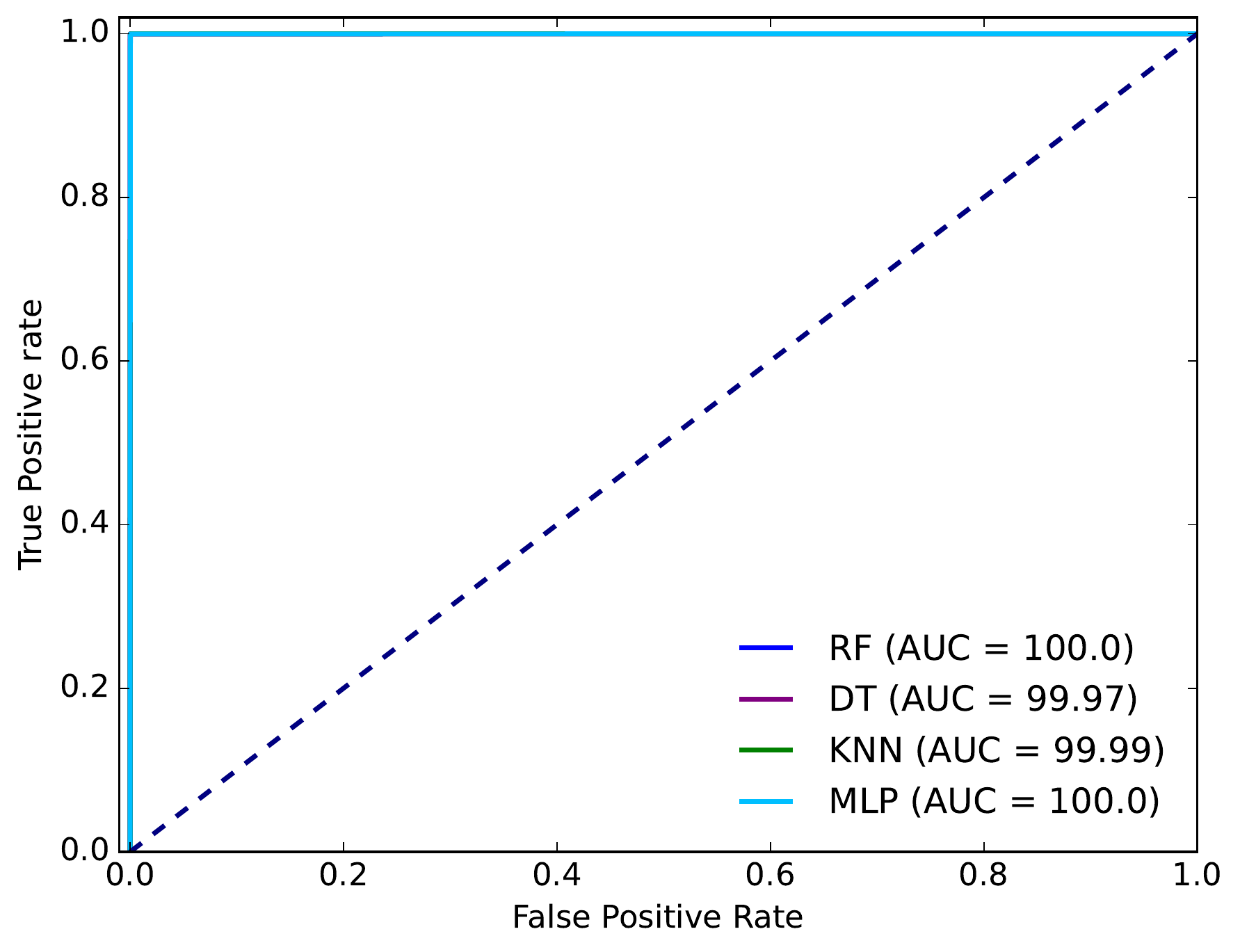}}\hspace{.1cm}
	\caption{Binary ROC Curve for KDDCUP'99.}
	\label{fig:kddcup_binary_roc_curve}
\end{figure*}

\begin{figure*}[!htbp]
	\centering
	\subfloat[Accuracy]{\includegraphics[scale=.50]{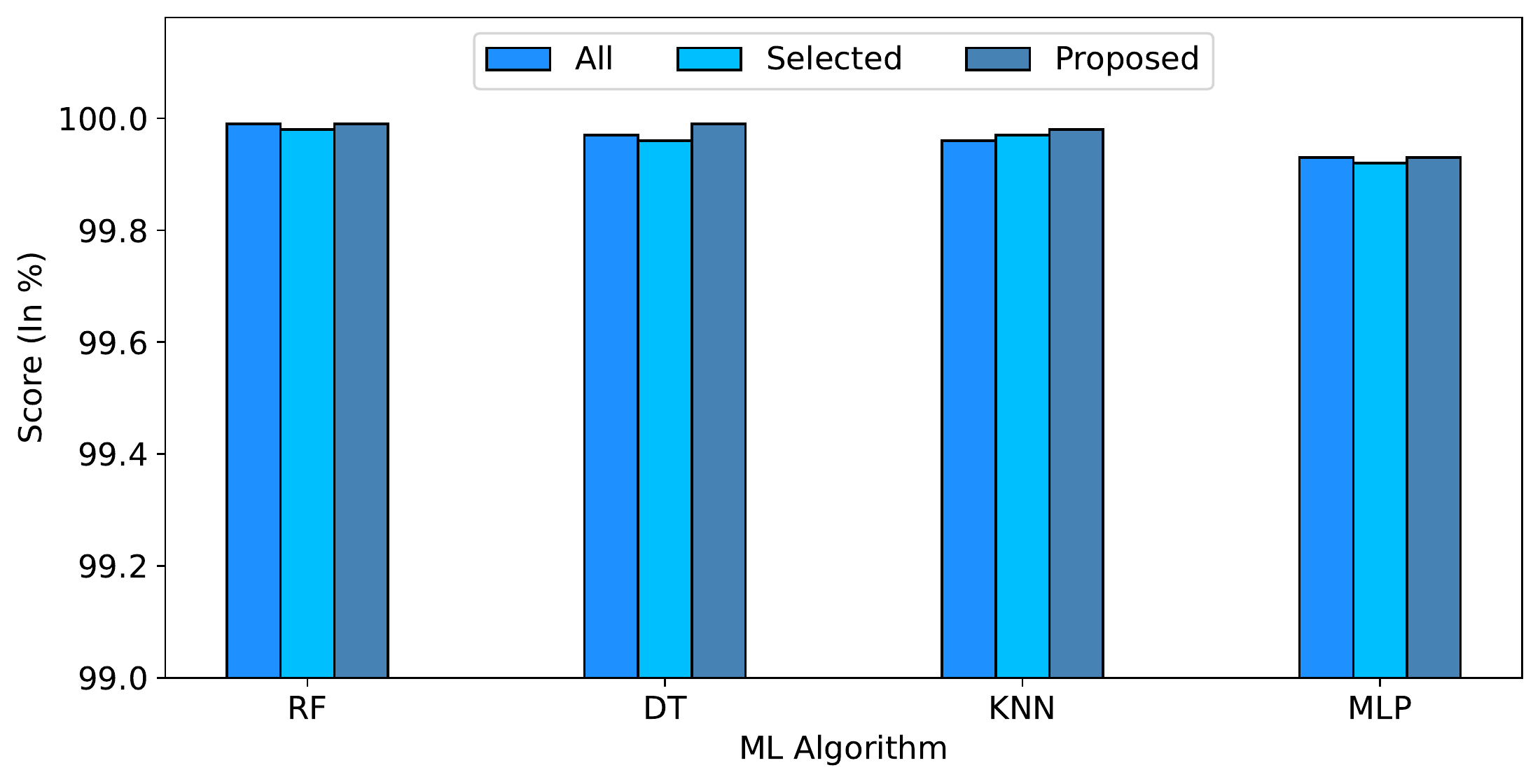}}\hspace{0.1cm}
	\subfloat[RMSE]{\includegraphics[scale=.50]{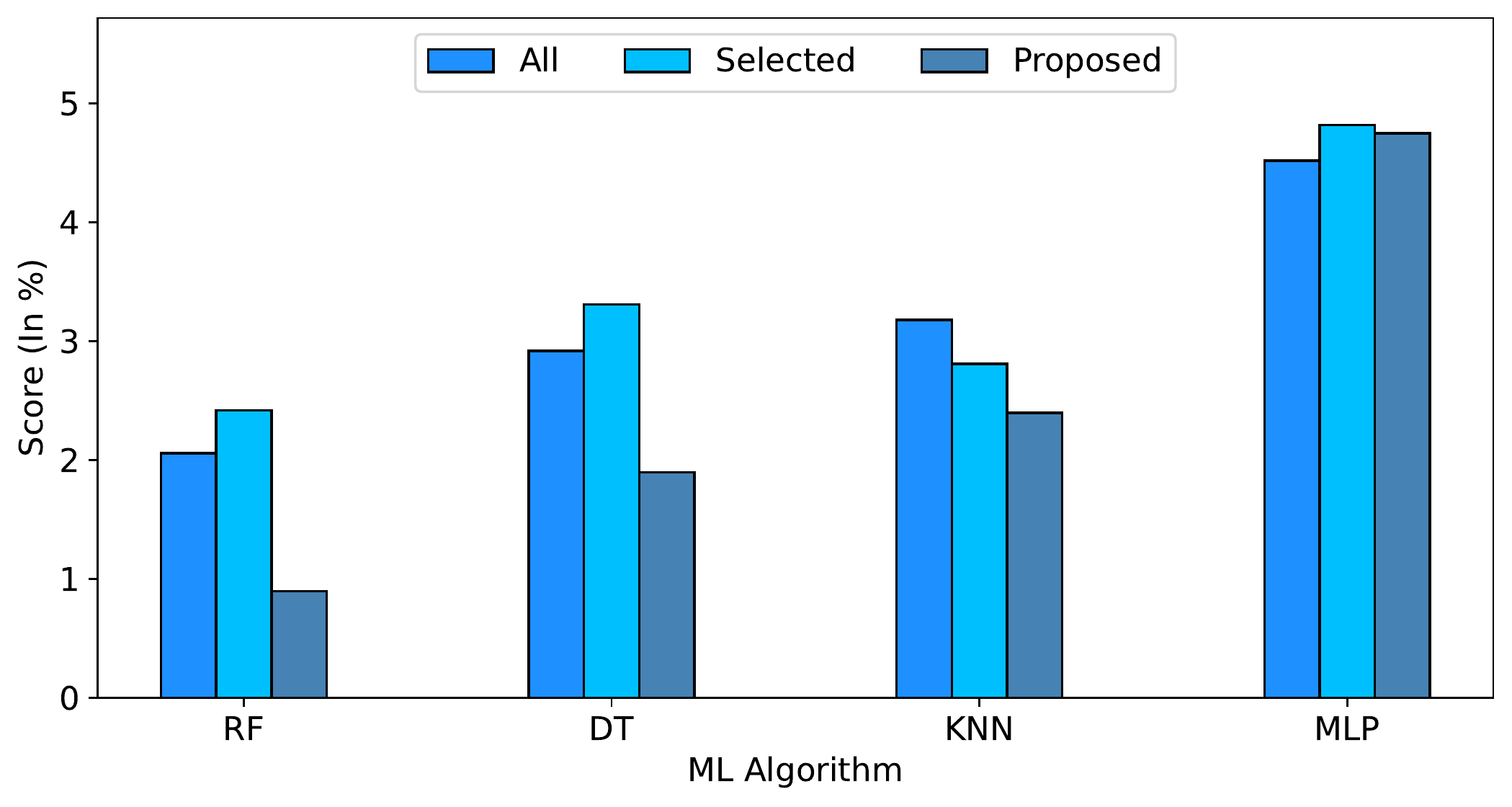}}
	\caption{Performance comparison graphs for multilabel classification.}
	\label{fig:kdd_comparision_multilabel_graph}
\end{figure*}

Fig \ref{fig:kddcup_binary_roc_curve} shows the Binary ROC Curve for KDDCUP'99, where the AUC score for RF, DT, KNN, and MLP are  100\%, 99.97\%, 99.99\%, and 100\%, respectively. Among all the RF and MLP give the highest accuracy rate. MLP consists of a series of layers made up of neurons and their connections. It has one or more hidden layers between the input and output layers where the neurons are arranged in layers and connections are often guided from lower to upper layers.
\\
The performance comparison results for multilabel classification for the KDDCUP'99 dataset are illustrated in Table \ref{tab:kdd_comparision_multilabel} and Fig. \ref{fig:kdd_comparision_multilabel_graph} in tabular and bar chart format, respectively. The experiment results for considering all features, 20 features without applying smote, and the proposed model’s performance is represented. Here, all indicate that the features that are not selected and not applied smote are just pre-proposed, scaled, and then applied to the machine learning algorithms to build models and evaluate the performance. On the other hand, the proposed model indicates all the proposed methodology processes by which evaluation has been measured.

\begin{table}[]
\centering
\resizebox{\textwidth}{!}{
\begin{tabular}{cccccccccccccccc}
\hline
\multirow{2}{*}{ML} & \multicolumn{3}{c}{Accuracy} & \multicolumn{3}{c}{Precision} & \multicolumn{3}{c}{Recall} & \multicolumn{3}{c}{F1-score} & \multicolumn{3}{c}{RMSE}  \\
                    & All  & Selected  & Proposed  & All   & Selected  & Proposed  & All  & Selected & Proposed & All  & Selected  & Proposed  & All & Selected & Proposed \\ \hline
RF  & 99.99 & 99.98 & 99.99 & 99.99 & 99.94 & 99.99 & 92.94 & 92.9  & 99.99 & 95.8  & 95.75 & 99.99 & 2.06 & 2.42 & 0.9  \\
DT  & 99.97 & 99.96 & 99.99 & 99.72 & 92.67 & 99.99 & 92.55 & 92.54 & 99.99 & 95.47 & 92.6  & 99.99 & 2.92 & 3.31 & 1.9  \\
KNN & 99.96 & 99.97 & 99.98 & 99.42 & 99.72 & 99.98 & 92.63 & 92.59 & 99.98 & 95.36 & 95.48 & 99.98 & 3.18 & 2.81 & 2.4  \\
MLP & 99.93 & 99.92 & 99.93 & 78.75 & 77.91 & 99.93 & 78.6  & 78.84 & 99.93 & 78.68 & 78.36 & 99.93 & 4.52 & 4.82 & 4.75 \\ \hline
\end{tabular}
}
\caption{Performance results for multilabel classification.}
\label{tab:kdd_comparision_multilabel}
\end{table}

The accuracy rate of our proposed scheme in multilabel classification is substantially higher, as shown in the bar chart. The accuracy rising rate among the three bar graphs is high, and the RMSE rate is significantly lower than all selected features.

From the multilabel comparison graph fig \ref{fig:kdd_comparision_multilabel_graph}, the accuracy of the proposed model for RF, DT, KNN, and MLP is 99.99\%, 99.99\%, 99.98\%, and 99.93\%, respectively. From graph \ref{fig:kdd_comparision_multilabel_graph}(a), it is clear that the accuracy increases by considering all the features with the proposed model are 0\%, 0.02\%, 0.02\%, 0\%; by considering selected features with the proposed models are 0.01\%, 0.03\%, 0.01\%, 0.01\% for RF, DT, KNN, MLP respectively that proves the effectiveness of the proposed model. \ref{fig:kdd_comparision_multilabel_graph}(b) shows a variation in the RMSE values for different features and the proposed models. By considering all the features with the proposed models the reduced RMSE error is 1.16\%, 1.02\%, 0.78\% and 0.23\%; considering selected features with the proposed models the reduced RMSE error is 1.52\%, 1.41\%, 0.41\%, 0.07\% for RF, DT, KNN, MLP respectively.

\begin{table}[]
\centering
\begin{tabular}{llllllll}
\hline
ML & Accuracy & Precision & Recall & F1-score & MAE & MSE & RMSE \\ \hline
CNN & 99.84 & 99.84 & 99.84 & 99.84 & 0.26 & 0.49 & 6.97 \\ 
ANN & 99.86 & 99.86 & 99.86 & 99.86 & 0.22 & 0.41 & 6.42 \\ \hline
\end{tabular}
\caption{Performance analysis for multilabel classification.}
\label{tab:kdd_multilabel_nn}
\end{table}

We also explore two neural network models, namely ANN and CNN, in our approach, wherein Table \ref{tab:kdd_multilabel_nn} shows that the performance of ANN is greater than CNN, which is 0.02\%. 

\begin{figure*}[!htbp]
	\centering
	\subfloat[Performance]{\includegraphics[scale=.50]{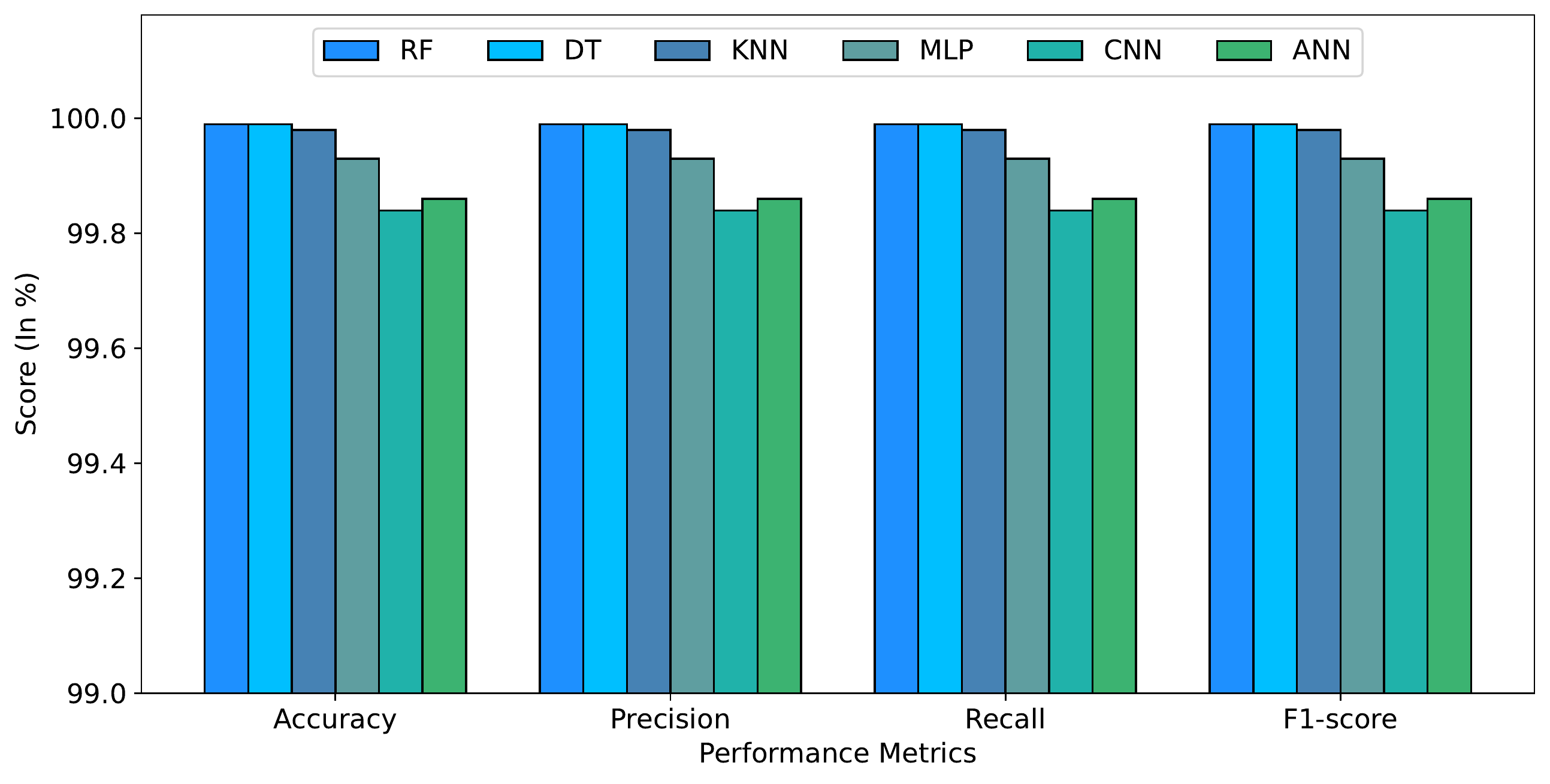}}\hspace{0.1cm}
	\subfloat[Error]{\includegraphics[scale=.50]{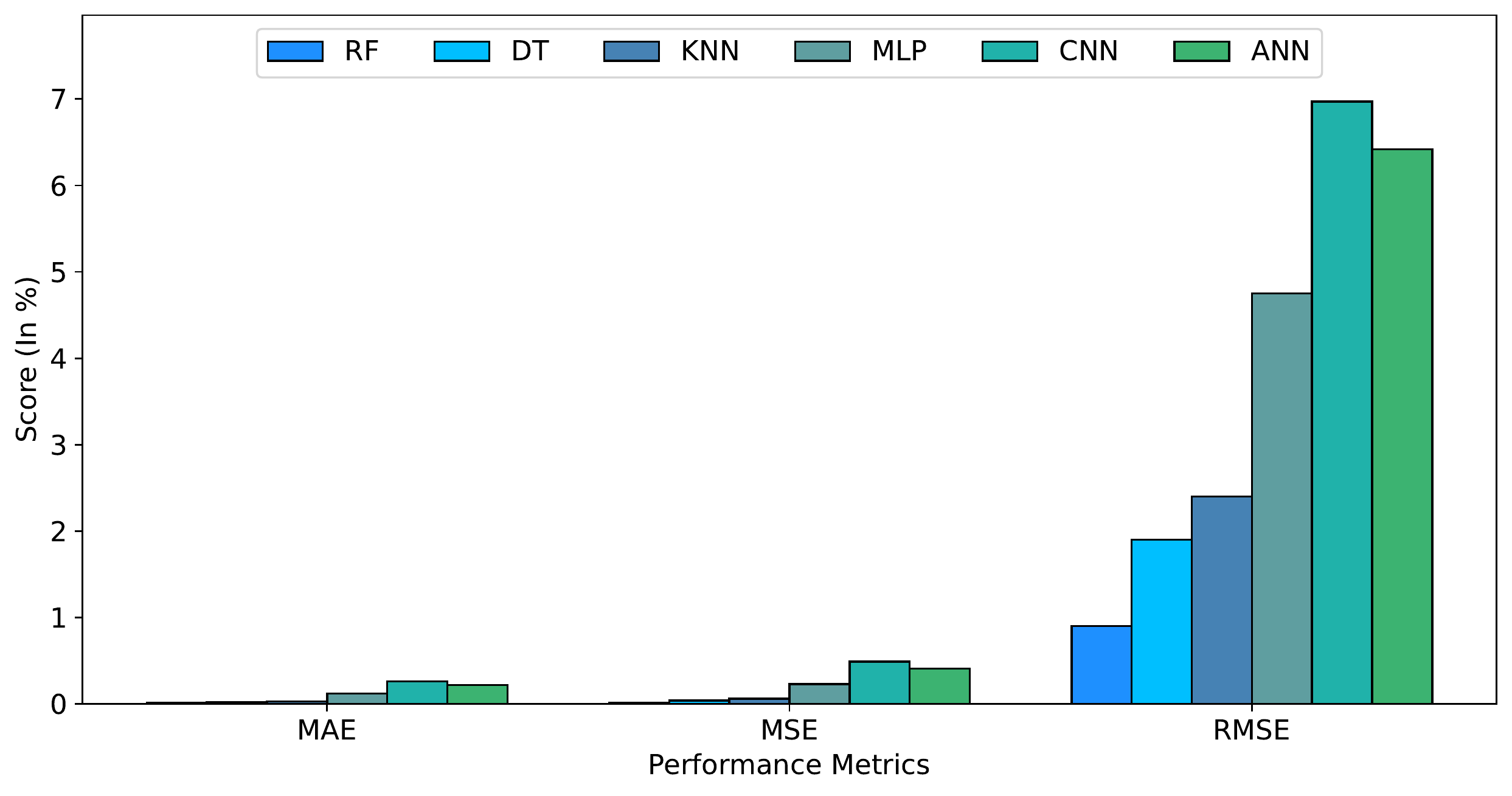}}
	\caption{Performance analysis graphs for multilabel classification.}
	\label{fig:kdd_performance_multilabel_graph}
\end{figure*}

The Performance analysis graphs for multilabel classification are shown in fig \ref{fig:kdd_performance_multilabel_graph} where the accuracy rate among all the algorithms RF and DT gives better performance, which is 99.99\%, and CNN gives lower accuracy, which is 99.84\%. The RMSE error rate for RF is 0.9\%, and ANN is 6.97\%.

\begin{figure*}[!htbp]
	\centering
	\subfloat[RF]{\includegraphics [scale=.2800]{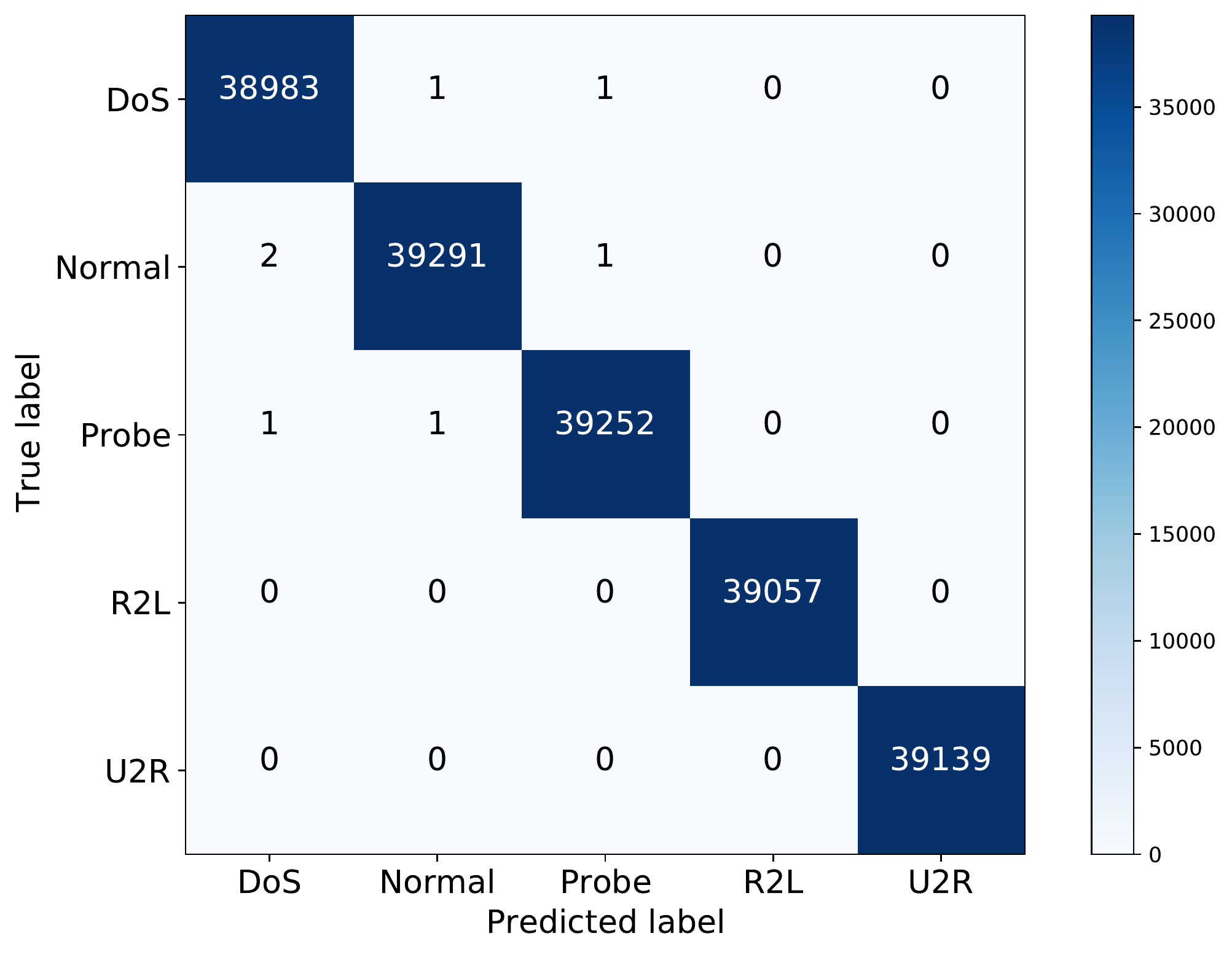}}\hspace{.1cm}
	\subfloat[DT]{\includegraphics [scale=.2800]{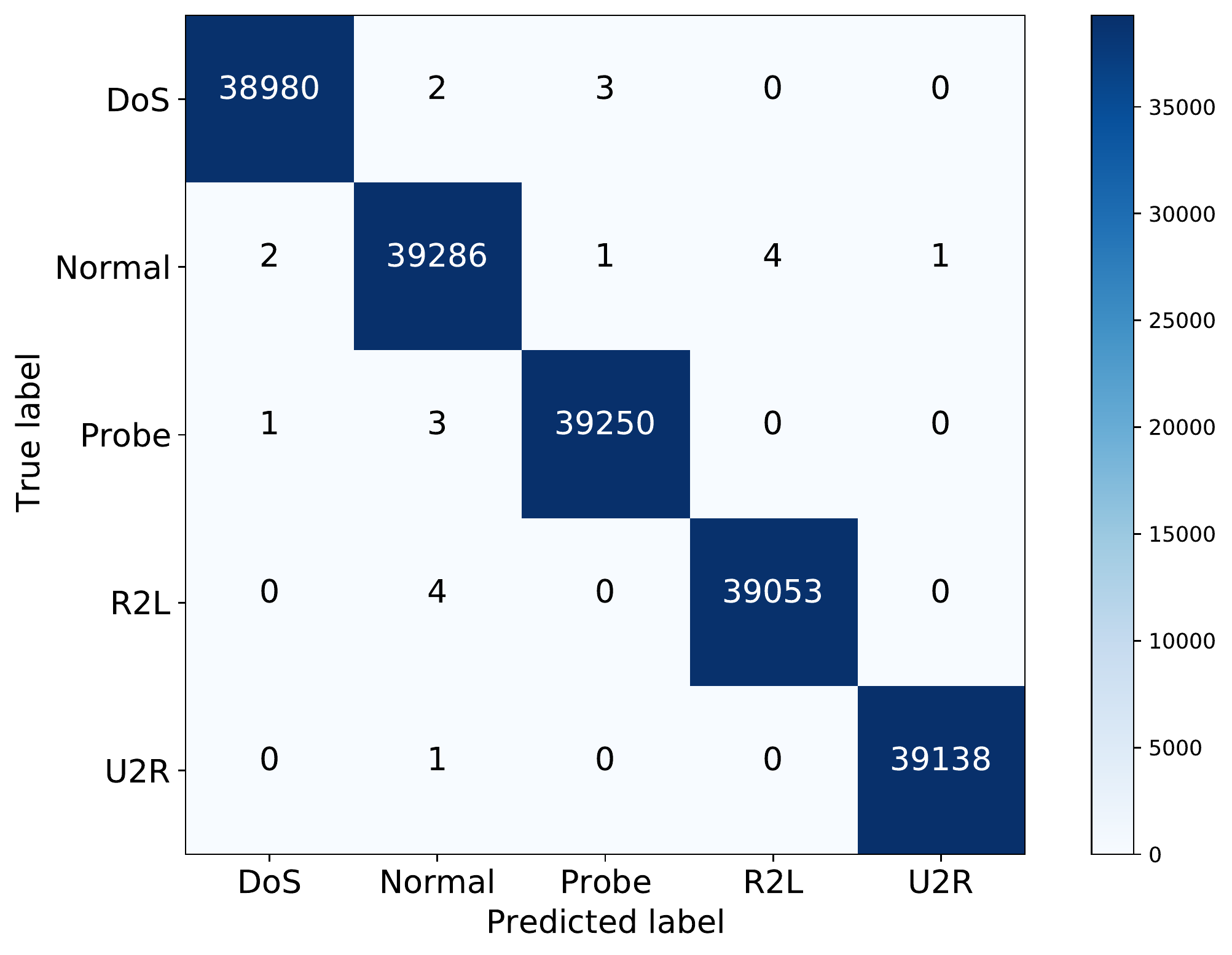}}\hspace{.1cm}
	\subfloat[KNN]{\includegraphics [scale=.2800]{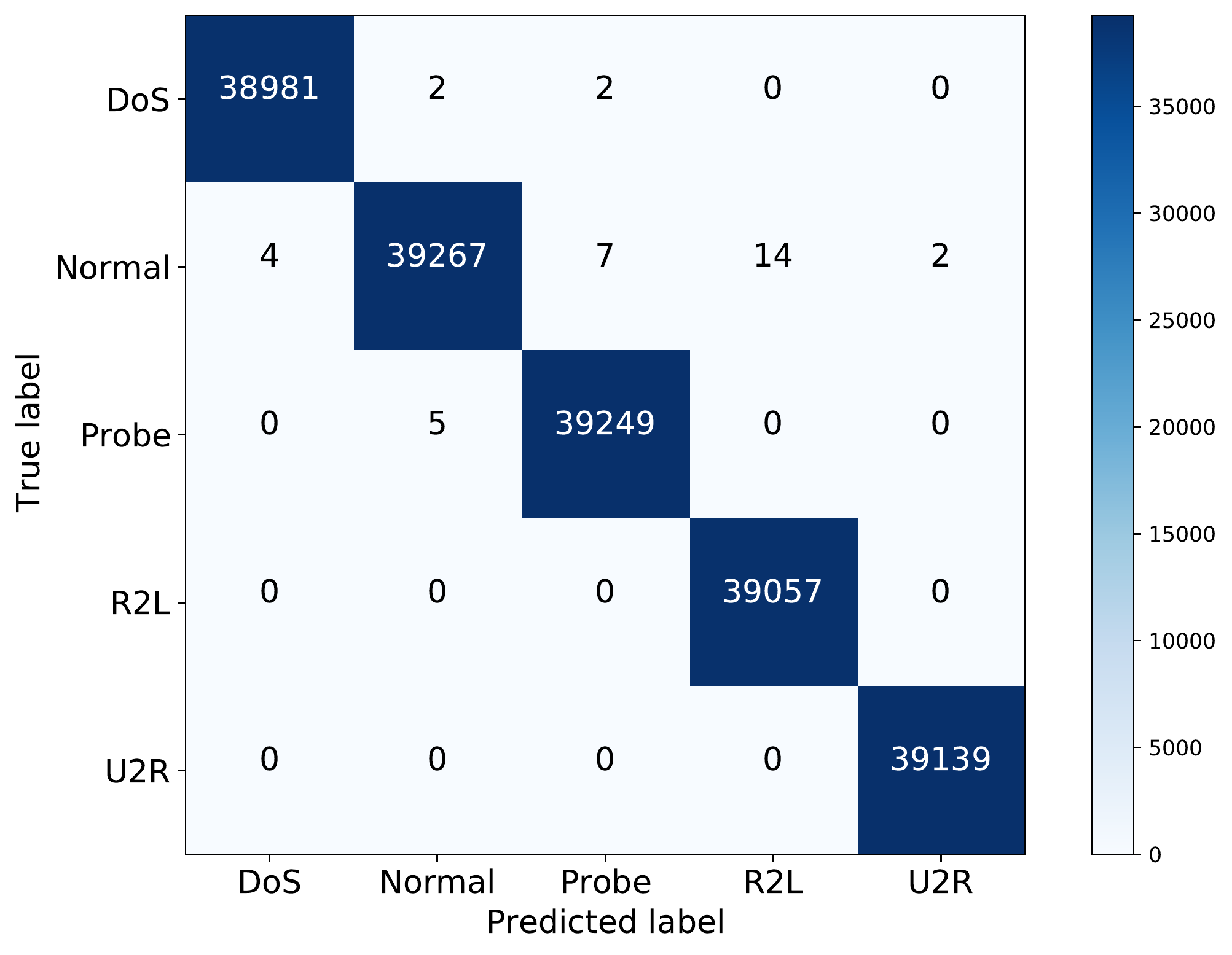}}\hspace{.1cm}
	\subfloat[MLP]{\includegraphics [scale=.2800]{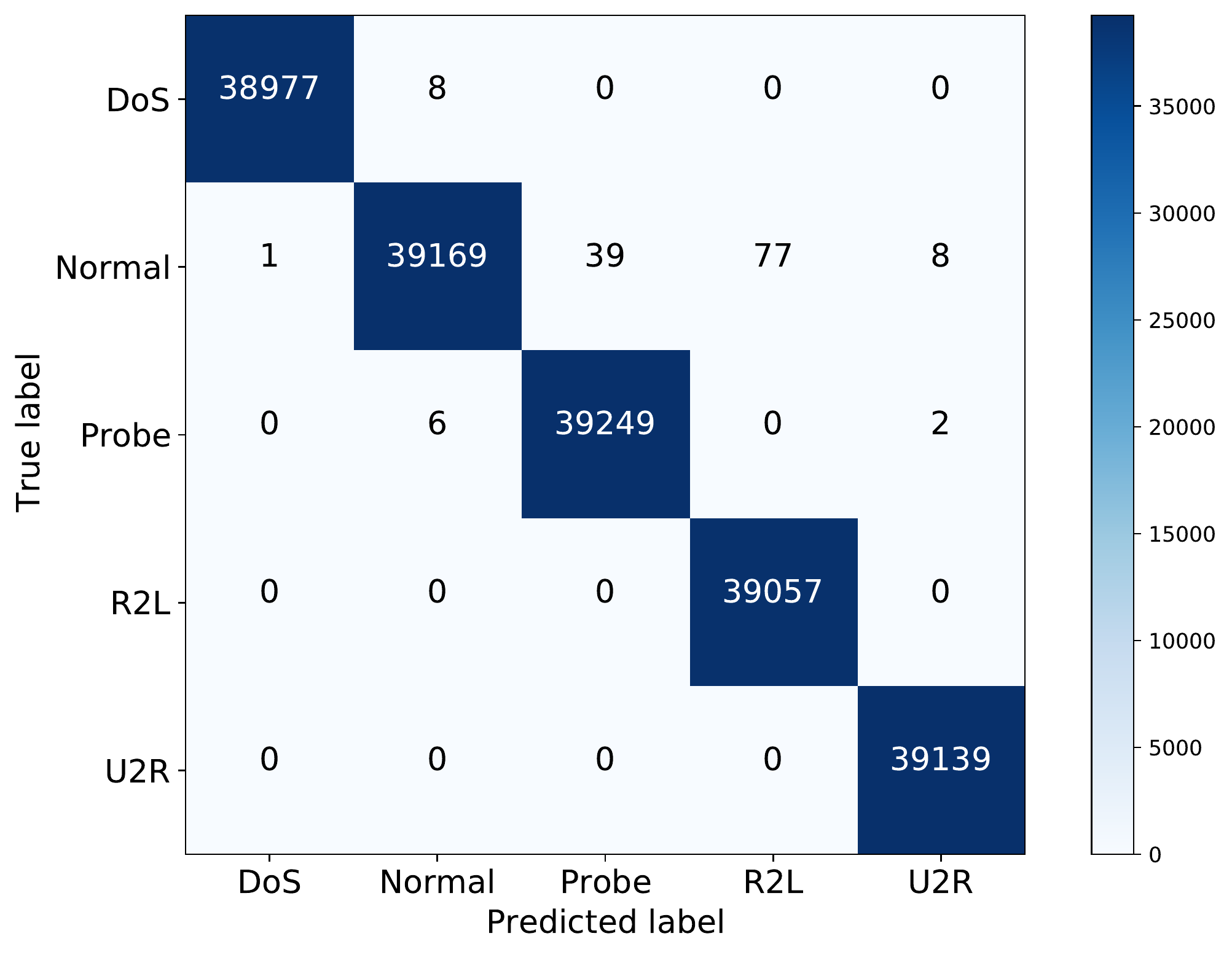}}\hspace{.1cm}
	\subfloat[CNN]{\includegraphics [scale=.2800]{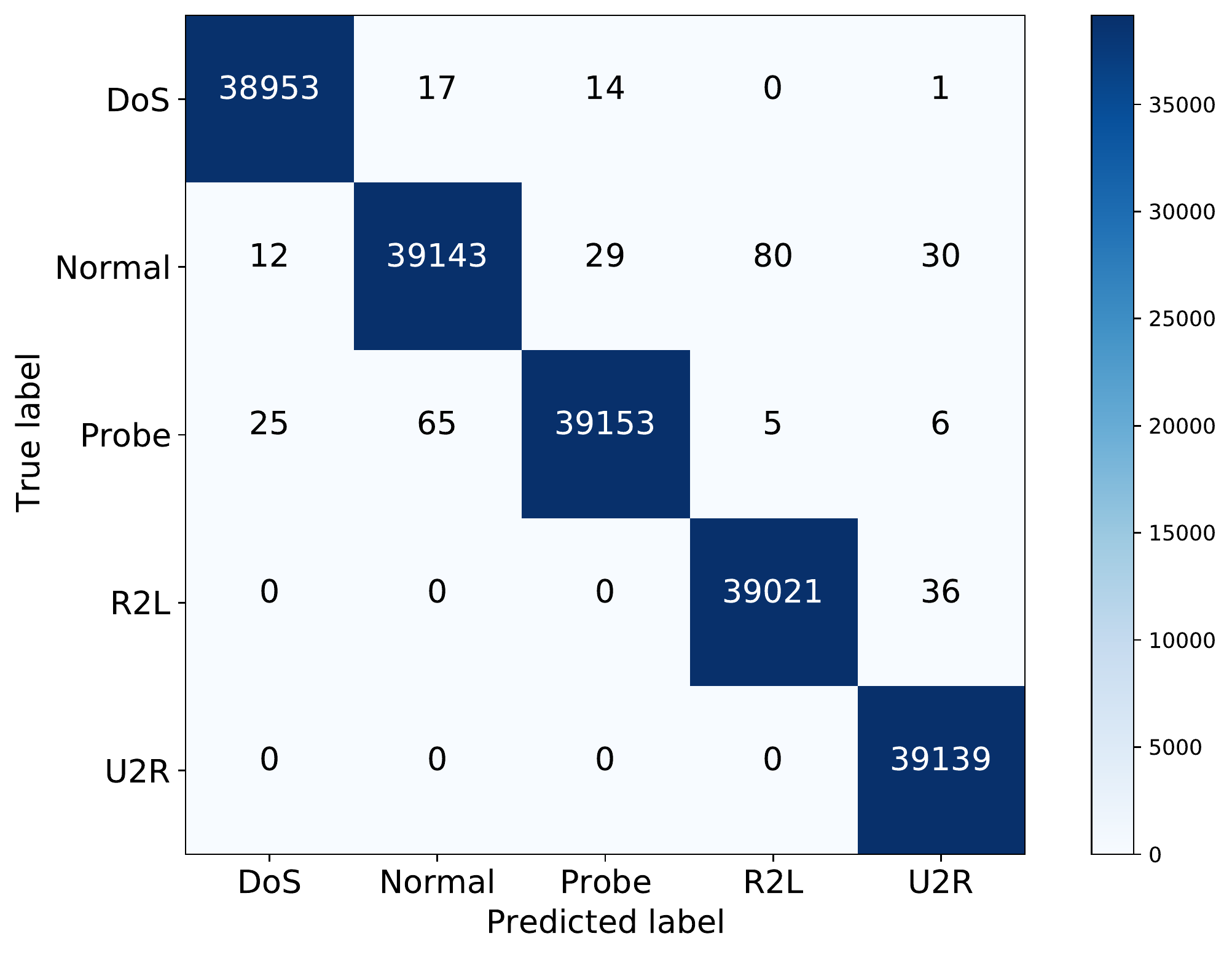}}\hspace{.1cm}
	\subfloat[ANN]{\includegraphics [scale=.2800]{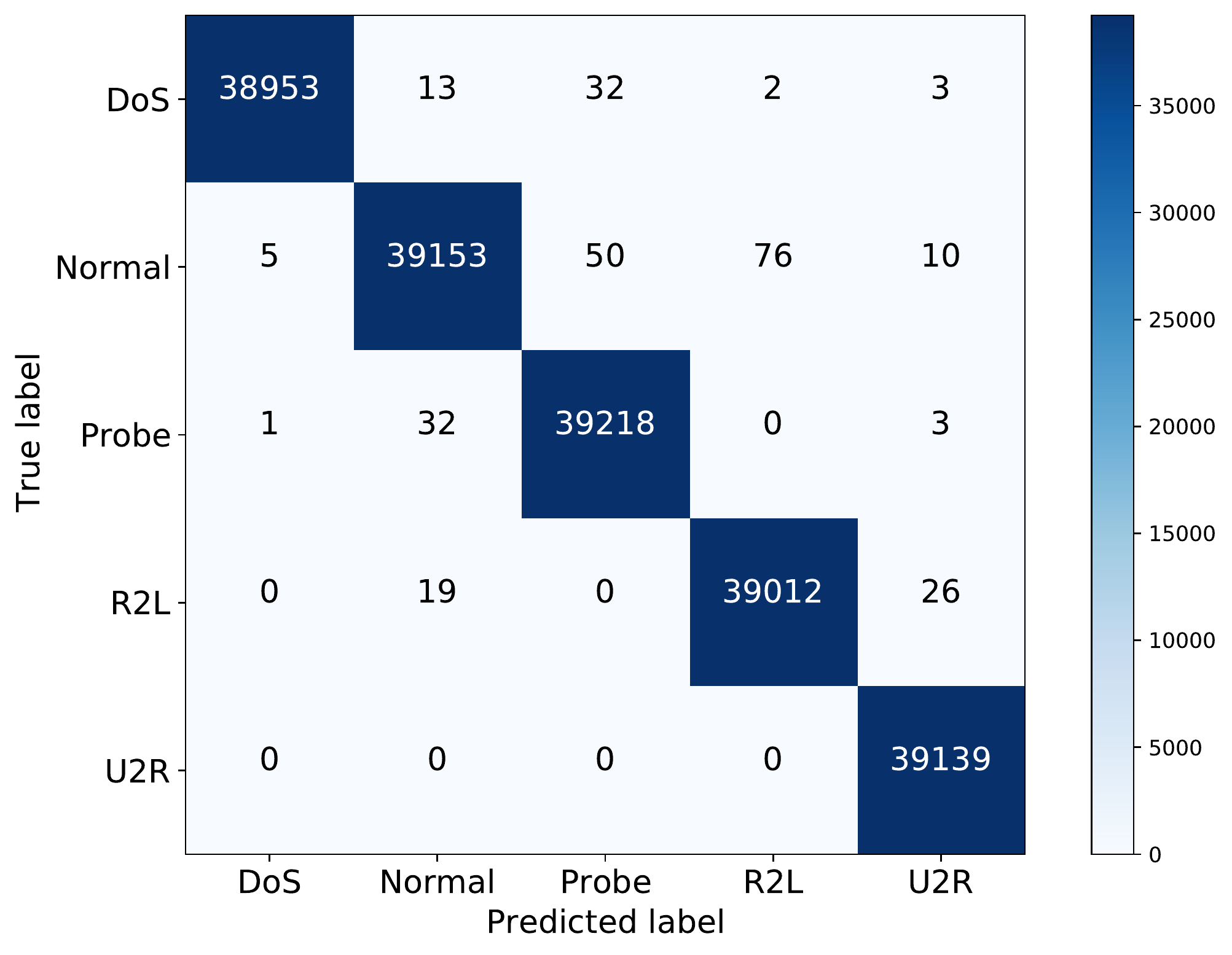}}\hspace{.1cm}
	
	\caption{Confusion matrix for multilabel classification of KDDCUP'99.}
	\label{fig:kdd_multilabel_confusion}
\end{figure*}
Fig \ref{fig:kdd_multilabel_confusion} shows the confusion matrix for all the proposed algorithms. Among all the confusion matrices, RF and DT produce a large number of TP, TN, and a very smaller number of FP, and FN rates, which gives a better performance of detecting intrusion detection with proper accuracy rate. For RF, the TP, TN, FP, FN rate are 19.92\%, 80.08\%, 0\%,	0\% ; 20.07\%, 79.92\%, 0\%, 0\% ; 20.05\%, 79.94\%, 0\%, 0\% ; 19.95\%, 80.04\%, 0\%, 0\% ; 20\%, 80\%,	0\%, 0\% ; considering DoS, Normal, Probe, R2L, U2R respectively. The TP, TN, FP, FN rate are 19.92\%, 80.08\%, 0\%, 0\% ; 20.07\%, 79.92\%, 0\%, 0.01\% ; 20.05\%, 79.94\%, 0\%, 0\% ; 19.95\%, 80.04\%, 0\%, 0\% ; 19.99\%, 80\%, 0\%, 0\% ; considering DoS, Normal, Probe, R2L, U2R respectively for DT. Random Forest produces superior findings, operates well on huge datasets, and can create estimates for missing data.  It is a collection of independent decision trees collaborating to form an ensemble. As a result, knowledge can be used to train several learner algorithms to reach maximum accuracy. On the other hand, DT is simple to use, comprehend and explain; it takes minimal time and effort to plan and produce; it may be used in conditional probability-based reasoning and can offer strategic responses to uncertain situations.

\begin{figure*}[!htbp]
	\centering
	\subfloat[ROC Curve]{\includegraphics [scale=.47]{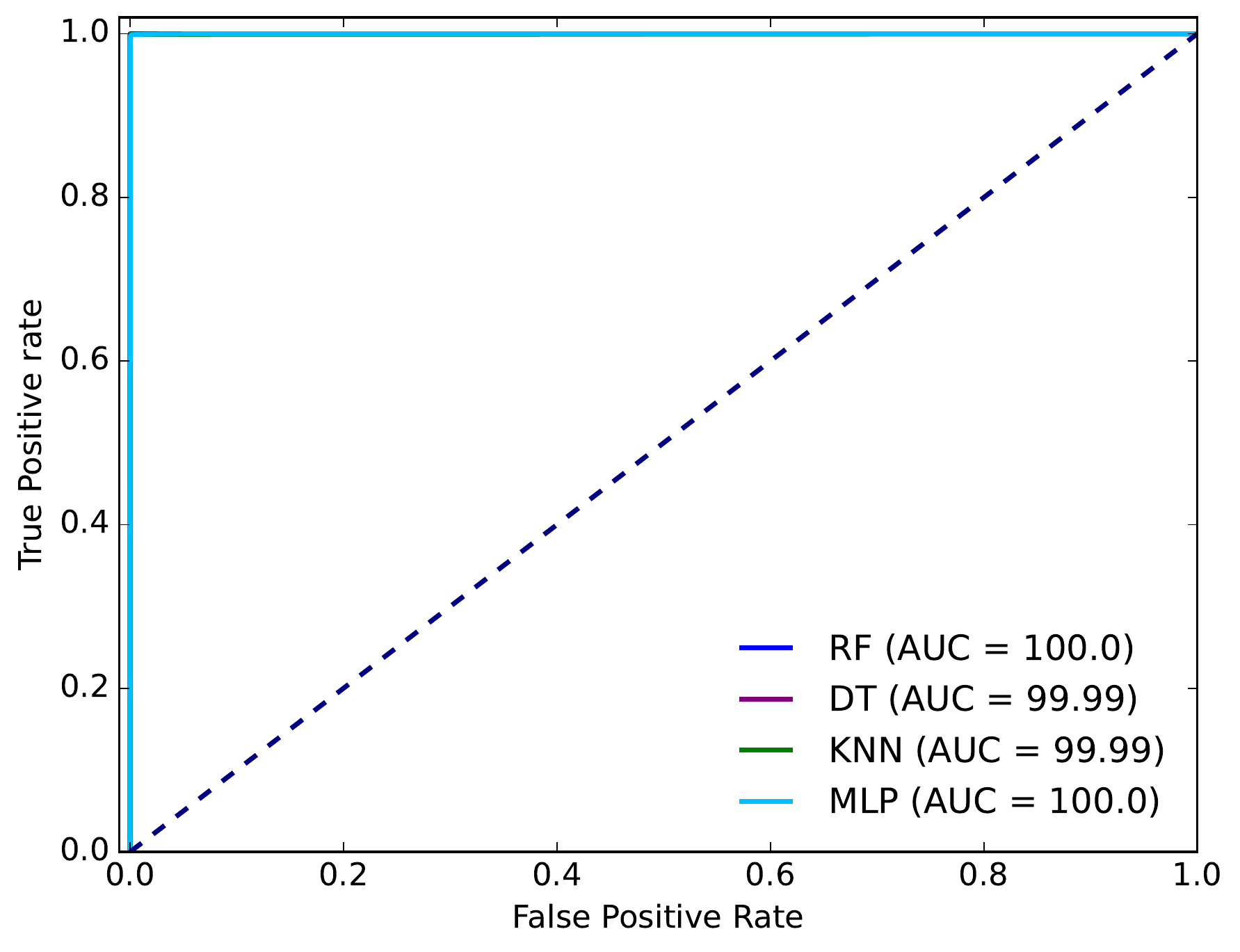}}\hspace{.1cm}
	\caption{Multilabel ROC Curve for KDDCUP'99.}
	\label{fig:kddcup_multilabel_roc_curve}
\end{figure*}

Fig \ref{fig:kddcup_multilabel_roc_curve} shows the multilabel ROC Curve for KDDCUP'99, where the AUC score for RF, DT, KNN, and MLP are  100\%, 99.97\%, 99.99\%, and 100\%, respectively. Among all the RF and MLP give the highest accuracy rate. MLP consists of a series of layers made up of neurons and their connections. It has one or more hidden layers between the input and output layers where the neurons are arranged in layers, and connections are often guided from lower to upper layers.

After analyzing all the ML and DL algorithms, we can find that RF outperforms other ML and DL algorithms both for binary and multilabel classifications to detect network intrusion.

\subsection{Performance Analysis of CIC-MalMem-2022 Dataset}
\begin{figure*}[!htbp]
	\centering
	\subfloat[Accuracy]{\includegraphics[scale=.60]{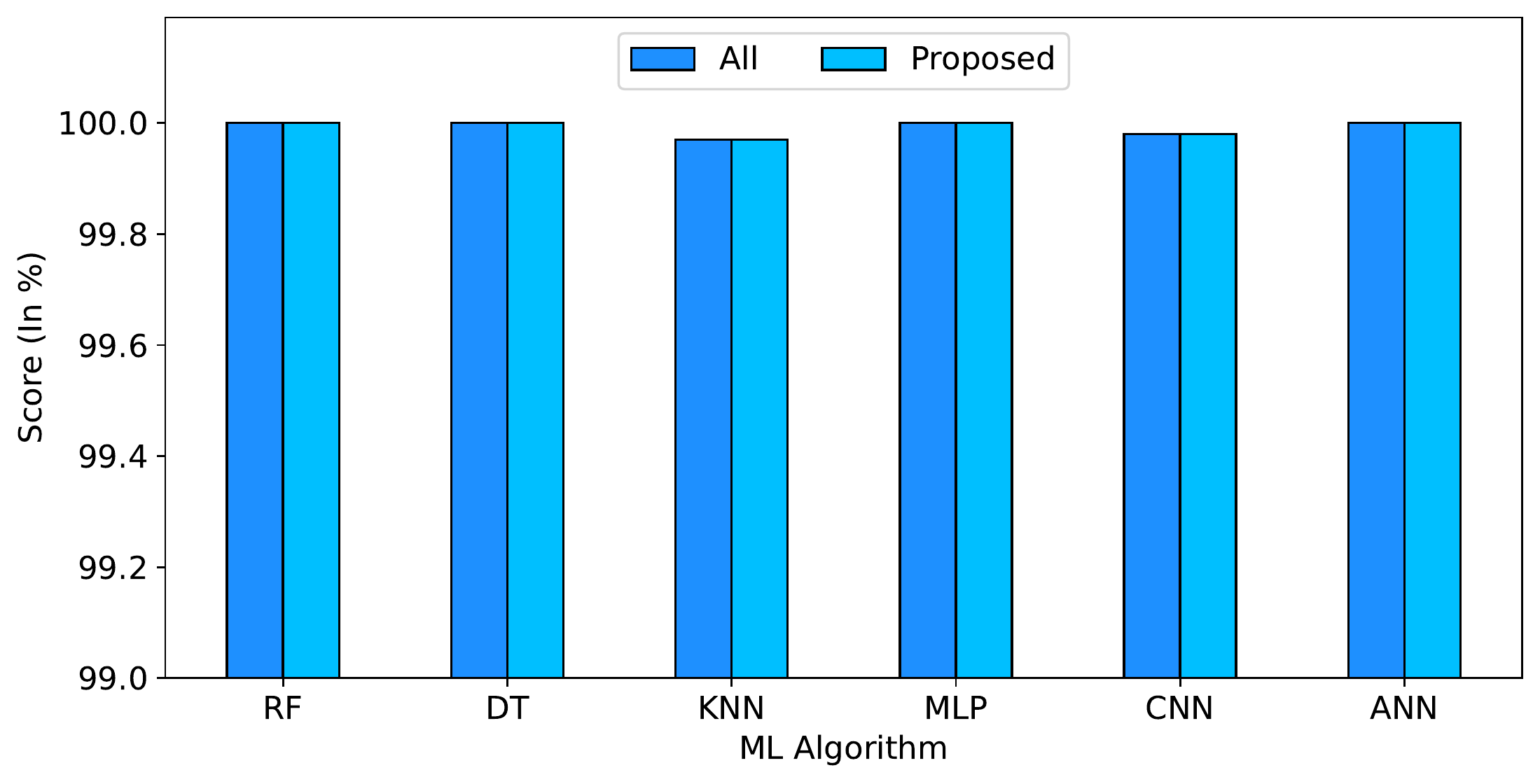}}\hspace{0.1cm}
	\subfloat[RMSE]{\includegraphics[scale=.60]{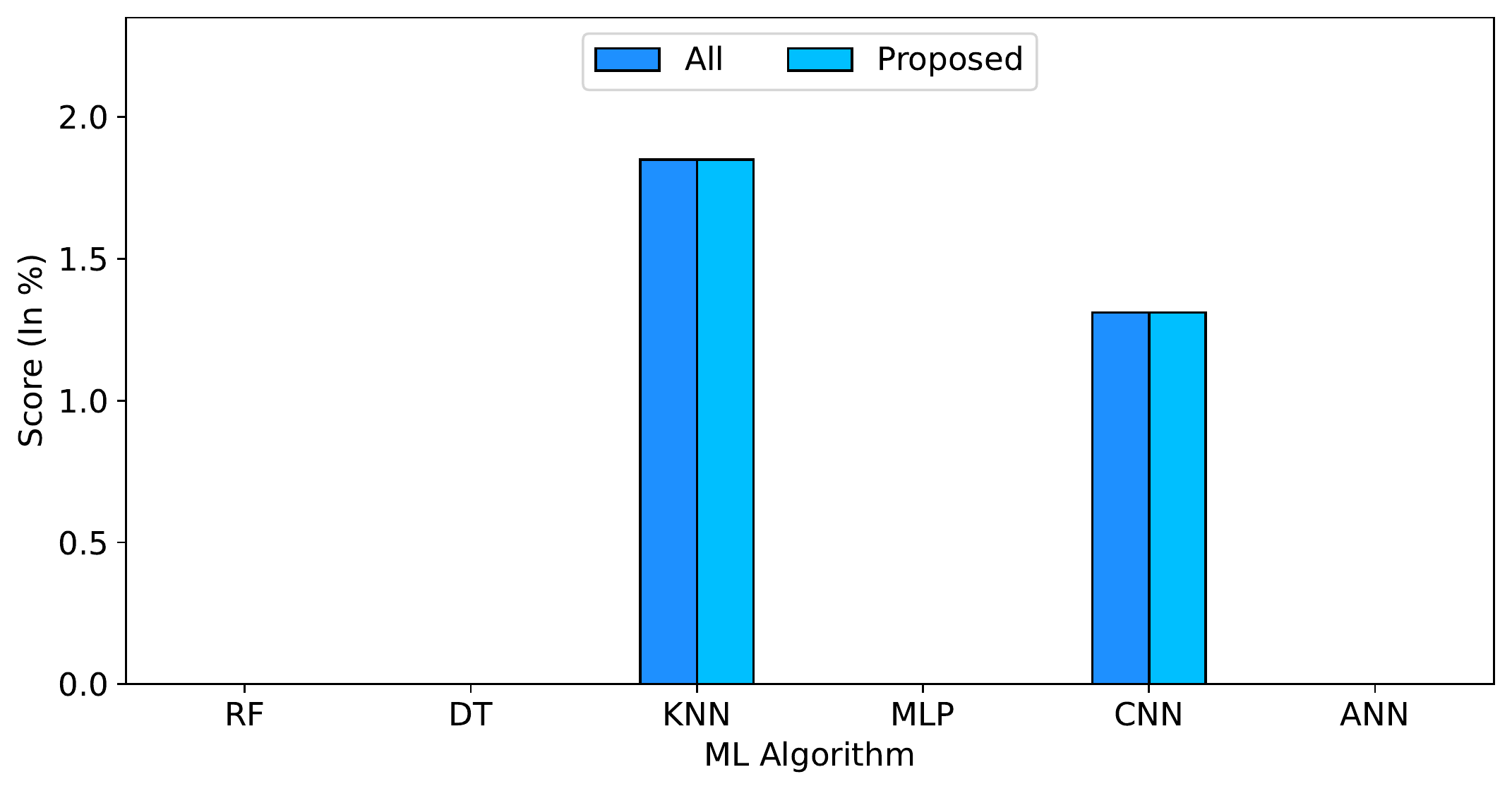}}
	\caption{Performance comparison graphs for binary classification.}
	\label{fig:malmem_comparison_graph}

\end{figure*}

The performance comparison results for binary classification for the CIC-MalMem-2022 dataset are illustrated in Table \ref{fig:malmem_comparison} and Fig. \ref{fig:malmem_comparison_graph} in tabular and bar chart format, respectively. The experiment results for considering all features and the proposed model’s performance are represented. Here, all indicate that the features are just pre-proposed, scaled, and then applied to the machine learning algorithms to build models and evaluate the performance. On the other hand, the proposed model indicates all the proposed methodology processes by which evaluation has been measured.

\begin{table}[]
\centering
\resizebox{\textwidth}{!}{
\begin{tabular}{lllllllllllllll}
\hline
\multirow{2}{*}{ML} &
  \multicolumn{2}{c}{Accuracy} &
  \multicolumn{2}{c}{Precision} &
  \multicolumn{2}{c}{Recall} &
  \multicolumn{2}{c}{F1-score} &
  \multicolumn{2}{c}{MAE} &
  \multicolumn{2}{c}{MSE} &
  \multicolumn{2}{c}{RMSE} \\
    & All   & Proposed & All   & Proposed & All   & Proposed & All   & Proposed & All  & Proposed & All  & Proposed & All  & Proposed \\ \hline
RF  & 100   & 100      & 100   & 100      & 100   & 100      & 100   & 100      & 0    & 0        & 0    & 0        & 0    & 0        \\
DT  & 100   & 100      & 100   & 100      & 100   & 100      & 100   & 100      & 0    & 0        & 0    & 0        & 0    & 0        \\
KNN & 99.97 & 99.97    & 99.97 & 99.97    & 99.97 & 99.97    & 99.97 & 99.97    & 0.03 & 0.03     & 0.03 & 0.03     & 1.85 & 1.85     \\
MLP & 100   & 100      & 100   & 100      & 100   & 100      & 100   & 100      & 0    & 0        & 0    & 0        & 0    & 0        \\
CNN & 99.98 & 99.98    & 99.98 & 99.98    & 99.98 & 99.98    & 99.98 & 99.98    & 0.02 & 0.02     & 0.02 & 0.02     & 1.31 & 1.31     \\
ANN & 100   & 100      & 100   & 100      & 100   & 100      & 100   & 100      & 0    & 0        & 0    & 0        & 0    & 0        \\ \hline
\end{tabular}
}
\caption{Performance comparison analysis for binary classification.}
\label{fig:malmem_comparison}
\end{table}

The accuracy rate of our proposed scheme in binary classification is substantially the same as all features, as shown in the Table and Bar Chart. The accuracy rate between the two bar graphs is almost the same, and the RMSE rate is also approximately the same as all features.

From the binary comparison graph fig \ref{fig:malmem_comparison_graph}, the accuracy of the proposed model for RF, DT, KNN, MLP, CNN, and ANN is 100\%, 100\%, 99.97\%, 100\%, 99.98\%, and 100\%, respectively. From the graph fig \ref{fig:malmem_comparison_graph}(a), it is clear that the accuracy increasing rate by considering all the features with the proposed model is the same for all ML algorithms that prove the effectiveness of the proposed model with a smaller number of features. Fig \ref{fig:malmem_comparison_graph}(b) shows the same RMSE values for all features and the proposed features. 

\begin{figure*}[!htbp]
	\centering
	\subfloat[Performance]{\includegraphics[scale=.50]{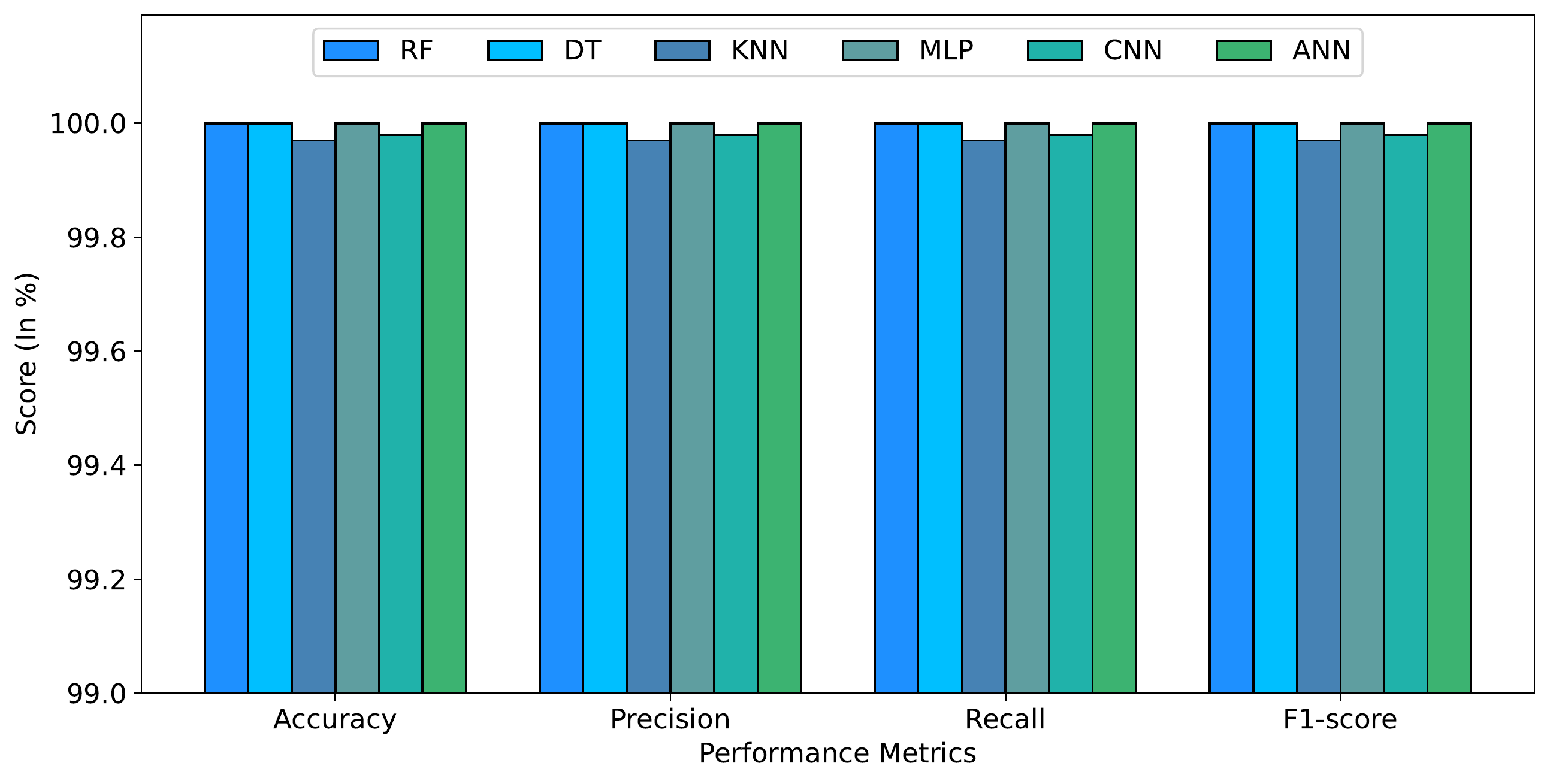}}\hspace{0.1cm}
	\subfloat[Error]{\includegraphics[scale=.50]{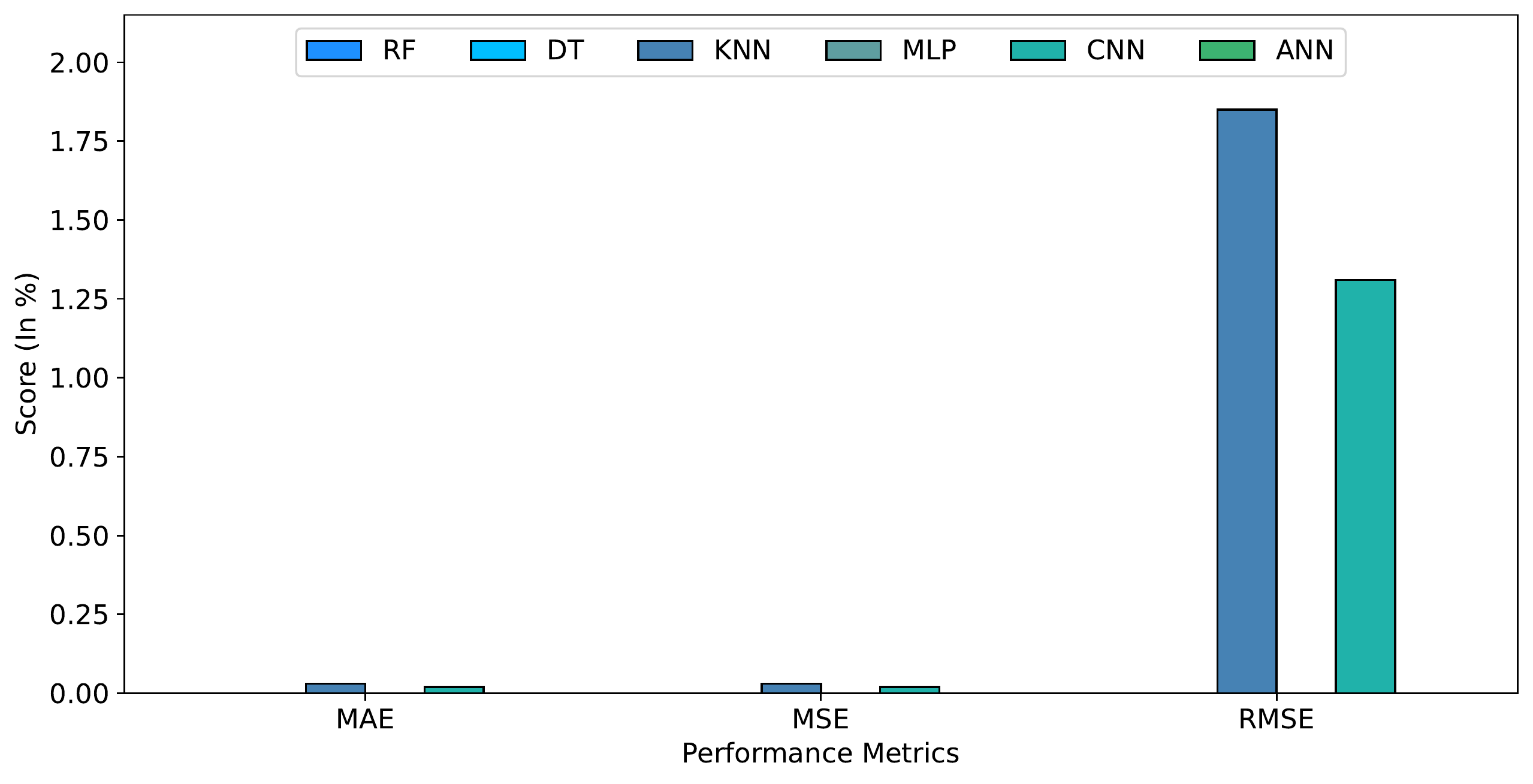}}
	\caption{Performance analysis graphs for binary classification.}
	\label{fig:malmem_performance_graph}

\end{figure*}

The Performance analysis graphs for binary classification are shown in fig \ref{fig:malmem_performance_graph} where, the accuracy rate among all the algorithms RF, DT, MLP, and ANN gives the better performance, which is 100\%, and KNN gives lower accuracy, which is 99.97\%. The RMSE error rate for RF,DT,MLP,ANN is 0\% and KNN is 1.85\%.

\begin{figure*}[!htbp]
	\centering
	\subfloat[RF]{\includegraphics[scale=.385]{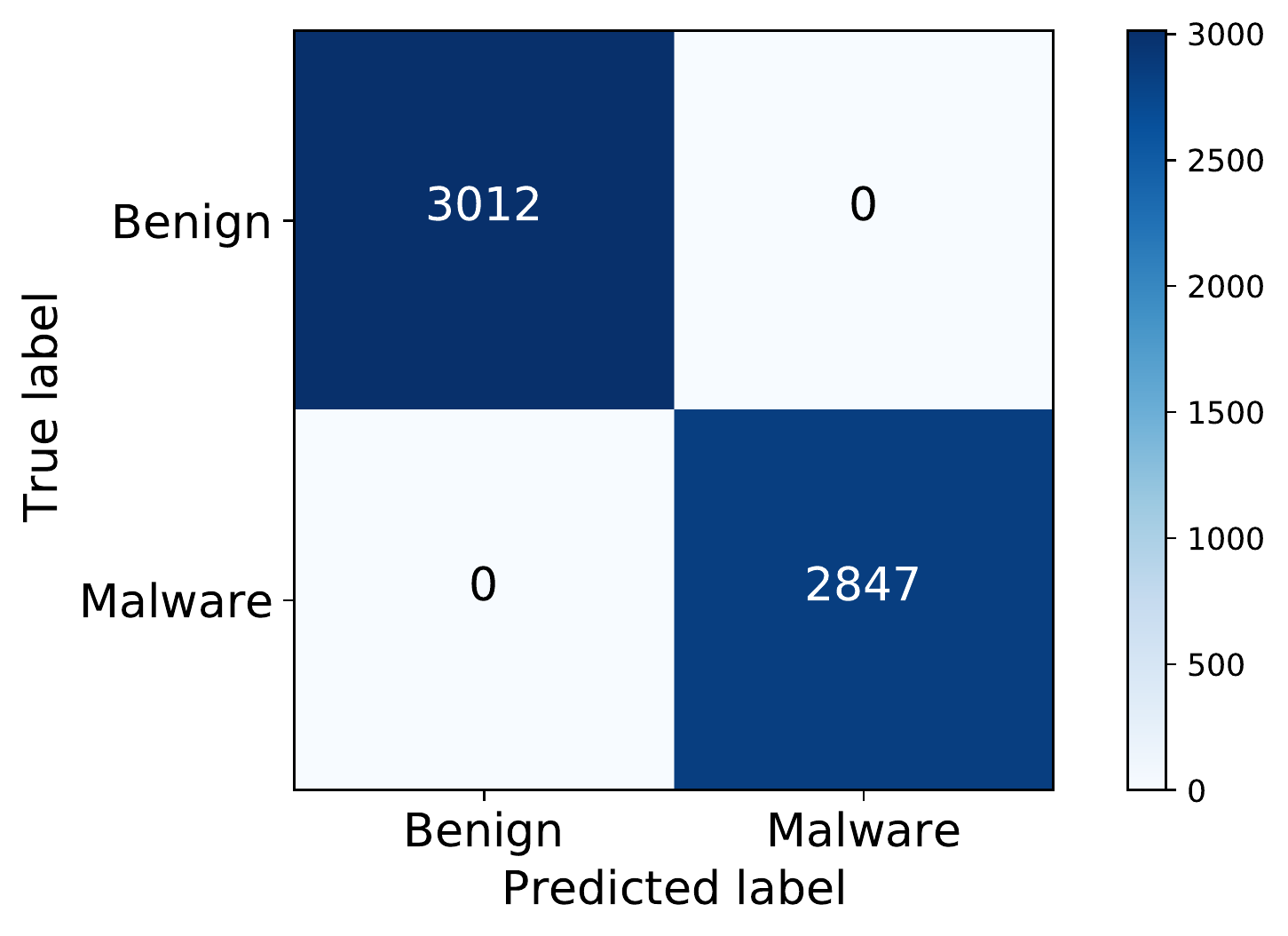}}\hspace{.1cm}
	\subfloat[DT]{\includegraphics[scale=.385]{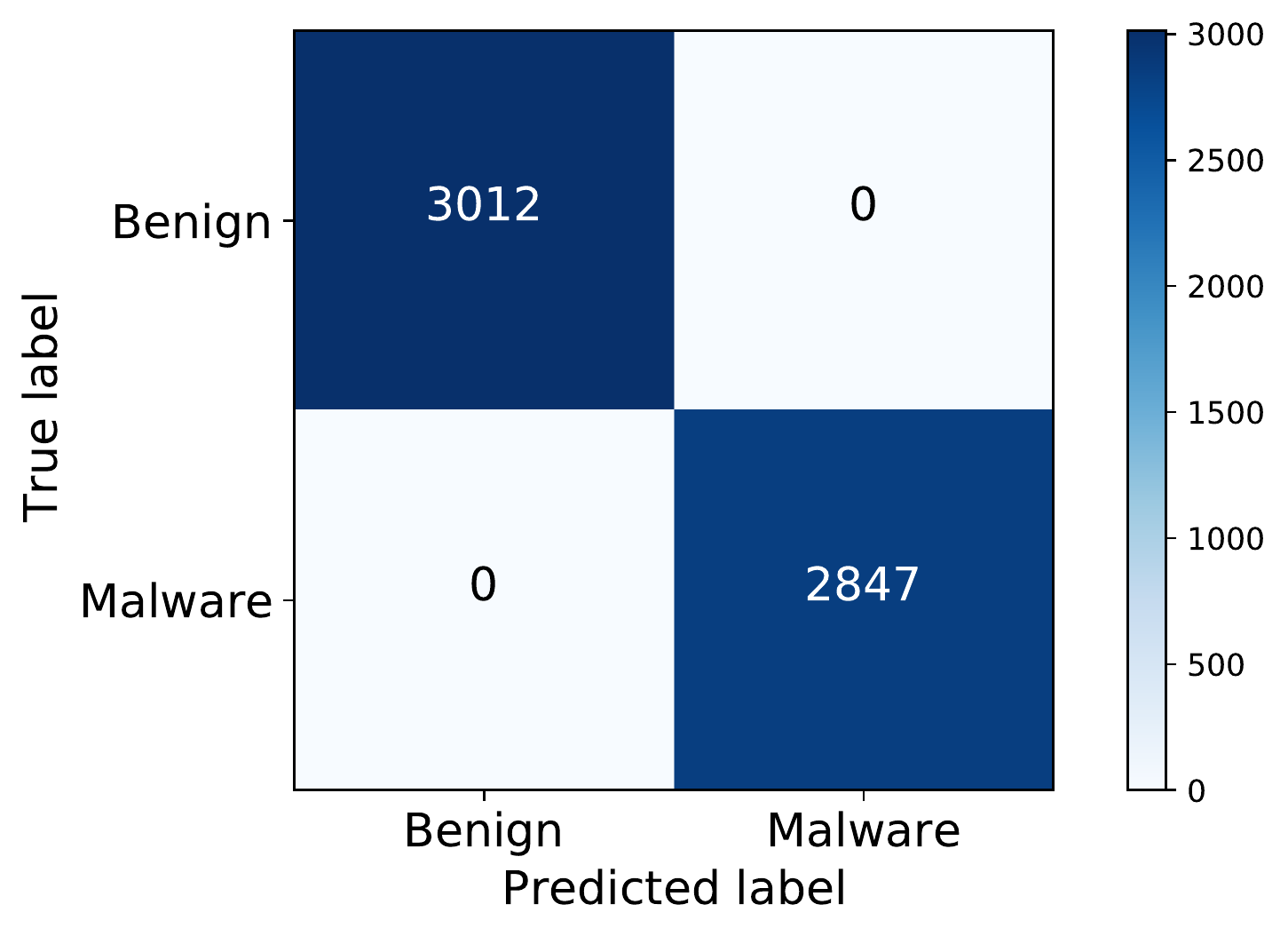}}\hspace{.1cm}
	\subfloat[KNN]{\includegraphics[scale=.385]{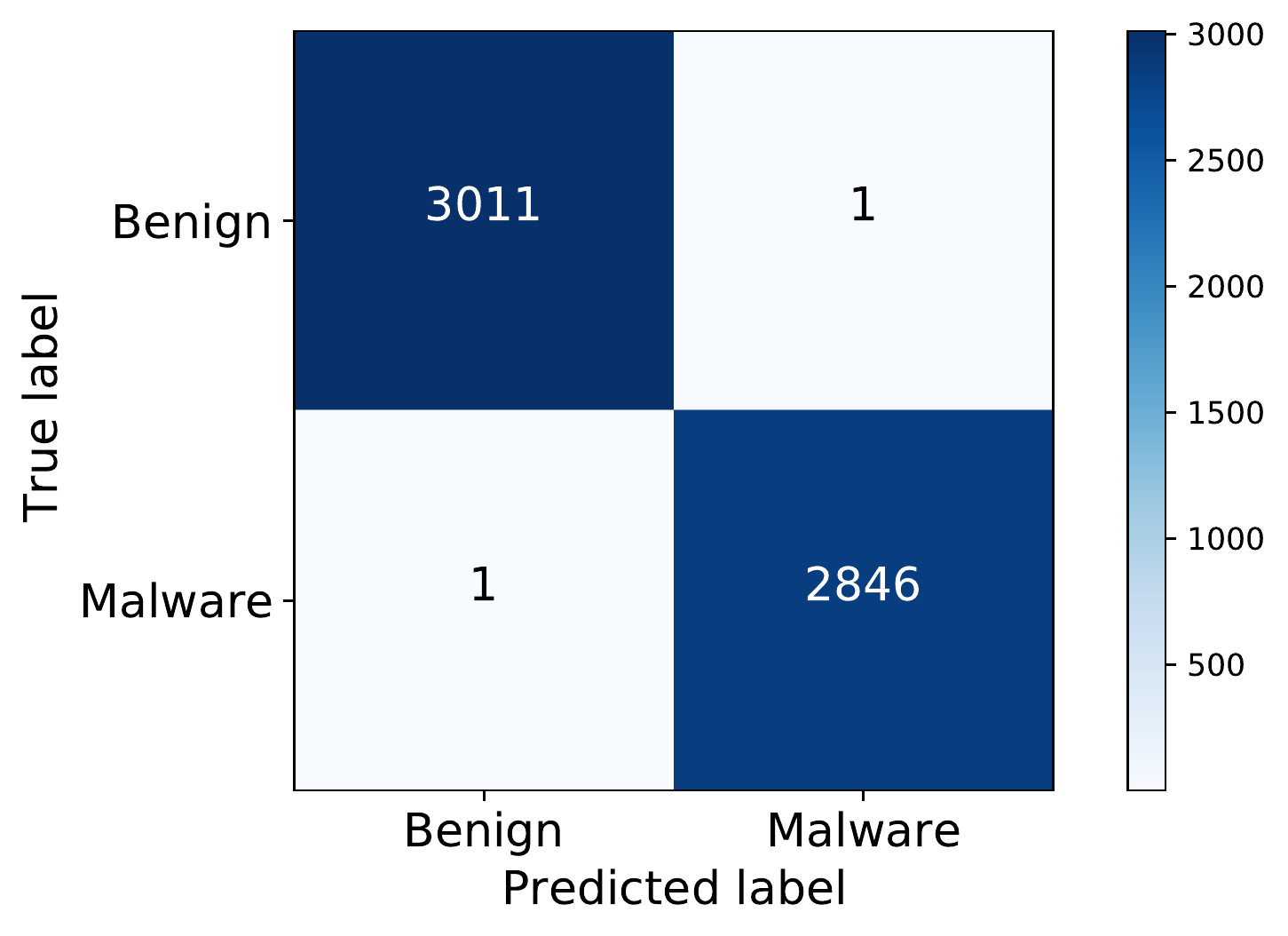}}\hspace{.1cm}
	\subfloat[MLP]{\includegraphics[scale=.385]{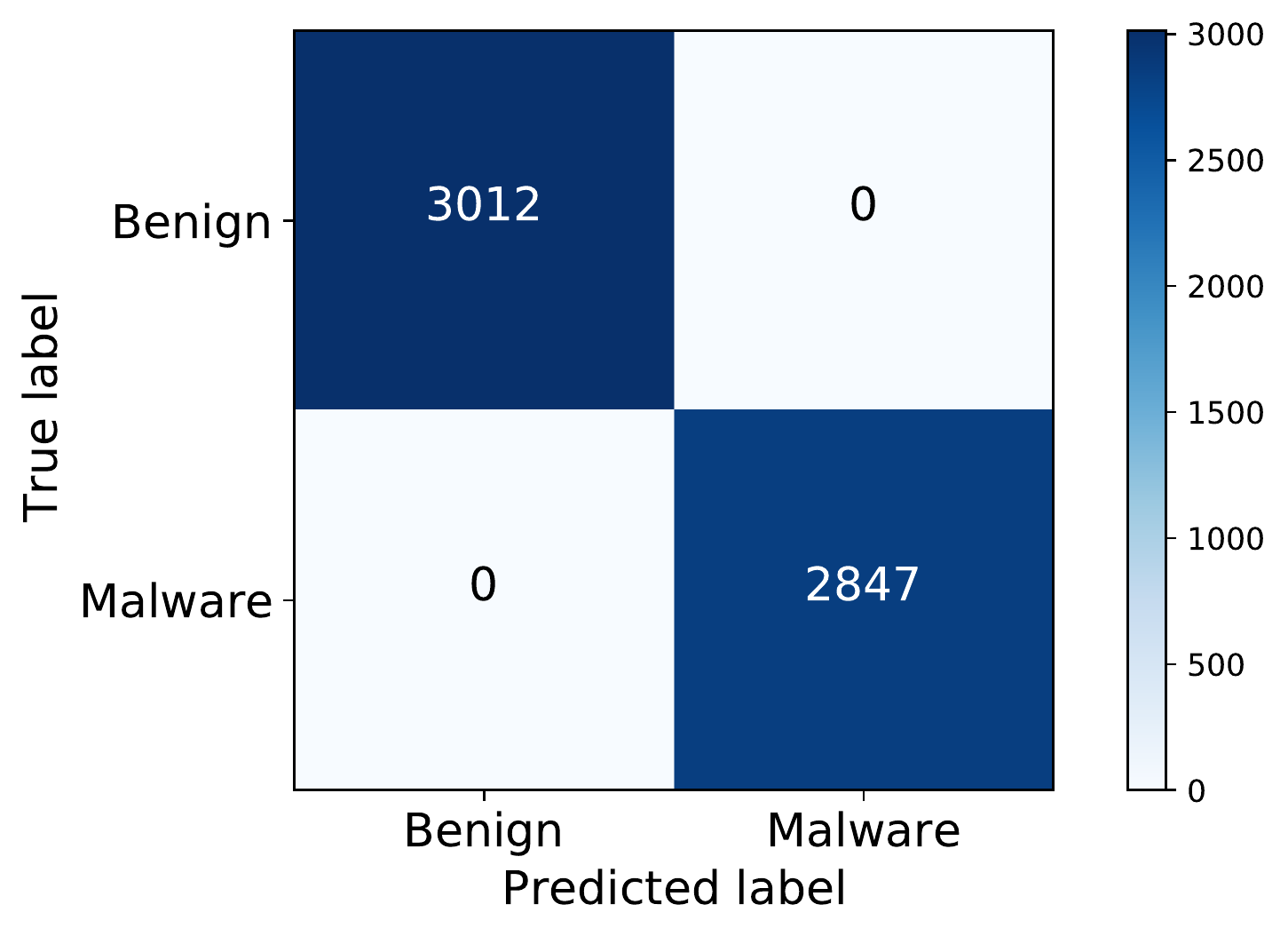}}\hspace{.1cm}
	\subfloat[CNN]{\includegraphics[scale=.385]{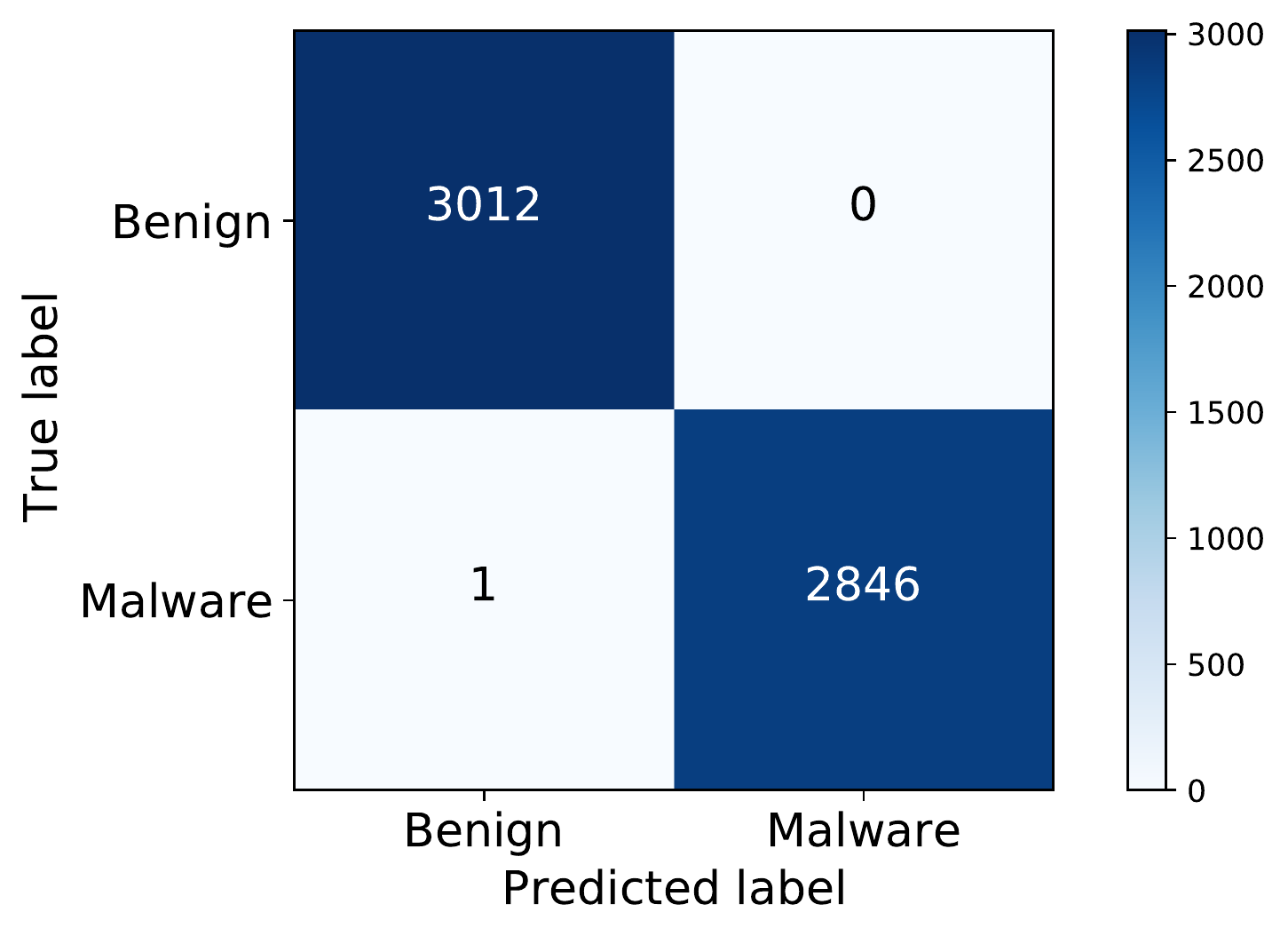}}\hspace{.1cm}
	\subfloat[ANN]{\includegraphics[scale=.385]{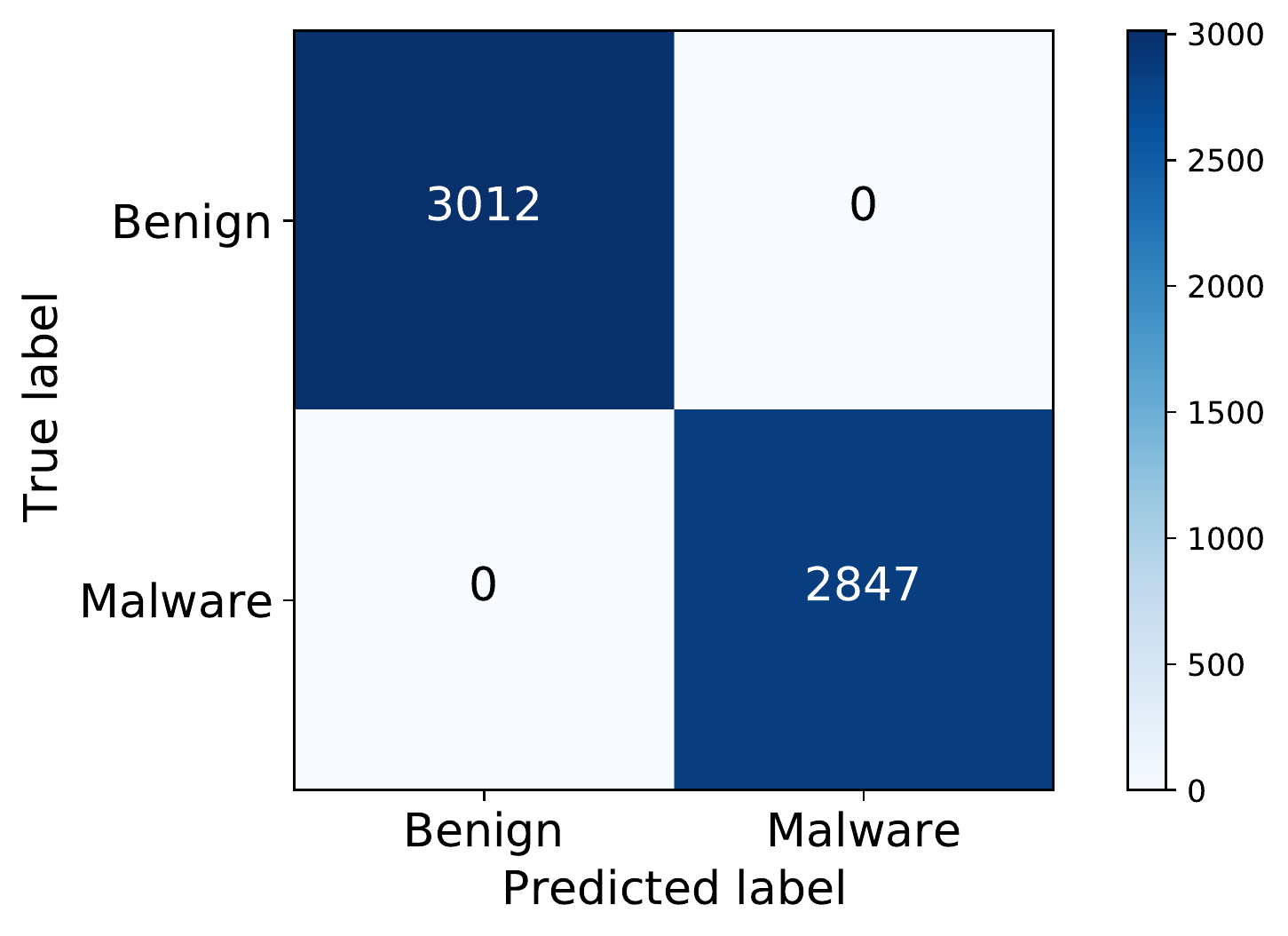}}\hspace{.1cm}
	\caption{Confusion matrix for multilabel classification of CIC-MalMem-2022.}
	\label{fig:malmem_confusion}
\end{figure*}

Fig \ref{fig:malmem_confusion} shows the confusion matrix for all the proposed algorithms. The TP, TN, FP, FN rates are 51.41\%, 48.59\%, 0\%, 0\% ; 51.41\%, 48.59\%, 0\%, 0\% ; 51.39\%, 48.57\%, 0.02\%, 0.02\% 51.41\%, 48.59\%, 0\%, 0\% ; 51.41\%, 48.57\%, 0.02\%, 0\% ; 51.41\%, 48.59\%, 0\%, 0\% ; for RF, DT, KNN, MLP, CNN and ANN respectively. The confusion matrix results show that RF, DT, KNN, and MLP outperform the other algorithms using 20 features. The TP and TN rates are very high, and FN and FN rate is zero for these algorithms, which provides a better confusion matrix for accurately detecting malware memory threats.

\begin{figure*}[!htbp]
	\centering
	\subfloat[ROC Curve]{\includegraphics[scale=.47]{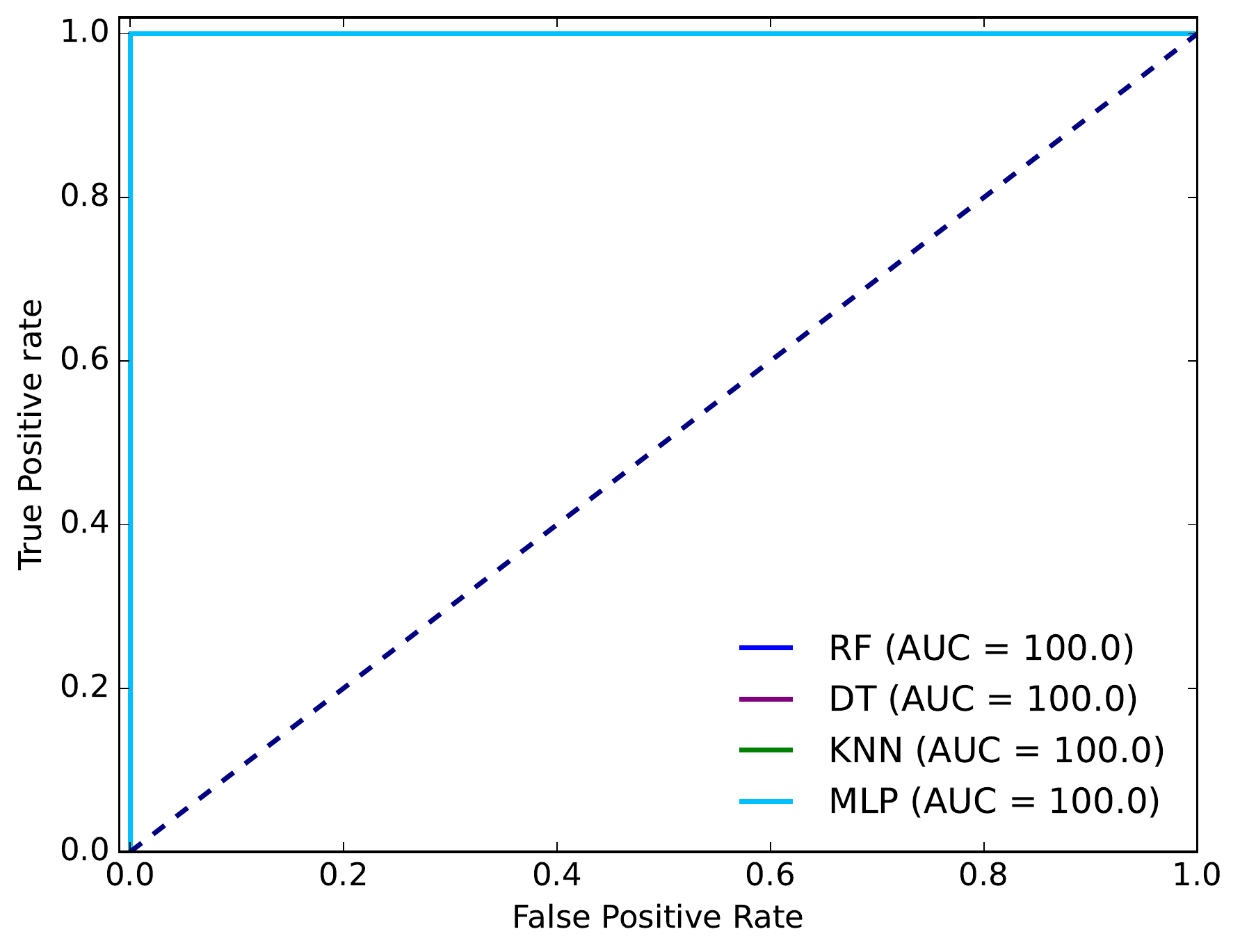}}\hspace{.1cm}
	\caption{ROC Curve for CIC-MalMem-2022.}
	\label{fig:malmem_roc_curve}
\end{figure*}

Fig \ref{fig:malmem_roc_curve} shows the Binary ROC Curve for CIC-MalMem-2022, where the AUC score for RF, DT, KNN, and MLP is 100\%. Here, all the ML performs at the highest accuracy rate. Random Forest produces superior findings, operates well on huge datasets, and can create estimates for missing data.  It is a collection of independent decision trees collaborating to form an ensemble. As a result, knowledge can be used to train several learner algorithms to reach maximum accuracy. DT is simple to use, comprehend and explain; it takes minimal time and effort to plan and produce; it may be used in conditional probability-based reasoning and can offer strategic responses to uncertain situations. KNN enables nonlinear solutions and performs instance-based learning. A well-tuned K can model large decision spaces with arbitrarily convoluted decision boundaries that are difficult to model by other "eager" learners who work in batches, modelling one group of early phases at a time. MLP consists of a series of layers made up of neurons and their connections. It has one or more hidden layers between the input and output layers where the neurons are arranged in layers, and connections are often guided from lower to upper layers. Hence, they give better performance in our proposed model.

After analyzing all the ML and DL algorithms, we select the RF and ANN to detect network intrusion for binary classifications.

\subsubsection{Dependability Analysis of our proposed approach}

This section examines the effectiveness study of our developed model's dependability. The aspects of availability, efficiency, and scalability are all considered in the reliability evaluation study. We choose features based on accuracy efficiency rate and then use our developed framework to reliably differentiate between benign and attack or malware scenarios without experiencing any loss, ensuring that our presented methodology remains available. Furthermore, comprehensive analysis and productivity assessments such as accuracy, precision, recall, f1-score, AUC score, ROC Curve, MAE, MSE, and RMSE indicate that the proposed framework is more efficient and has revealed improved results than many other existing techniques with a lower error rate and computational loss.

\begin{figure*}[!htbp]
	\centering
	\subfloat[Accuracy]{\includegraphics[scale=.40]{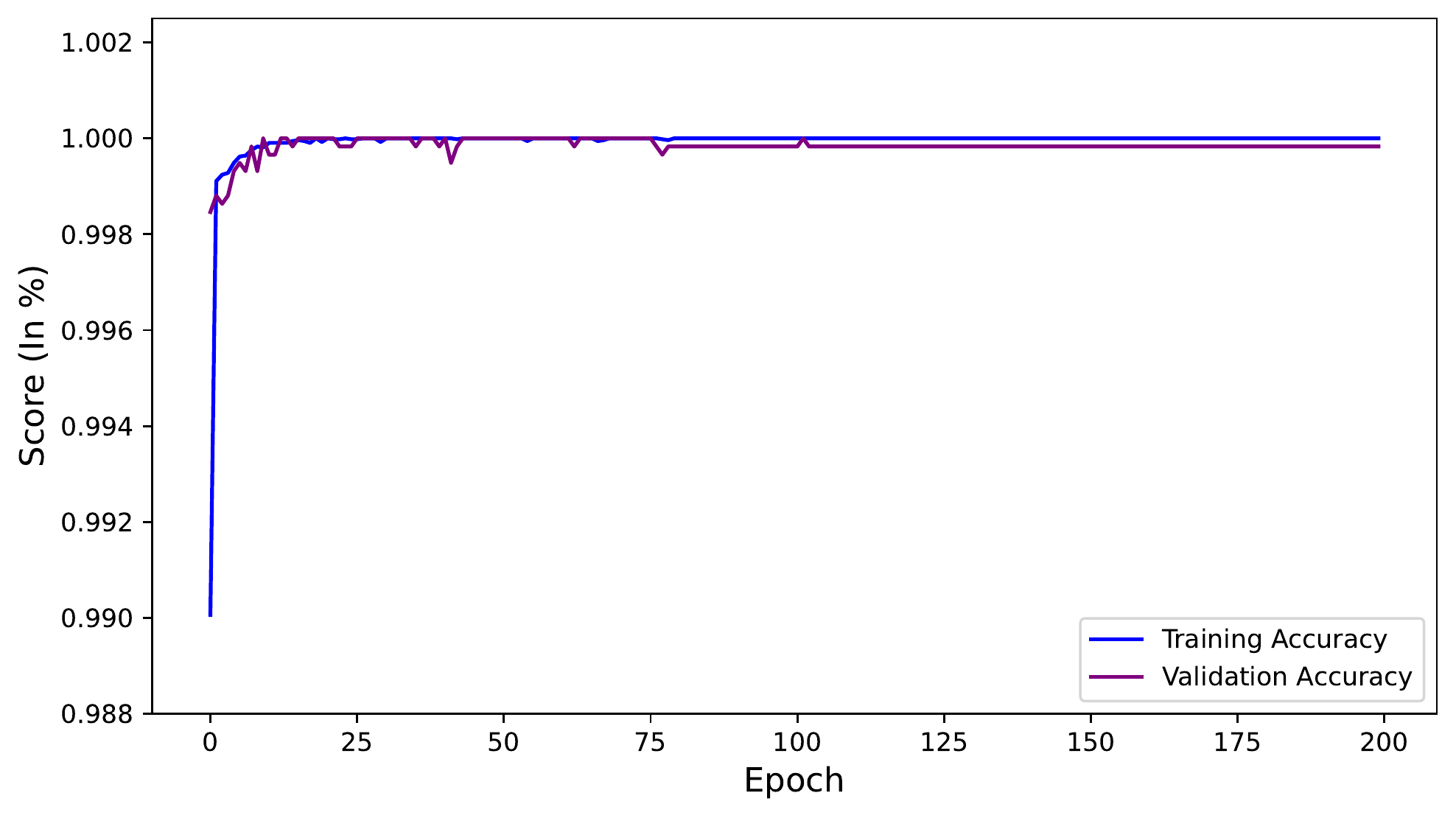}}\hspace{.1cm}
	\subfloat[Loss]{\includegraphics[scale=.40]{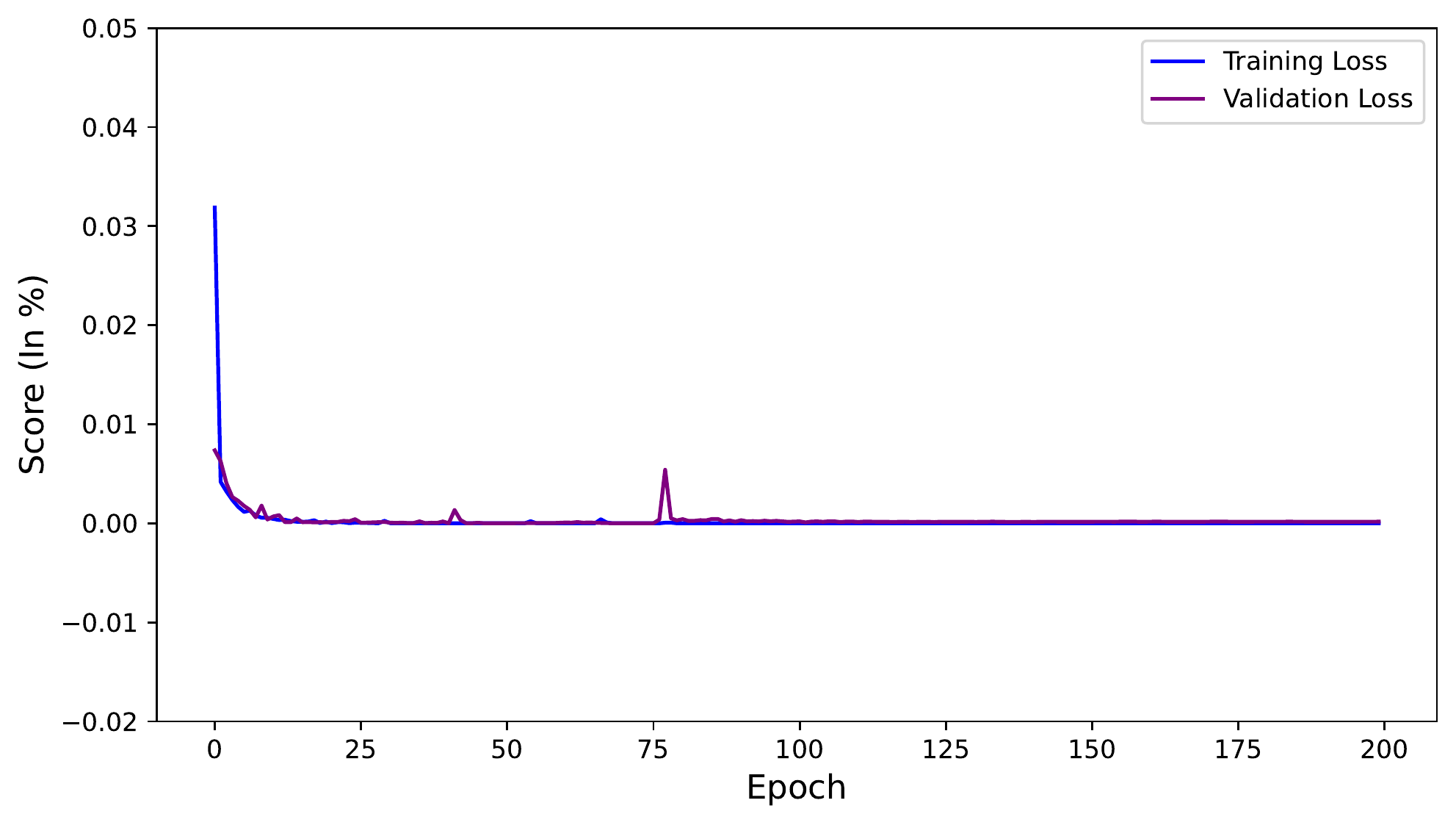}}\hspace{.1cm}
	\caption{The effectiveness and computational loss.}
	\label{fig:ec_loss}
\end{figure*}

Figure \ref{fig:ec_loss} depicts the developed model's effectiveness and computational loss. Ultimately, we found that incorporating numerous incompatible reliable datasets in the training sample with optimum reliability, which were gathered from a broader variety of IoT sensors, improved the scalability qualities of our developed framework. Consequently, when we extended the epoch quantity from 125 to 200, our proposed approach's accuracy performance rate remained nearly unchanged, confirming its flexibility or scalability.

\begin{figure*}[!htbp]
	\centering
	\subfloat[Bar chart]{\includegraphics[scale=.40]{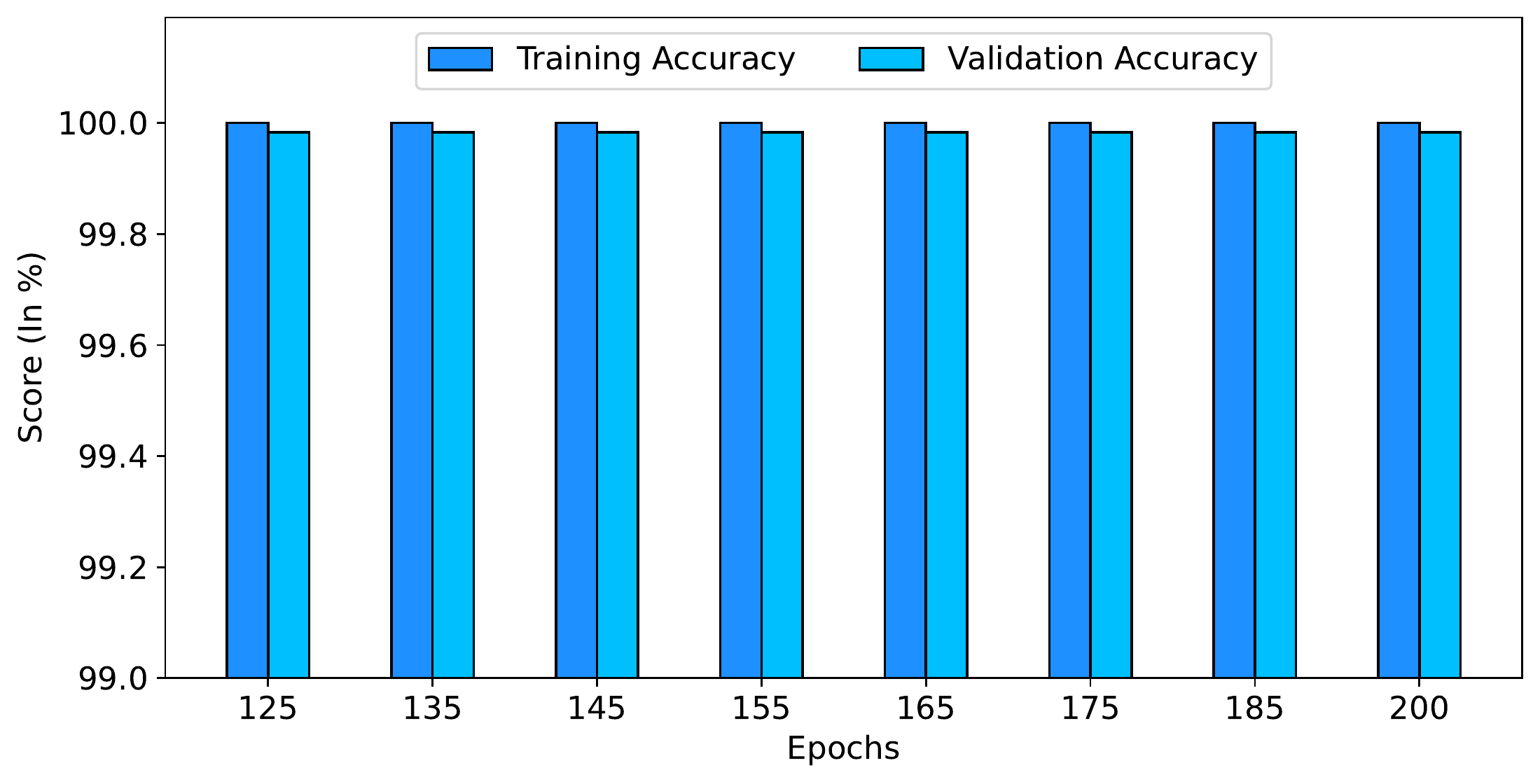}}\hspace{.1cm}
	\subfloat[Line chart]{\includegraphics[scale=.40]{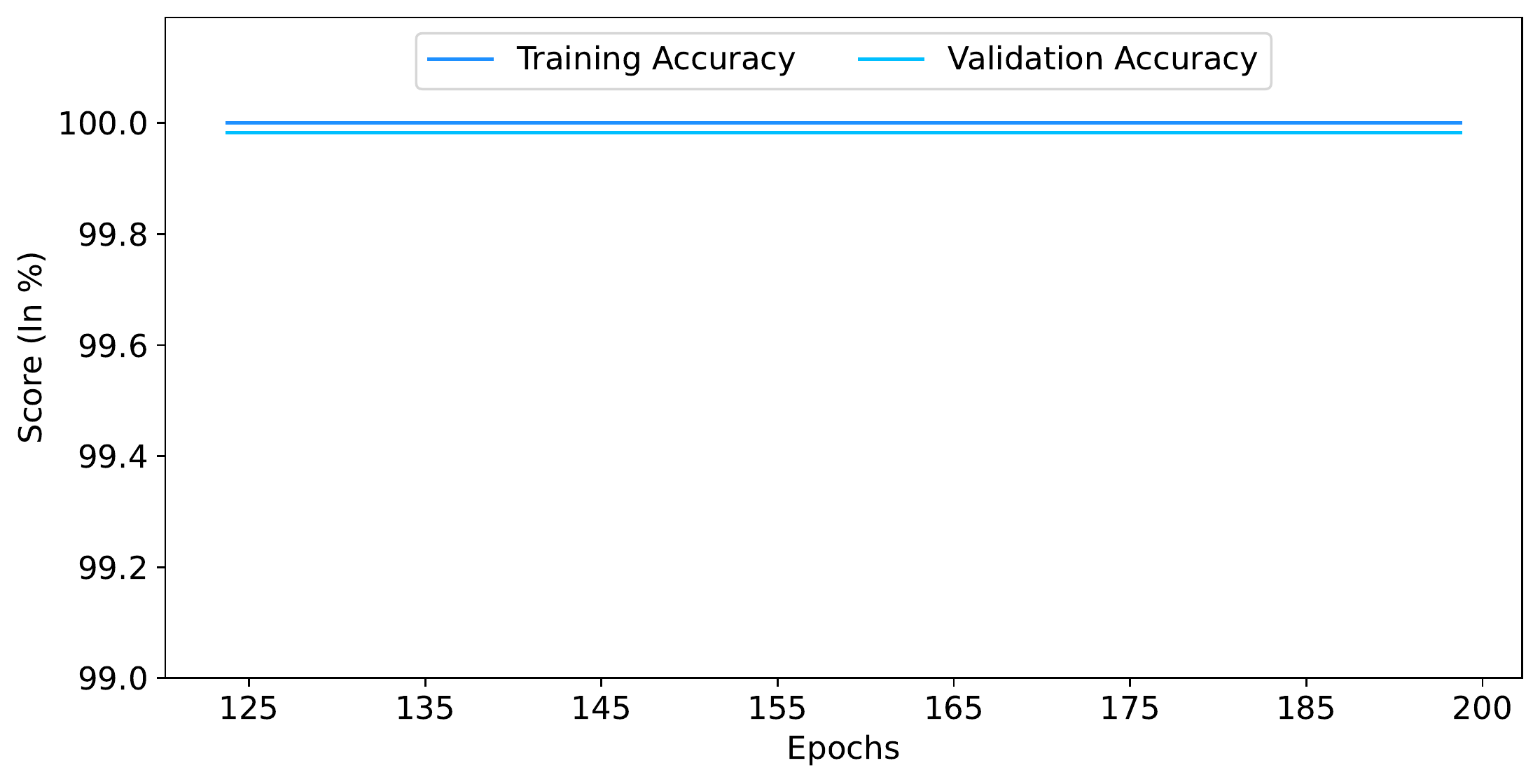}}\hspace{.1cm}
	\caption{The scalability of our proposed model.}
	\label{fig:ec_scalability}
\end{figure*}

Figure \ref{fig:ec_scalability} illustrates the proposed model's scalability.

\subsubsection{Discussion}

The comparison analysis with existing research work and our proposed model is shown in Table \ref{fig:kddcup_comparision_analysis}and Table \ref{fig:malmem_comparision_analysis} for KDDCUP'99 and CIC-MalMem-2022 datasets, respectively. The comparison results of KDDCUP'99 show that our proposed model provides outperforms other existing works considering both binary and multilabel classification. Similarly, for the CIC-MalMem-2022 dataset, the comparison results with existing work prove that our proposed model provides a better performance accuracy rate to detect malware memory attacks. 

Hence, the performance results of the proposed model are compared to the recent research work ensuring that our proposed model gives a higher detection rate as well as provides the robustness and effectiveness to detect intrusion. Moreover, the aforementioned results showed that proper data preparation could assist to improve the accuracy of ML and DL algorithms, while the XGBoost technique can quickly discover the most promising features with greater accuracy outcomes.

\begin{table}[]
\centering
\resizebox{\textwidth}{!}{
\begin{tabular}{llllll}
\hline
SL.No. & Author & Feature Selection Method & Classification Algorithm & Selected Features & Performance (Accuracy In \%) \\ \hline
1 & \cite{tan2019wireless} & - & SMOTE+RF & All & 92.57 \\ 
2 & \cite{bhati2021improved} & - & XGBoost & All & 99.95 \\ 
3 & \cite{choudhary2020analysis} & - & DNN & All & 91.50 \\  
4 & \cite{hu2021identification} & -  & CNN + LSTM & All & 98.48 \\
5 & \cite{alqahtani2020cyber} & - & DT & All & 94.00 \\
6 & \cite{mahhizharuvi2021effective} & - & EMRFT & All & 96.56 \\
7 & \cite{norwahidayah2021performances} & PSO & ANN & 20 & 98.00 \\ 
8 & \cite{li2021feature} & CR & DNN & 30 & 99.40 \\ 
9 & \cite{narayanasami2021biological} & BAT & SVM & 25 & 94.12 \\
10 & \cite{mohammadi2019cyber} & FGCC+CFA & DT & 10 & 95.03 \\ 
11 & \cite{kshirsagar2021efficient} & IGR+CR+ ReF+SCS & PART & 12 & 99.32 \\ 
12 & \cite{mugabo2021intrusion} & EFS & RF & 15 & 93.90 \\ 
13 & \cite{talita2021naive} & PSO & NB & 38 & 99.12 \\ 
14 & Proposed Method (Binary) & SMOTE + XGBoost & RF & 20 & 99.99 \\ 
15 & Proposed Method (Multilabel) & SMOTE + XGBoost & RF & 20 & 99.99 \\ \hline
\end{tabular}%
}
\caption{Comparison analysis for KDDCUP'99 dataset.}
\label{fig:kddcup_comparision_analysis}
\end{table}

\begin{table}[]
\centering
\resizebox{\textwidth}{!}{%
\begin{tabular}{lllllll}
\hline
SL.No. & Author & Dataset & Feature Selection Method & Classification Algorithm & Selected Features & Performance (Accuracy In \%) \\ \hline
1 & \cite{icissp22} & CIC-MalMem-2022 & - & Stacked Ensemble & - & 99.00\\ 
2 & \cite{dener2022malware} & CIC-MalMem-2022 & - & LR & - & 99.97 \\ 
3 & \cite{louk2022tree}  & CIC-MalMem-2022 & - & RF & - & 100 \\ 
4 & Proposed Method & CIC-MalMem-2022 & XGBoost & RF, ANN & 20 & 100 (RF), 100 (ANN) \\ \hline
\end{tabular}
}
\caption{Comparison analysis for the CIC-MalMem-2022 dataset.}
\label{fig:malmem_comparision_analysis}
\end{table}


Our proposed model is compatible with both contemporary network topologies with NIDS system and also that enable intelligently centralised network management, such as software-defined networks (SDN). In SDN, the application layer consists of network applications associated with data security, where we can employ our proposed IDS model \citep{li2021challenge}. It, therefore, can interact with network resources via the Northbound Interface (NBI) to defend the infrastructure.
Control layer SDN operates as a centralised SDN controller hosted on a server to manage network traffic and policy \citep{yan2015software}. This controller is based on the model used in the application layer. By customising the network switch's security rule, malicious traffic can be regulated to meet the needs of different business networks with little effort. This permits flexible and fruitful regulation of network congestion. Our model will make SDN highly adaptable, manageable, flexible, and cost-effective. Therefore, it can be used in both high-bandwidth and dynamic systems for network security.



One dataset in this paper is concerned with a network attack (KDDCUP'99), while the other is concerned with a malware memory attack (CIC-MalMem-2022), so their characteristics couldn't be more dissimilar. With 44 features, the KDDCUP'99 dataset contains 494021 data points. The dataset is multiclass labeled and has some imbalances. In contrast, the CIC-MalMem-2022 dataset includes 58,596 data points across 57 features. There is a binary class label in this balanced dataset.

If we compare the proposed XGBoost-based feature selection process to others like PSO, BAT, IGR, CRC, ReF, EFS, and FGLCC+CFA, etc., we can find that it provides more value. When compared to XGBoost, the current feature selection process doesn't make use of its second derivative to reduce model error and L1 \& L2 regularisation to make it more generic \citep{chen2016xgboost, wang2020imbalance}. Furthermore, it's performance is enhanced by the stochastic gradient boosting algorithm \citep{kiangala2021effective}. Therefore, XGBoost aids in achieving superior feature selection capabilities, which in turn boost the model's performance more than competing methods.

\section{Conclusion}

The purpose of this study is to develop a novel, dependable, and effective network intrusion detection system. In order to achieve that, we developed a hybrid machine learning approach that included data balancing with SMOTE and the extraction of dominant features with XGBoost. ML and DL algorithms like RF, DT, KNN, MLP, CNN, and ANN were used to test and evaluate the proposed method in order to find the best model for detecting network intrusion. 
Several performance metrics were used to determine how well the binary and multiclass attack algorithms worked. The performance results show that among all ML and DL algorithms, RF has the highest accuracy rate of 99.99\% with the chosen features for the KDDCUP'99 dataset and 100\% for the CIC-MalMem-2022 dataset. 
Furthermore, performance comparison with previous research demonstrates the model's dependability and robustness over others. 

Overall, the key findings of this paper are summarized as follows: 
\begin{itemize}

 \item Our proposed hybrid model performs significantly better than traditional ML and DL-based models because it integrates effective preprocessing, feature scaling, and feature selection.

 \item In the traditional approach, often one class predicts more accurately than others due to data imbalance, resulting in an imbalanced classification model. SMOTE data balance, in our approach, helps produce a better prediction model by removing the negative effect of class imbalance.

 \item SMOTE helps balance the dataset, which leads to a more accurate prediction model by reducing type-1 (false positive) and type-2 (false negative) errors. In general, the negative effect of a model's performance due to the data imbalance is clear when examining the confusion matrix. When true positives and negatives are larger than false positives and false negatives, this demonstrates that the model can only predict a particular class label. If new data is tested on the class imbalance model, performance will be significantly reduced due to this issue. Using SMOTE in our approach reduces type-1 and type-2 errors.

 \item The use of XGBoost extracts dominant features to improve the accuracy. XGBoost uses extra precise approximate to find the best model tree. It also uses the second derivative to minimize model error and L1 \& L2 regularization to generalize it. The stochastic gradient boosting algorithm within it helps to improve accuracy. XGBoost also has an inbuilt process to get feature importance for feature selection by relating multiple variables to feature importance. Overall, it contributes to achieving excellent feature selection capabilities to improve the model's performance.
 
 \item Such a hybrid pipeline can reliably offer a superior detection rate along with the model's availability and scalability. 

\end{itemize}

Overall, our proposed approach is dependable and can be implemented in real-time by installing this model on internet-connected IDS devices. 
In future, we will investigate our model’s performance with any emerging threats upon the availability of the latest datasets. To further enhance the performance of the intrusion detection system, we will also compare our model's results with those of the ensemble feature selection method, which uses the union and intersection of features. This will assist in identifying relationships for selecting the most important features to use in the deep neural network.



\bibliographystyle{vancouver}
\bibliography{bibs}

\end{document}